# On thermal effects in solid state lasers:

# the case of ytterbium-doped materials

Sébastien Chénais[1], Frédéric Druon, Sébastien Forget[1], François Balembois and Patrick Georges.

Laboratoire Charles Fabry de l'Institut d'Optique, CNRS, Université Paris-Sud
Campus Polytechnique
RD128
91127 Palaiseau cedex
Tel : +(33) 1 69 35 88 52; fax : +331 6935 8807 ;

e-mail : chenais@galilee.univ-paris13.fr, frederic.druon@institutoptique.fr

**Abstract:** A review of theoretical and experimental studies of thermal effects in solid-state lasers is presented, with a special focus on diode-pumped ytterbium-doped materials. A large part of this review provides however general information applicable to any kind of solid-state laser. Our aim here is not to make a list of the techniques that have been used to minimize thermal effects, but instead to give an overview of the theoretical aspects underneath, and give a state-of-the-art of the tools at the disposal of the laser scientist to measure thermal effects.

After a presentation of some general properties of Yb-doped materials, we address the issue of evaluating the temperature map in Yb-doped laser crystals, both theoretically and experimentally. This is the first step before studying the complex problem of thermal lensing (part III). We will focus on some newly discussed aspects, like the definition of the thermo-optic coefficient: we will highlight some misleading interpretations of thermal lensing experiments due to the use of the dn/dT parameter in a context where it is not relevant. Part IV will be devoted to a state-of-the-art of experimental techniques used to measure thermal lensing. Eventually, in part V, we will give some concrete examples in Yb-doped materials, where their peculiarities will be pointed out.

---

[1] Current address : Laboratoire de Physique des Lasers, Institut Galilée, Université Paris Nord, 99 avenue Jean-Baptiste Clément, 93430 Villetaneuse, France.

# Contents:

# I. Introduction

## *I.1. The Yb$^{3+}$ breakthrough*

Diode-pumped solid-state laser (DPSSL) technology has become a very intense field of research in Physics [1,2]. The replacement of flash-lamp pumping by direct laser-diode pumping for solid-state materials has brought a very important breakthrough in the laser technology in particular for high power lasers [3-4]. In fact, the better matching between absorption wavelength and material's absorption spectra brought by the use of laser diode emission — compared to the broad one of flash-lamps — has lead to a significant benefit in efficiency and subsequently in simplicity, compactness, reliability and cost. This progress has substantial implications on laser applications such as fundamental and applied research, laser processing, medical applications …

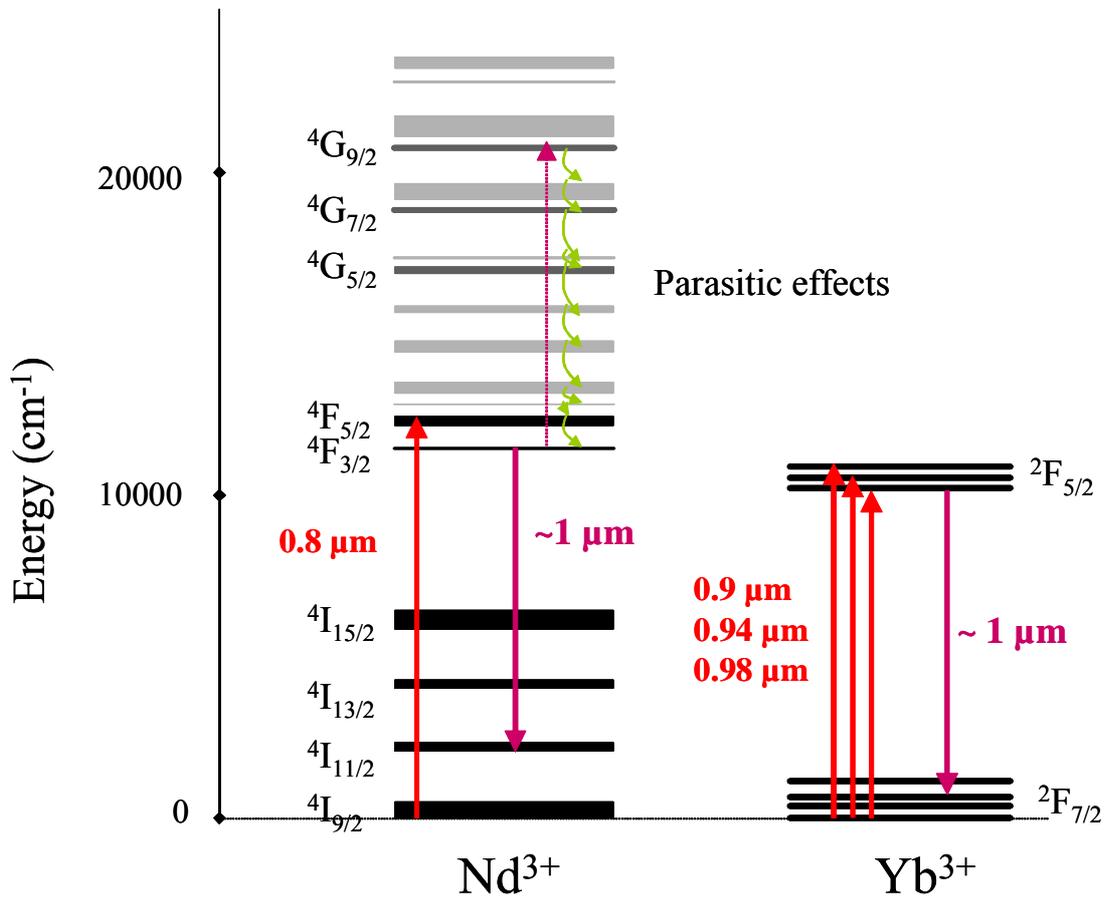

**Figure 1:** *Energy levels of Yb and Nd ions. Typical laser transition lines are represented for both pump absorption and laser emission. Lines of high excited states are also represented including the lines involved in the deleterious effects (up-conversion, excited-state absorption or concentration quenching).*

In the realm of high average power DPSSL, two rare-earth ions dominate: neodymium and ytterbium [5-6]. Actually they can be efficiently pumped, respectively at 808 nm with InGaAsP/GaAs or AlGaAs/GaAs diodes for neodymium, and between 900 and 980 with InGaAs/GaAs diodes for ytterbium (fig. 1). In both case the standard laser emission is around 1 μm, corresponding to the transition between the $^4F_{3/2}$ and $^4I_{11/2}$ lines for the $Nd^{3+}$ and between the $^2F_{5/2}$ and $^2F_{7/2}$ for the $Yb^{3+}$. At the beginning of the high-power-laser development, the Nd-doped materials were preferred to the Yb-doped ones mainly because of the four level nature and their many absorption lines, which are more convenient as far as flash-lamp pumping is concerned. However it seems obvious, for more than one decade now, that Yb-doped materials are more suited for very efficient and very-high-average-power diode-pumped lasers. The main reason for this is the very simple electronic level structure of the $Yb^{3+}$ ion, which consists on two manifolds as shown in figure 1. This singular property allows avoiding *most* of the parasitic effects such as upconversion, cross relaxation or excited-state absorption which are present in Nd-doped materials [7] because of the existence of higher excited-state levels ($^4G_{9/2}$ for the 1-µm-laser emission). These deleterious effects have two main consequences. First, they increase the thermal load and subsequently the thermal problems [8] because the main desexcitation paths of the high-excited state levels are non-radiative (as represented in fig. 1). Secondly, they also alter the gain because they can induce strong depopulation of the $^4F_{3/2}$ level implicated in the laser inversion population. Another advantage of Yb-doped materials compared to their neodymium doped counterparts is the very low quantum defect (again due to the 2-manifold based electronic structure). In fact, when pumped at 980 nm the quantum defect of ytterbium is around 5 % compared to 30 % for neodymium (in YAG). This is a real benefit for reducing the thermal problems and thus to attain very high average powers. As an example of comparison between Nd and Yb doped materials, we summarized in table 1, the different parameters for the same well-known matrix host: YAG ($Y_3Al_5O_{12}$) [**9-13**].

**Table 1:** *comparison between Nd:YAG and Yb:YAG*

| Crystal | Nd:YAG | Yb:YAG | |
|---|---|---|---|
| Emission line<br>Wavelength<br>Cross section<br>Broadness (FWHM) | 1064 nm<br>28 $10^{-20}$ cm$^{-2}$<br>0.8 nm | 1031 nm<br>2.1 $10^{-20}$ cm$^{-2}$<br>9 nm | |
| Lifetime | 230 µs | 951 µs | |
| Saturation fluence | 0.67 J/cm$^2$ | 9.2 J/cm$^2$ | |
| Maximum doping rate | 2 % | 100 % | |
| Absorption line<br>Wavelength<br>Cross section<br>Broadness (FWHM) | 808 nm<br>67 $10^{-20}$ cm$^{-2}$<br>2 nm | 968 nm<br>0.7 $10^{-20}$ cm$^{-2}$<br>4 nm | 942 nm<br>0.75 $10^{-20}$ cm$^{-2}$<br>18 nm |
| Quantum defect | 32 % | 6.5 % | 9.5 % |

## *I.2. Strategy on the matrix host*

Another advantage of Yb-doped versus Nd-doped materials is the longer lifetime which may allow a better storage of the pump energy; and the last but not the least advantage is the generally broader bandwidth of the emission lines. This last advantage leads to a potential for femtosecond pulse generation which, in the current state-of-the-art has never been demonstrated with neodymium. The emergence of ytterbium-based lasers has allowed crucial progress in the ultrashort-pulsed laser technology. These materials have been actually the key point for the development of the latest generation of "ultrafast" lasers: the all-solid-state femtosecond lasers [14-36]. Applications for such lasers are abundant and excite a great interest in the scientific community.

However Yb-doping brings several drawbacks or difficulties. The first one is the very strong influence of the matrix on the spectral properties. Actually, as the two levels $^2F_{5/2}$ and $^2F_{7/2}$ are split in manifolds by the Stark effect due to the electric crystalline field of the host matrix, the ion environment strongly models the spectrum. In a simple way, the spectral broadening can be directly related to the level of disorder of the matrix [37-48]. On the first hand, if the matrix is relatively simple and well-ordered, the spectra would reveal relatively narrow and intense lines (which are a strong disadvantage for short pulse generation). Though, a simple matrix structure generally implies a high thermal conductivity which is a key point for developing high power lasers. An example of such an Yb-doped material is given in figure 2 with Yb:YAG. On the second hand, if the disorder of the host matrix is high, the spectrum will be large and suitable for very-short pulse generation but at the expanse of thermal conductivity. An example of such an Yb-doped material is given in figure 3 with Yb:SYS. The numerous advantages of the $Yb^{3+}$ ion have led to a strong interest for many host matrices but in general favouring either short pulses or high power applications. Table 2 represents this diversity of already studied Yb-doped host matrices and their principal properties.

**Table 2:** *Comparison between Yb-doped crystals*

| Material (name and formula) | Emission line | | | Lifetime (ms) | Absorption line Usual wavelength (nm) | Thermal conductivity (undoped) (W/m/K) |
|---|---|---|---|---|---|---|
| | Wavelength (nm) | Cross section $10^{-20}$ cm$^2$ | Broadness (nm) | | | |
| Yb:YAG Yb:Y$_3$Al$_5$O$_{12}$ | 1031 | **2.1** | 9 | 0.951 | 942 968 | **11** |
| Yb:GGG Yb:Gd$_3$Al$_5$O$_{12}$ | 1025 | **2** | 10 | 0.8 | 971 | **8** |
| **Yb:Y$_2$O$_3$** | 1076 | 0.4 | 14.5 | 0.82 | 979 | **13.6** |
| **Yb:Sc$_2$O$_3$** | 1041 | **1.44** | 11.6 | 0.8 | 979 | **16.5** |
| **Yb:CaF$_2$** | 1045 | 0.25 | **70** | 2.4 | 979 | **9.7** |
| **Yb:YVO$_4$** | 1020 | 0.9 | **40** | 0.25 | 985 | **5.1** |
| Yb:LSO Yb:Lu$_2$SiO$_5$ | 1040 | 0.6 | 35 | 0.95 | 978 | **5.3** |
| Yb:YSO Yb:Y$_2$SiO$_5$ | 1042 | 0.6 | **40** | 0.67 | 978 | 4.4 |
| Yb:YLF Yb:YLiF$_4$ | 1030 | 0.81 | 14 | 2.21 | 940 | 4.3 |
| Yb:KGW Yb:KGd(WO$_4$)$_2$ | 1023 | **2.8** | 20 | 0.3 | 981 | 3.3 |
| Yb:KYW Yb:KY(WO$_4$)$_2$ | 1025 | **3** | 24 | 0.3 | 981 | 3.3 |
| Yb:SYS Yb:SrY$_4$(SiO$_4$)$_3$O | 1065 | 0.44 | **73** | 0.82 | 980 | 2 |
| Yb:GdCOB Yb:Ca$_4$GdO(BO$_3$)$_3$ | 1044 | 0.35 | **44** | 2.6 | 976 | 2.1 |
| Yb:BOYS Yb:Sr$_3$Y(BO$_3$)$_3$ | 1060 | 0.3 | **60** | 1.1 | 975 | 1.8 |
| Yb:glass (phosphate glass) | 1020 | 0.05 | **35** | 1.3 | 975 | 0.8 |

Another drawback of the Yb-doped materials is due to the quasi-3-level structure of these lasers. As it is apparent on the spectra of figure 2 and 3, there is an overlap between the emission and the absorption bands which leads to strong re-absorption effects and to a reduction of the effective emission band broadness. Moreover, since the splitting due to Stark effect is relatively small (between 200 and 1000 cm$^{-1}$), the high energy levels within the $^2$F$_{7/2}$ manifold (corresponding to the different possible low-energy levels of the laser transition) are somewhat populated at thermal equilibrium. This implies two deleterious effects when temperature increases: first, a reduction of the laser inversion population, second, an increase of the reabsorption at the laser wavelength. A special care concerning the thermal load and thermal management will be then necessary to develop efficient lasers based on ytterbium-doped materials, especially in the high power regime.

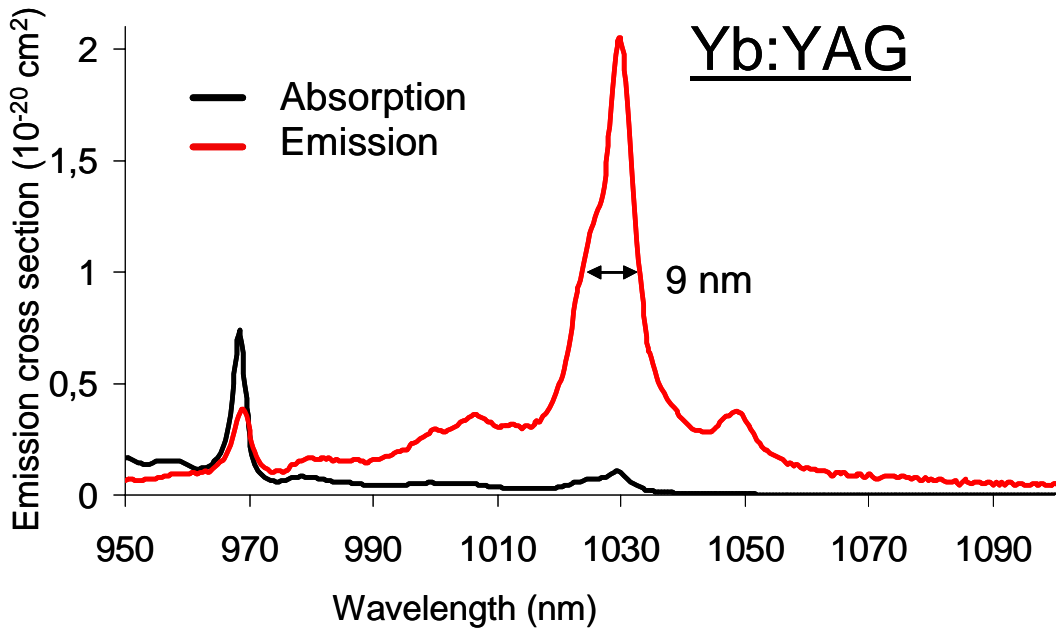

**Figure 2** : *Absorption and emission spectra of Yb:YAG.*

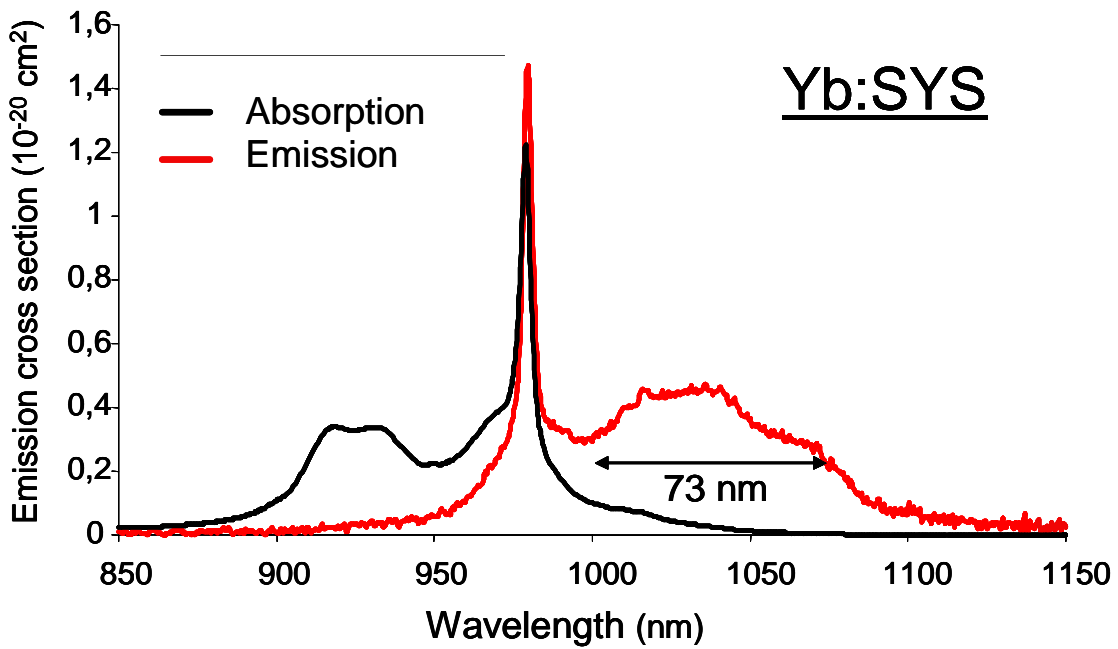

**Figure 3 :** *Absorption and emission spectra of Yb:SYS.*

As a first conclusion, Ytterbium-doped crystals are particularly suitable for directly diode-pumped, solid-state high-power and/or femtosecond lasers. The emergence of new ytterbium-doped laser crystals has allowed crucial progress in the DPSSL technology. Nevertheless, a very special care has to be done concerning the thermal properties and thermal effects in the Yd-doped materials because of their strong influence on the laser performance. In this paper, we then propose to make a review of different thermal-effect studies made on ytterbium-doped laser crystals.

# II. Temperature profile of an ytterbium-doped crystal under diode pumping

In this part we present a review about how to *calculate* and *measure* the temperature distribution in an end-pumped laser crystal. We explain in which cases it is possible to obtain an analytical expression (otherwise a finite-element analysis would be necessary), and how these well-known results have to be corrected when we deal with *ytterbium-doped crystals*, because absorption saturation cannot be ignored in this case. Then, we investigate the role of the thermal contact at the boundaries, which is an essential parameter for the knowledge of the temperature. This will be illustrated, in the last part of this section, by experimental absolute temperature maps, obtained with an infrared imaging camera.

## II.1. Theoretical aspects

### II.1.1. The steady-state heat equation

A study of thermal effects in crystals first requires the calculation of the temperature field at any point of the crystal. One has to solve the heat equation:

$$\rho c_p \frac{\partial T(x,y,z,t)}{\partial t} - K_c \nabla^2 T(x,y,z,t) = Q_{th}(x,y,z,t) \qquad \text{(II.1.1.)}$$

with :  - T = T(x,y,z,t) : temperature in K;

- $\rho$ : density in kg.m$^{-3}$;

- $c_p$ : specific heat in J.kg$^{-1}$. K$^{-1}$;

- $K_c$ : thermal conductivity in $W.m^{-1}.K^{-1}$;

- $Q_{th}$ : thermal power (or thermal load) per unit volume in $W.m^{-3}$.

The specific heat affects the temperature variation in the pulsed regime or in the transient regime: we will ignore it in the following work since we will only consider CW lasers. The thermal conductivity governs the temperature *gradient* inside the crystal, and will have a crucial importance for the thermal lens magnitude. The heat transfer coefficient H is arising when writing the boundary conditions, and has then an influence only on the absolute value of the temperature inside the crystal. We will discuss at the end of this part how to measure it, and some ways to improve its value.

In order to obtain analytical expressions, some assumptions have to be made. We will assume the following: (1) The pump profile is axisymmetric. End-pumping by a fiber-coupled diode is a good example of such a profile; (2) the thermal conductivity $K_c$ is a scalar quantity, not a tensor; this means that we restrict our discussion to glasses and cubic crystals [49] (in practice however the anisotropy of the thermal conductivity tensor is often weak); (3) the cooling is isotropic in the z-plane, which means that the crystal mount does not favour one given direction of cooling. (4) At last we assume that the thermal conductivity is not significantly dependant on temperature, so that it can be considered as a constant. This approximation is very realistic in YAG around room temperature, following the study of Brown *et al.* [50], and we assume that it is also true in other crystals. However this approximation would not be valid anymore at cryogenic temperatures.

The heat equation, including all these assumptions, becomes:

$$\frac{1}{r}\frac{\partial}{\partial r}\left(r\frac{\partial T}{\partial r}\right)+\frac{\partial^2 T}{\partial z^2}=-\frac{Q_{th}(r,z)}{K_c} \qquad (II.1.2.)$$

when r is the radial coordinate of a point inside the crystal, measured with respect to the pump distribution axisymmetric axis. To simplify a bit further the equations, the crystal will be considered cylindrical, whose axis corresponds to the pump symmetry axis, with a radius $r_0$ and a length L (see figure 4)

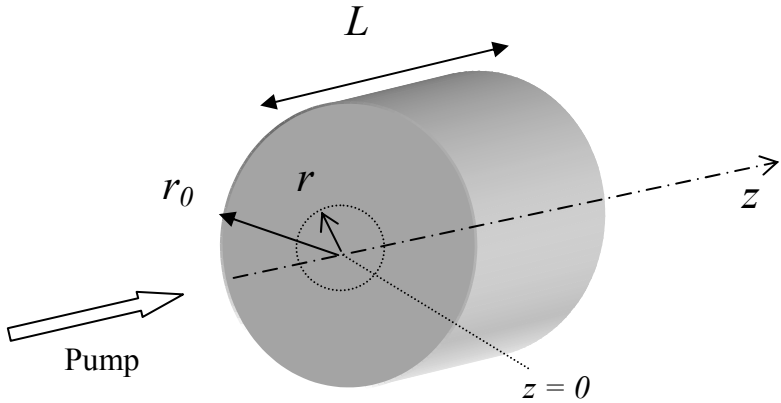

**Figure 4**: *geometry of the crystal, taken for all calculations. z=0 is the input face.*

## II.1.2. A review of analytical solutions of the steady state heat equation

An analytical expression of the temperature distribution inside a crystal is calculable only in a limited number of simple systems. For more complex geometries, one has to use finite element codes. We present in this subsection a brief review of the cases where such an analytical treatment is feasible.

The first study of thermal effects in crystals has been provided by Koechner [51] in the early seventies. He considered flash-pumped Nd:YAG rods, within which the thermal load is uniform :

$$Q_{th} = \frac{P_{th}}{\pi r_0^2 L} \quad \text{(II.1.3.)}$$

where $P_{th} = \eta_h P_{abs}$ is the thermal power (in W) dissipated into the rod. $P_{abs}$ is the absorbed pump power and $\eta_h$ is the fractional thermal load. The solution writes:

$$T(r) = T(r_0) + \frac{\eta_h P_{abs}}{4\pi r_0^2 L K_c}(r_0^2 - r^2) \tag{II.1.4.}$$

where T(r$_0$) is the temperature at the edge of the crystal, which will be estimated later thanks to the boundary conditions. It is useful to write the temperature shift between the centre and the edge of the rod:

$$\Delta T = T(0) - T(r_0) = \frac{\eta_h P_{abs}}{4\pi L K_c} \tag{II.1.5.}$$

We note that ΔT is independent of the radius *r$_0$* of the crystal, but scales inversely with L.

The previous results can not be applied to end-pumping configurations, because in these latter cases the thermal load is localized within a small volume inside the crystal. In a vast majority of practical circumstances, the pump beam profile is axisymmetric and can be described by a super-gaussian shape. The general solution of the heat equation for a super-gaussian beam of any order has been derived by Schmid *et al.* [52].

The situation is even simpler in most cases: indeed, the pump is often either a near-diffraction-limited Gaussian beam (laser or single mode diode), or a "top-hat" beam (that is a super-gaussian profile of infinite order). The latter description corresponds quite well to fiber-coupled diode laser array pumping.

The solution of the heat equation in the specific case of Gaussian-beam pumping is treated in [53] and [54]. The case of the top-hat shape has been derived by Chen *et al.* [55].

Hereafter we give the solution for a top-hat beam profile. Assuming that the thermal load in each slice at z coordinate is a disk of radius *w$_p$*(z), the temperature shift with respect to the edge temperature is:

$$T(r,z) - T(r_0,z) = \frac{\eta_h P_{abs}}{4\pi K_c} \frac{\alpha_{NS} e^{-\alpha_{NS} z}}{1 - e^{-\alpha_{NS} L}} \times \begin{cases} ln\left(\frac{r_0^2}{w_p^2(z)}\right) + 1 - \frac{r^2}{w_p^2(z)} & r \leq w_p(z) \\ ln\left(\frac{r_0^2}{r^2}\right) & r > w_p(z) \end{cases} \quad (\text{II.1.6.})$$

where:

- $\alpha_{NS}$ is the (non-saturated) linear absorption coefficient; saturation absorption is then *not* taken into account in this formula ;

- The axial heat flux (along z) is ignored, which means in other words that $\frac{\partial^2 T}{\partial z^2}$ is neglected in the heat equation. We'll see in the next section the reasons why we can neglect axial heat flux. The formula is then not valid for thin disks.

- As a consequence of the latter point, the temperature can be computed inside each "slice" of thickness *dz* of the crystal, as if the surrounding slices did not exist.

The temperature gradient inside the pumped volume is of particular interest because the laser beam is usually (and preferably) smaller than the pump volume. One obtains:

$$\Delta T(r,z) = T(r=0,z) - T(r,z) = \frac{\eta_h P_{abs}}{4\pi K_c} \frac{\alpha_{NS} e^{-\alpha_{NS} z}}{1 - e^{-\alpha_{NS} L}} \frac{r^2}{w_p^2(z)} \quad (\text{II.1.7.})$$

The temperature shift turns out to be independent of the crystal radius $r_0$ and of the crystal length L, which was not the case in Koechner's simple model (equation II.1.4). It makes sense since the important parameter here is the absorption length $L_{abs} = 1/\alpha_{NS}$, and not the whole length L of the crystal.

In figure 5 we plotted the normalized temperature distribution for a typical ratio $w_p/r_0 = 0.1$.

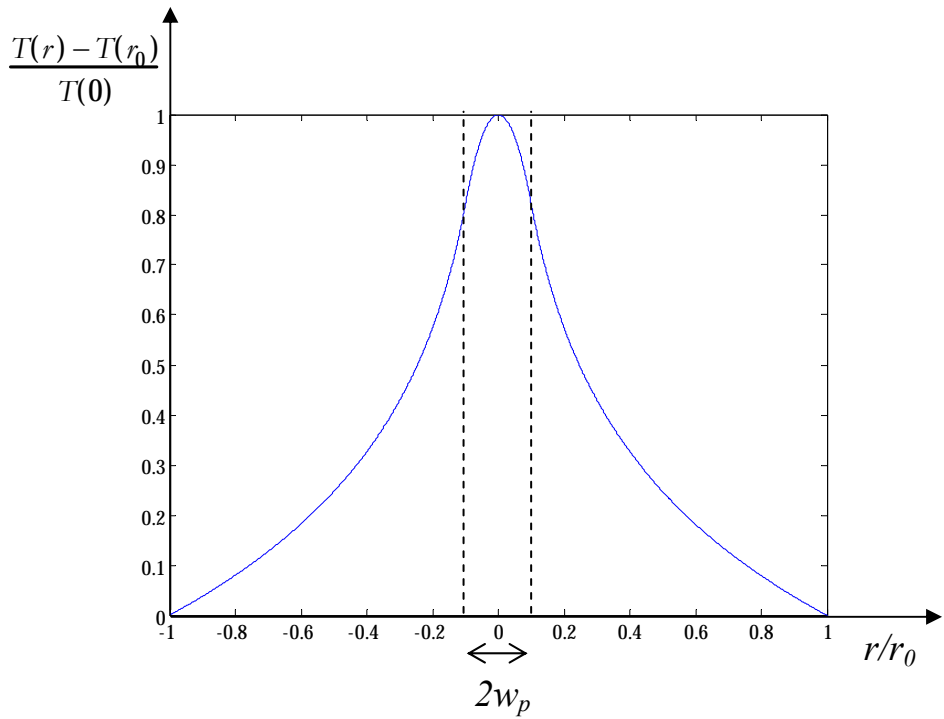

**Figure 5:** *Normalized temperature distribution (in a plane perpendicular to the propagation axis) for a crystal pumped by a top-hat-profile fiber-coupled laser diode with $w_p/r_0 = 0.1$).*

**II.1.3. What is special about ytterbium-doped materials? The influence of absorption saturation in the temperature distribution.**

An ytterbium-doped material, especially when pumped at the zero-line wavelength (i.e. around 980 nm), has many common points with a saturable absorber. The absorption rate due to the pump is counterbalanced by the spontaneous emission rate, but also (which is far from being negligible) by the stimulated emission rate at the pump wavelength. It is essential to take absorption saturation effects into account; otherwise the absorbed pump power can be dramatically overestimated. As a result, the absorbed pump power is lower under nonlasing than under lasing conditions, since lasing provides (hopefully) a very efficient path to carry the excited population back to the fundamental level.

It is noteworthy that most Finite Element Analysis (FEA) codes (primarily designed for 4-level laser systems in which absorption saturation is not a problem) basically assume an exponential decay for the pump power inside the crystal. It can lead to large errors, as we illustrate below.

Let $P_p(z)$ be the pump power through a plane in the crystal at $z$ coordinate. The thermal load density generated into a disk of radius $w_p(z)$ and thickness $dz$ is:

$$Q_{th}(z) = \frac{-\eta_h dP_p(z)}{\pi w_p^2(z) dz} \tag{II.1.8}$$

where $-dP_p(z)$ represents the absorbed pump power in the thin slice of thickness $dz$.

The temperature field is:

$$T(r,z) - T(r_0,z) = \frac{-\eta_h \frac{dP_p(z)}{dz}}{4\pi K_c} \times \begin{cases} \ln\left(\frac{r_0^2}{w_p^2(z)}\right) + 1 - \frac{r^2}{w_p^2(z)} & r \leq w_p(z) \\ \ln\left(\frac{r_0^2}{r^2}\right) & r > w_p(z) \end{cases} \tag{II.1.9.}$$

Inside the pumped volume the temperature shift writes as follows:

$$T(0,z) - T(r,z) = \frac{-\eta_h \frac{dP_p(z)}{dz}}{4\pi K_c} \frac{r^2}{w_p^2(z)} \tag{II.1.10.}$$

Absorption saturation issues are taken into account, under nonlasing conditions, by the following equation for the pump irradiance $I_p$ (which is the pump power divided by the pump spot area):

$$\frac{dI_p}{dz} = \frac{-a_{NS} I_p}{1 + \frac{I_p}{I_{p_{sat}}}} \tag{II.1.11.}$$

Where $\alpha_{NS}$ is the absorption coefficient in the non saturated regime.

The pump saturation irradiance $I_{p_{sat}}$ is calculated from the spectroscopic properties of the material:

$$I_{p_{sat}} = \frac{hc}{\lambda_p [\sigma_{abs}(\lambda_p) + \sigma_{em}(\lambda_p)] \tau} \tag{II.1.12.}$$

where $\sigma_{abs}$ is the absorption cross section, $\sigma_{em}$ is the emission cross section, $\lambda_p$ the pump wavelength, and $\tau$ the radiative lifetime.

The pump power $P_p(z)$ obeys the following equation, for a top hat beam profile (one may find the equivalent formulation for a gaussian pump profile in [56]):

$$\frac{dP_p(z)}{dz} = \frac{-a_{NS} P_p(z) I_{p_{sat}} \pi w_p^2(z)}{P_p(z) + I_{p_{sat}} \pi w_p^2(z)} \qquad (II.1.13.)$$

A practical way to study absorption saturation issues, and to check the assumptions made so far, is to perform fluorescence imaging experiments in a pumped crystal. Using a crystal whose one of the edge surfaces (in practice one side not facing the radiator) has been polished, one can make an optical image of the fluorescence, under lasing or nonlasing conditions, with a CCD camera and an interference filter at a long wavelength (at 1064 nm for instance), required to completely eliminate the scattered light at the pump wavelength, as well as to prevent detection of fluorescence photons which could have experienced reabsorption. This simple experiment allows visualizing absorption saturation (the fluorescence intensity, integrated along the depth of focus of the imaging system, does not decay exponentially) and also to measure what is the optimum location for the pump spot inside the crystal (figure 6). The experiments we performed with different Yb-doped materials taught us that the optimum focus (the one for which the measured laser efficiency was the highest) was always located at about one third of the whole crystal length from the input face. This parameter is taken into consideration in the following.

The low brightness of the diode pump beam (compared to the brightness of the laser beam) makes the effective Rayleigh distance of the pump beam considerably shorter than the crystal length. For this reason, the divergence of the pump beam inside the crystal *must* also be considered, in order to correctly account for saturation issues. Here we describe the pump radius evolution by a relation of the type:

$$w_p(z) = w_{p_0}\sqrt{1+\left(\frac{M^2\lambda_p(z-z_0)}{n\pi w_{p_0}^2}\right)^2} \qquad (\text{II.1.14.})$$

where $w_{p_0}$ is the pump beam waist radius. The $M^2$ factor is determined experimentally. In our case we used a 200μm-diameter core fiber-coupled diode (HLU15F200-980 from LIMO GmbH), whose the $M^2$ was measured to be around 80.

Results shown in figure 6 show experimental data and theoretical predictions in a 15%-at. doped Yb:GdCOB crystal [57]. The theoretical profiles are computed assuming that: 1) the pump volume has a top-hat profile, and 2) the imaging objective has a very low numerical aperture, so that the rate of spontaneous photons detected by one pixel can be calculated by integrating the fluorescence yield over one vertical line underneath. The good match between theory and experiments show incidentally that the "top hat" hypothesis for the pump beam profile is well justified.

**Without saturation absorption :**
**At low power**
($P_{inc}$ = 1 W
$P_{abs}$ =200 mW)

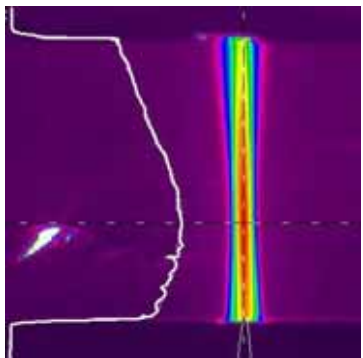
experiment

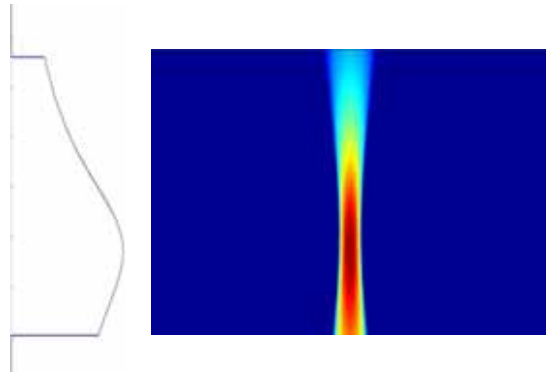
simulation

**With saturation absorption :**
**At high power**
($P_{inc}$ = 13.7 W
$P_{abs}$ =6 W
$I_{psat}$ =4.1 kW/cm$^2$)

*Experimental profile measured along the symmetry axis*

*Theoretical profile computed along the symmetry axis*

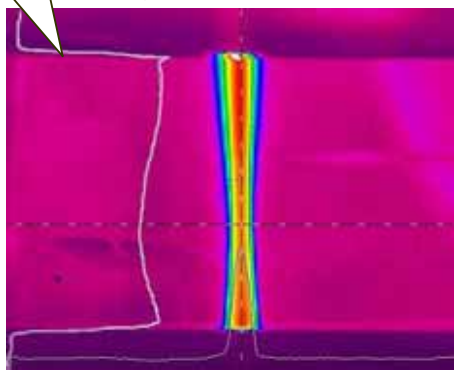
experiment

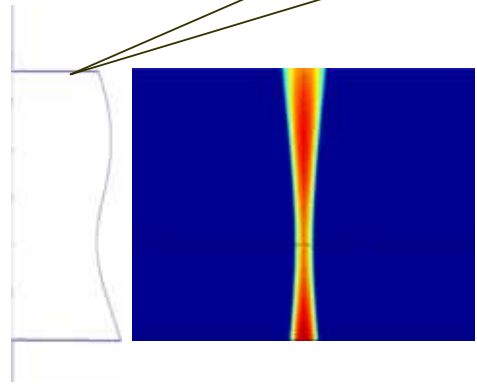
simulation

**Figure 6:** *Fluorescence detected @ 1064 nm on a crystal pumped at 980 nm at low power (top) and at high power (bottom), through the optically-polished top face. The influence of absorption saturation is clearly visible: at low pump power, the fluorescence yield is higher at the pump waist location, as expected provided that both absorption coefficient and absorption saturation are weak; on the contrary, when absorption saturation becomes non negligible, the amount of fluorescence photons is minimum at the pump waist. Theory and experiments agree very well, except near the exit face of the crystal, a discrepancy which could be related to the fact that far from the waist, the pump beam is no longer "top hat".*

**Fig 7a)**

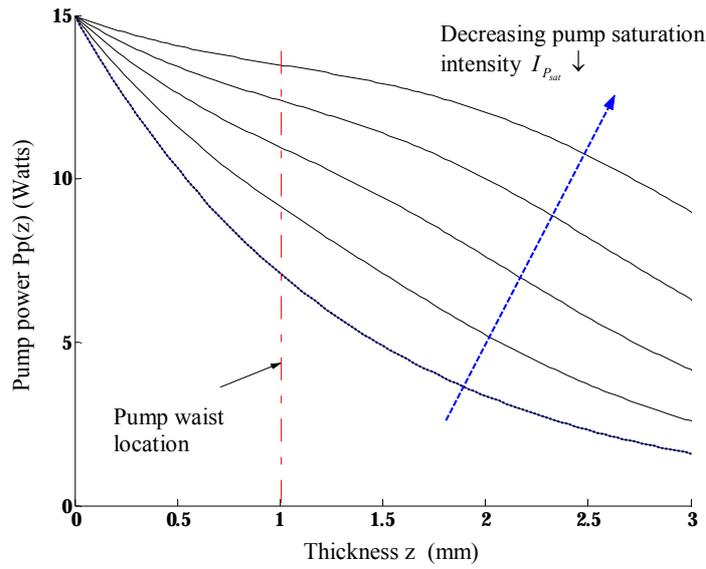

**Fig 7b)**

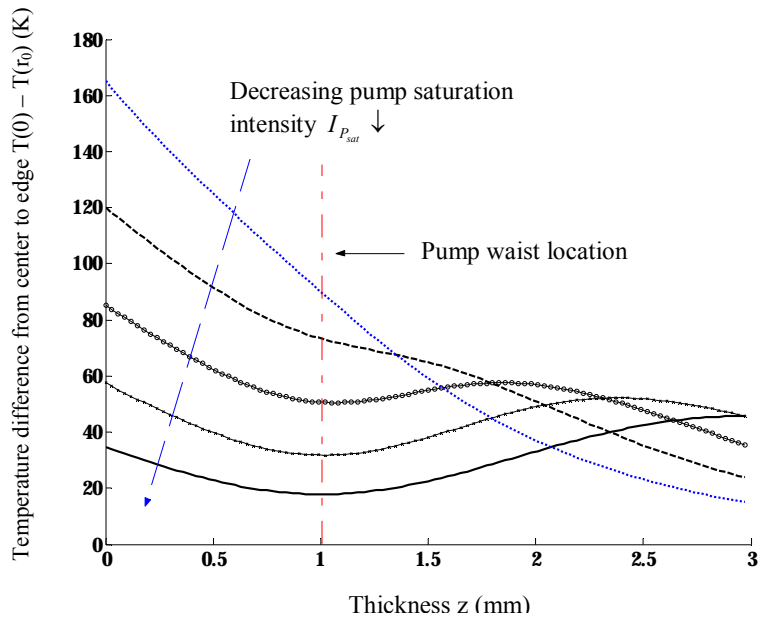

**Figure 7:** *Evolution of pump power (Fig. 7a) and temperature difference T(0) –T(r₀) (Fig. 7b) versus crystal thickness z. The pump saturation intensities values are: $I_{p_{sat}} = \infty$ - 50 - 20 - 10 - 5 kW/cm² (for these curves $\eta_h$= 0.065 et $K_c$ = 2 W.m⁻¹.K⁻¹, corresponding to the parameters of Yb:GdCOB).*

**Fig 8a)**

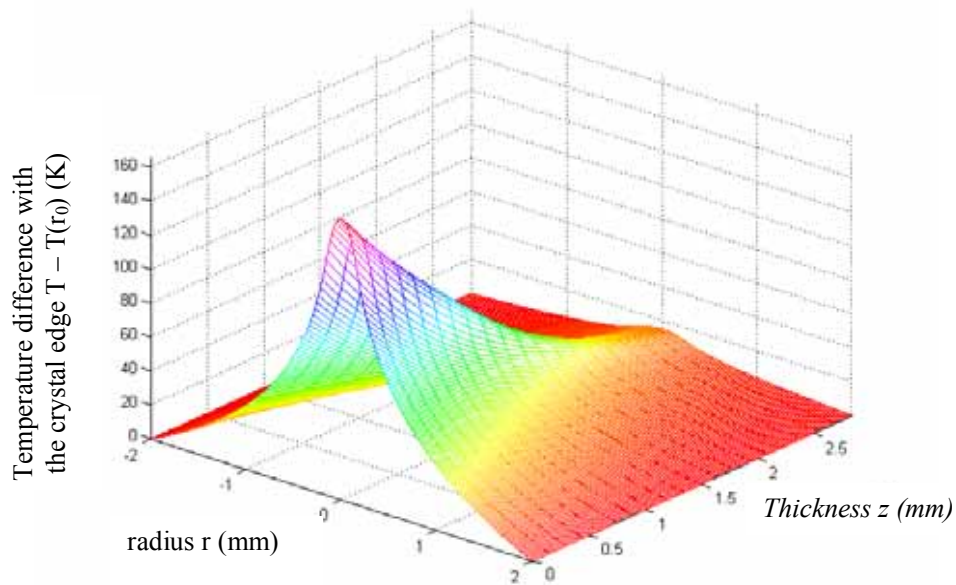

**fig. 8b)**

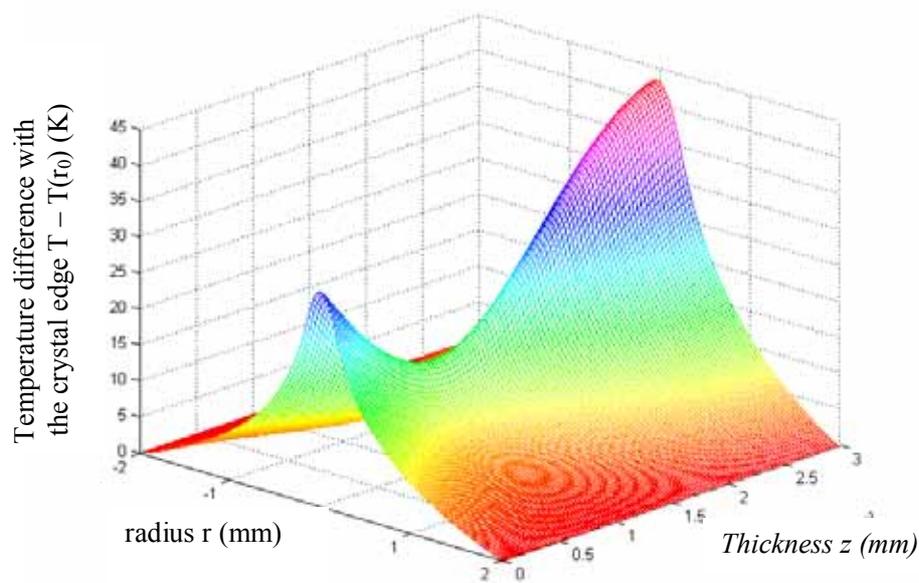

**Figure 8:** *Temperature distribution under nonlasing condition. The pump beam divergence inside the crystal is taken into account ($M^2 = 80$). In Fig 8a) the saturation of absorption is ignored; Fig*

*8b) pump absorption saturation is taken into account ($I_{P_{sat}}$ = 4.1 kW/cm²). The parameters used are form Yb:GdCOB.*

Equations (II.1.13) and (II.1.14.) can be solved numerically and injected in (II.1.9.) to obtain the temperature distribution. Figure 7 shows the evolution of pump power (fig. 7a) and temperature (fig. 7b) at the center of the rod versus crystal thickness, for various values of the pump saturation intensity.

Here we assumed that the pump beam waist was located at $z_0 = L/3$, which is experimentally well verified, as far as the laser output is optimized (see figure 6 and above text). In absence of saturation ($I_{p_{sat}}$ infinite), both the pump power and the temperature experience an exponential decay as expected; but for lower values of the pump saturation intensity, the temperature reaches a local minimum at the pump beam waist. Figure 8 shows a 3D view of temperature distribution without saturation (fig. 8a) and in presence of strong saturation (fig 8b.) corresponding to $Ip_{sat}$ = 4.1 kW/cm², that is the value for Yb:GdCOB. It appears in the latter case that the region where the pump density is the strongest (near the pump beam waist) is not the region where the temperature is the highest (near the faces of the crystal). Pump beam divergence appears to be an important parameter: it makes, for this example, the temperature higher at the exit face than at the entrance face of the crystal.

In presence of laser extraction, the pump intensity evolution through the crystal is given by:

$$\frac{dI_p}{dz} = -\frac{\alpha_{NS}^p + \alpha_{NS}^l \frac{I}{I_{min}}}{1 + \frac{I_p}{I_{p_{sat}}} + \frac{I}{I_{L_{sat}}}} I_p \quad \text{(II.1.15.)}$$

where

$$\alpha_{NS}^l = \sigma_{abs}(\lambda_l) N \quad \text{(II.1.16.)}$$

and

$$I_{L_{sat}} = \frac{hc}{\lambda_l \left[\sigma_{abs}(\lambda_l) + \sigma_{em}(\lambda_l)\right]\tau} \qquad (II.1.17.)$$

are the non-saturated absorption coefficient at laser wavelength, and the laser saturation intensity, respectively.

When the intracavity laser intensity $I$ largely exceeds $I_{L_{sat}}$, and if reabsorption at laser wavelength is small, one can show that (II.1.15) simply becomes:

$$\frac{dI_p}{dz} = -\alpha_{NS}^p \, I_p \qquad (II.1.18.)$$

which means that the ground manifold is repopulated so that absorption is not saturated any more. In real cases, as a matter of fact, the absorbed pump power under lasing conditions is *intermediate* between the non saturated regime and the saturated (non lasing) regime: in a first approximation it is possible to ignore saturation effects only if the laser extraction is efficient.

### *II.1.4. Determining the absolute temperature: the boundary conditions*

In this subsection we deal with the boundary problem. For the moment we have established expressions for the temperature gradient, but we have no idea of the absolute temperature inside the crystal. Let us assume that the four edge faces of the crystal are in "contact" with a radiator, which will be in most cases a piece of cooled copper. The first boundary condition expresses the continuity of the thermal flux across these contacts:

$$K_{crystal}\left(\frac{\partial T}{\partial n}\right)_{inside\ the\ crystal} = K_{copper}\left(\frac{\partial T}{\partial n}\right)_{inside\ copper} \qquad (II.1.19.)$$

where K is the thermal conductivity, **n** is the surface normal vector, and $\partial/\partial n$ the normal derivative. Common metals (Copper or indium, the latter being used as an intermediate contacting material) have thermal conductivities that are several orders of magnitude higher than the usual conductivities of laser crystals: 400 W.m$^{-1}$.K$^{-1}$ for copper and 820 W.m$^{-1}$.K$^{-1}$ for indium. This means that the temperature gradients inside these metals will always be negligible, so that we consider in the following that the temperature inside the radiator is uniform and is noted $T_c$.

Let's see now the second boundary condition. In many papers and FEA codes, the temperature at the edge of the crystal is set equal to $T_c$:

$$T(r_0) = T_c \qquad (II.1.20.)$$

This is actually true only for an ideal contact [58]. But even for flat and polished surfaces pressed one against another, this relation is far from reality [59].

The most realistic condition is surprisingly a Newton-type law of cooling, even if we indeed deal with *conduction* problems here:

$$\mathbf{j_q} \cdot \mathbf{n} = -K_c \frac{\partial T}{\partial n} = H(T(r_0) - T_c) \qquad (II.1.21.)$$

where $\mathbf{j_q}$ is the thermal density flux. *H* is the heat transfer coefficient or surface conductance (W.cm$^{-2}$.K$^{-1}$). *H* is of course infinite for ideal thermal contact.

Carslaw *et al.* [58] have shown that the physical origin of a temperature gap between the edge of the rod and the mount was due to the presence of a thin oxide (or air or grease) layer, which acts as a very large thermal resistance.

Measuring the heat transfer coefficient is usually difficult and not found easily in the literature: we present in the next section a simple and accurate method to perform this measurement.

What about the end faces, which are in contact with air most of the time? The heat can flow out of the crystal through the two end faces by both convection in free air and thermal radiation. Cousins [60] calculated the equivalent H coefficient for the two processes and has shown that both coefficients were of the order of 10$^{-3}$ W.cm$^{-2}$.K$^{-1}$.

Since the measured H coefficients for conduction are typically in the range 1-10 W.cm$^{-2}$.K$^{-1}$, this is the proof that the assumption of *pure radial heat flux* made on the previous subsection is correct.

Using (II.1.9.) et (II.1.21.) one can calculate the temperature gap between the radiator and the edge of the crystal:

$$T(r_0) - T_c = \frac{K_c}{H}\left|\frac{\partial T}{\partial r}(r_0)\right| = \frac{K_c}{H}\frac{2(T(0) - T(r_0))}{r_0\left(1 + 2\ln\frac{r_0}{w_p}\right)} \qquad \text{(II.1.22.)}$$

The parameter of interest here is the normal derivative of the temperature at the interface. This explains why the quality of the thermal contact has a tremendous impact on side or edge-pumped slabs or rods; since in these configurations the temperature distribution is described by a formula of the type II.1.4, that is a parabolic dependence. In contrast, in end-pumped configurations, where the temperature profile is described by equation II.1.6., the thermal gradient at the periphery is smaller and the requirement of a good thermal contact can be loosen.

It is also interesting to know the maximum temperature reached inside the crystal. In order to obtain a easy-to-handle scaling formula, we make the strong assumption that absorption saturation is absent, and we ignore the divergence of the pump beam inside the crystal. We have:

$$T_{max} = T_c + \frac{\eta_h P_{inc}\alpha}{2\pi}\left[\frac{1}{H.r_0} + \frac{1}{2K_c}\left(1 + 2\ln\frac{r_0}{w_p}\right)\right] \qquad \text{(II.1.23.)}$$

As a conclusion for this section, we will list some conclusions one can make from these two last equations, like a list of recipes to reduce the temperature $T_{max}$ :

- *Increase $w_p$* : obvious and efficient, but at the expense of laser efficiency.
- *Increase H*. As shown in the next experimental section, reducing H does not affect the temperature gradient, and *will not help to reduce the thermal lens magnitude.* All we can get is a uniform decrease of temperature. However reducing the absolute temperature is actually more interesting in an Yb-doped crystal than in an Nd-doped material for example, in virtue of reabsorption losses that are highly temperature-dependant. A better contact can also help reducing fracture risks but this is not directly linked to a decrease of the temperature either: it is because a good contact can induce radial components to the stress tensor at the periphery, or also because it will decrease the density of high spatial frequency alterations of the surface which are the ultimate causes of crack-induced propagating fractures [61-63].

- *Decrease $T_c$*: if the radiator temperature is decreased but still remains around the room temperature (that is with a standard thermoelectric or water-flow cooling), the effect is the same as increasing H: we only play on the temperature pedestal, not on the gradient. However, if the mount is cooled far below room temperature (at cryogenic temperatures for instance), the thermal conductivity of the crystals significantly increases, which is highly positive for the thermal gradient. This approach has been successively applied to reduce the thermal lens in high-energy femtosecond laser chains [64] or in Nd:YAG rods [65]

- *Decrease the crystal size?* The crystal size has no influence on the temperature gradient. To understand its influence on the maximum temperature, $T_{max}$ is plotted versus $r_0$ for different values of H in figure 9. We observe that when the radius of the crystal exceeds roughly 10 times the pumped area radius $w_p$, the temperature becomes independent of $r_0$. In practice, it is possible to reduce the absolute temperature using small crystals, providing they are *really* small (see for example [66]). It is practically very difficult to cut and polish crystals whose size is smaller than 2 mm: one can then conclude that the transverse section of a crystal is not a parameter on which one can play efficiently. Besides, the effect of a bad thermal contact is visible only for crystals whose size would be on the order of the pump spot size. As illustrated by figure 9, a small crystal with bad cooling (for example $r_0$ = 0.5 mm and H = 0.1 W.cm$^{-2}$.K$^{-1}$) is far worse than a « reasonably » sized crystal with correct cooling ($r_0$ = 2 mm ; H = 1 W.cm$^{-2}$.K$^{-1}$) since the temperature difference between the two configurations reaches 200 °C.

- *Add an axial component to the heat flux*: it does not appear in equation II.1.23 because it has been derived with the assumption of a purely radial heat flux. However, one can add a large axial heat flux by putting either a « transparent radiator » in front of the input face (this is the principle of composite bondings [67] ) or by using thin disks (i.e. L« $r_0$) that are very efficiently cooled through the face in contact with the radiator [68].

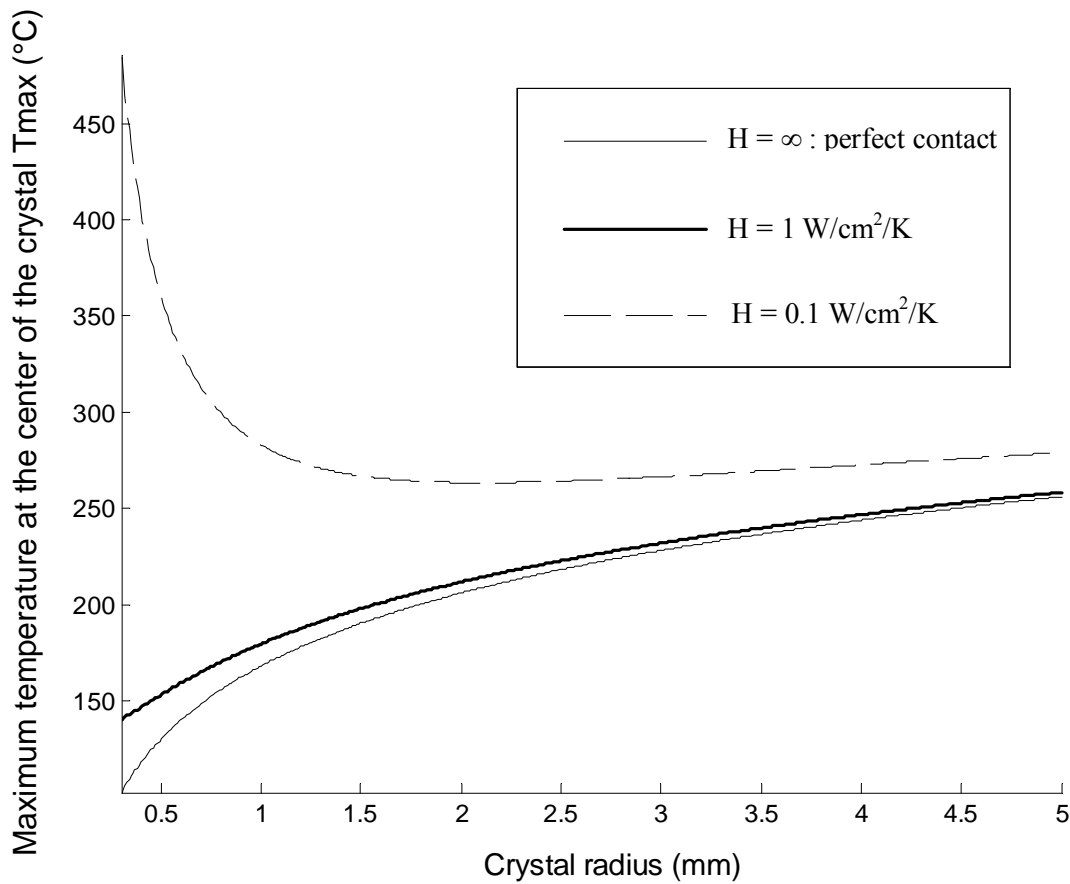

**Figure 9:** *Maximum temperature at the center of the input face of the crystal T(r=0, z=0) versus the crystal radius $r_0$. The absorption saturation is neglected, as well as the divergence of the pump beam inside the crystal. The parameters are: $T_c$=15°C, $\eta_h$=6.5 %, $\alpha_{NS}$ = 7.4 $cm^{-1}$, $K_c$=2.1 W/m/K (values for GdCOB), $P_{inc}$=15 W, $w_p$=100 μm.*

## II.2. *Experimental absolute temperature mapping and heat transfer measurements using an infrared camera*

### II.2.1. Introduction

As depicted in the previous paragraph, the temperatures obtained by solving the heat equation are only relative temperature distributions, expressed with respect to the rod surface temperature. The latter depends on the boundary conditions and is then very difficult to predict. Direct temperature mapping could consequently be a helpful measurement to understand pump-induced thermal effects. Moreover, we have shown that one of the crucial parameter to uniformly decrease the temperature inside the crystal (which can be useful to reduce fracture risks, see above paragraph) is the thermal contact between the crystal and its surrounding mount. Consequently, the knowledge of *quantitative* and *experimentally measured* information as the heat transfer coefficient H is of practical importance for high power laser development.

We herein report on a very simple experimental setup, based on an infrared camera that can perform spatially resolved analysis of the absolute temperature on the entrance face of the crystal, where temperature reaches generally a very high value (in any case higher than at the beam waist, as explained in the subsection II.1). We can also experimentally measure the heat transfer coefficient H between the crystal and its surrounding for different types of commonly used thermal contacts.

We first describe the experimental setup that allows such measurements, and illustrate it with the well-known Yb:YAG crystal [8].

### II.2.2. *Experimental setup for direct temperature mapping*

The experimental setup is presented on figure 10. A fiber-coupled laser diode was focussed inside an Yb:YAG laser crystal; the infrared emission of the entrance face of the crystal was observed with an infrared camera. A dichroic Zinc selenide (ZnSe) plate was used as a dichroic mirror: it was High Reflectivity (HR) coated for 960-1080 nm on one face (at 45° angle of incidence) to direct the pump beam into the crystal, and also coated for High Transmission (HT) in the 8-12 μm spectral range on both faces to let the thermal radiation reach the camera.

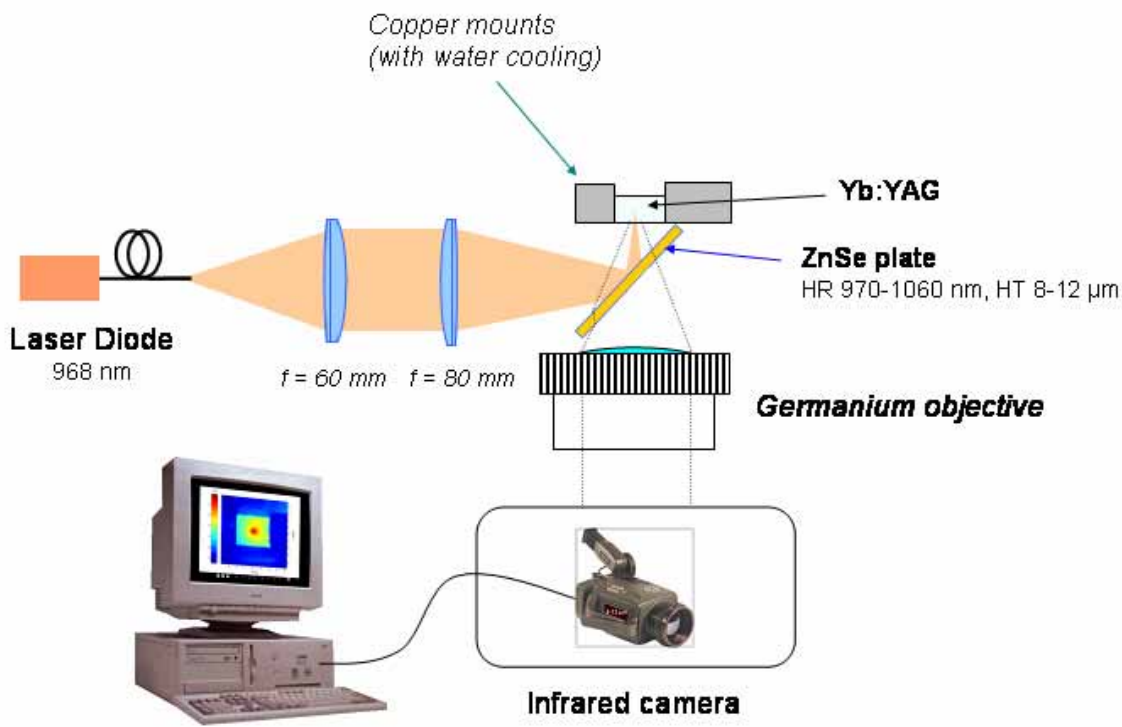

**Figure 10:** *Experimental setup for absolute temperature measurements.*

A germanium objective (focal length 50 mm, N.A. 0.7, aberration-corrected for infinite conjugation) was appended close to the ZnSe plate to create the intermediate thermal image with high spatial resolution. The camera was an AGEMA 570 (*Flir Systems Inc.*) consisting of 240x320 microbolometers working at room temperature. The measured noise equivalent temperature difference (NETD) of the camera is 0.2 °C. The numerical aperture of the whole imaging system in the object plane being around 1, a theoretical spatial resolution of about 10 μm could be achieved; however, the resolution is here limited to 60 μm by the size of the pixels of our camera. The crystal used here was a 2-mm long, 4x4 mm$^2$ square cross section, 8-at. % doped Yb:YAG crystal. It was AR-coated on its faces (the lateral ones are polished). Its thermal conductivity, which is lower than that of an undoped YAG crystal, was measured to be 7 W.m$^{-1}$.K$^{-1}$ (11 W.m$^{-1}$.K$^{-1}$ for the undoped crystal). The pump source was a high power fiber-coupled diode array (HLU15F200-980 from LIMO GmbH) emitting 13.5 W at 968 nm. The fiber had a core diameter of 200 µm and a numerical aperture of 0.22. The output face was imaged onto the crystal to a 270-µm-diameter spot via two doublets. The crystal absorbed 5.4 watts of pump power in this case. The crystal was clamped in a copper block by its four side faces. In addition, on the top surface of the crystal, a frictionless copper finger allowed us to apply a well-controlled pressure on the crystal by the use of a set of known weights put upon the finger. The heat is finally evacuated from the copper block by a flow of circulating water.

The key issue of infrared absolute temperature measurements is the correct calibration of the system. Indeed, neither the crystal nor the copper mount has an infrared luminance which equals that of a blackbody at the same temperature. The signal V detected by one pixel for a portion of crystal (or copper mount) at temperature T is:

$$V(T) = G \int_{\Delta\lambda} S_r(\lambda) \left[ Tr_{opt} \left( \varepsilon(T) \frac{dL_{BB}^T}{d\lambda} + L_t \right) + L_r \right] d\lambda \qquad \text{(II.2.1.)}$$

where G is the geometric extent; $S_r(\lambda)$ is the spectral sensitivity; $Tr_{opt}$ is the whole transmission coefficient of the ZnSe plate, Germanium objective and camera optics; $\frac{dL_{BB}^T}{d\lambda}$ is the spectral luminance of a blackbody at temperature T, ε(T) is the emissivity; $L_r$ denotes the infrared luminance of the camera itself (and its close surroundings) which is reflected back into it by the Germanium objective and by the polished surface of the crystal; $L_t$ is the luminance transmitted through the crystal: it is zero in the 8-12 µm range since the crystal is highly opaque in this spectral region. $L_r$ is nonzero and makes polished objects look brighter than blackbodies: if $L_r$ is ignored it leads to overestimation of the temperature around room temperature. Inversely, the emissivity is less than one and makes objects radiate less than a blackbody. Since the parameters ε and $L_r$ are dramatically dependant on the surface quality and flatness, all the visible parts of the heat sink were covered with lustreless black painting. Moreover, the evaluation of all those parameters is not straightforward. We propose to calibrate the whole system as follows: the crystal and the copper mount were heated together to a set of given temperatures using a thermoelectric (Peltier) element, and we then compare with the temperature given by the camera to apply the adequate correction. This careful calibration allows rigorous and absolute measurement of the temperature with a spatial resolution large enough to study with sufficient accuracy the thermal behaviour on the crystal's input face.

*II.2.3 Results and measurement of heat transfer coefficients*

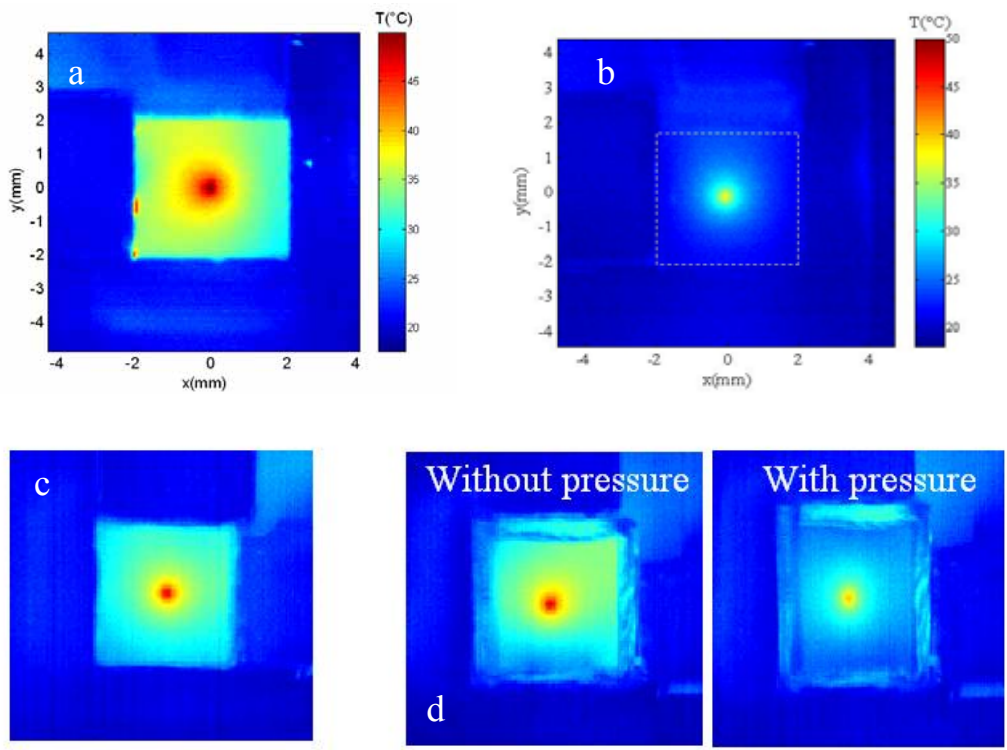

**Figure 11:** The temperature map obtained when the crystal is clamped by its four edge faces by bare contact with copper without thermal joint (a), with heat sink grease (b), with a thin graphite layer (c), and with pressured and non-pressured indium (d).

Figure 11 shows the temperature map obtained when the crystal is clamped by its four edge faces by bare contact with copper without thermal joint (a), with a thin graphite layer (b), with indium (c) and with heat sink grease (d).

Figure 12 is an enlargement of the two extreme cases, namely the bare contact and heat sink grease contact, with a transverse profile (y = 0) that shows the temperature evolution along the crystal lateral dimension.

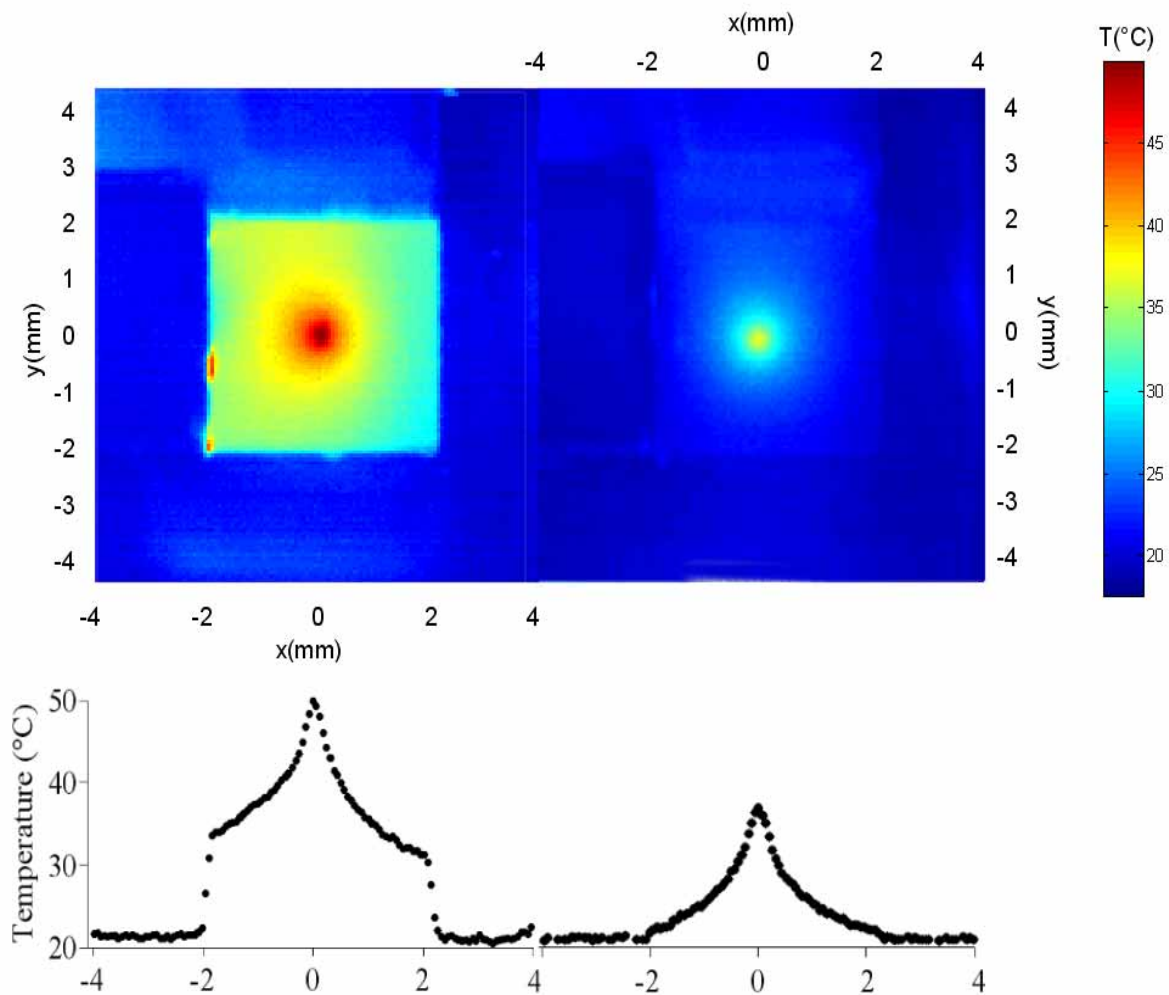

**Figure 12 :** *Temperature mapping of the crystal (front view) and lateral profile at y=0 for two different types of thermal contact (direct copper-crystal contact on the left, with grease on the right).*

In the "bare contact" case, a clear gap is noticeable between the temperatures of the mount and at the edge of the crystal. The temperature distribution is parabolic inside the pumped region and then experiences a logarithmic decay until the edge of the crystal, in good agreement with the theory described in the previous section in the case of fibre-coupled diode pumping (see equation II.1.9 and figure 5). As already mentioned, the quality of the heat transfer at the interface between the crystal and its mount has an influence on the value of the temperature but not on the thermal gradient.

We consequently studied more in detail the heat contact. The heat transfer coefficient H is defined by equation (II.1.22), where the thermal gradient is considered normal to the surface.

Our system provides a space-resolved temperature mapping of the crystal, with a spatial resolution which is far below the crystal size: it then allows the measurement of H.

By performing a linear fit of the temperature versus position on the points that are closer to the crystal edge, the heat flux can be determined: by applying the equation (II.1.22), one can then infer the value of H. We found for instance a value of 0.25 $W.cm^{-2}.K^{-1}$ in the case of bare contact. We estimate that the uncertainty on H is about 15%. The order of magnitude obtained is consistent with the values evoked by Carslaw [58] and Koechner [69]. The hot spot that can be noticed in figure 12 betrays the poor contact between the polished face of the crystal and the copper surface. The heat transfer is primarily a question of how much two surfaces are in contact with respect to each other; we checked experimentally that the temperature inside the crystal did not depend on the applied pressure: we did not observe any noticeable variation of the temperature when changing the applied pressure in absence of thermal joint between the crystal and the copper mount.

We summarized in table 3 the results obtained for the different thermal joints used in our set of experiments, namely graphite layer, indium foil and heat sink grease (CT40-5 from Circuitworks®). Graphite layer (around 0.5 mm thick) does not modify significantly either the maximum temperature or the heat transfer coefficient, but it was noticed that the contact was much more uniform than in the case of bare contact: in particular no hot spot appeared any more and the contact was somewhat independent of the applied pressure. It is not the case with indium foil. For this experiment the crystal was wrapped within a 1-mm thick indium foil. Since Indium is a soft material, the quality of the contact is greatly dependant on the applied pressure. The temperature at the center of the pumped region experiences a 7°C decrease while the pressure increased from 1.5 kg/cm$^2$ to 22 kg/cm$^2$ as shown in figure 13 (note that in this case the H coefficient is measured across the surface where the pressure is applied and is then an "effective" heat coefficient that takes into account the transfer from crystal to indium and then from indium to copper.)

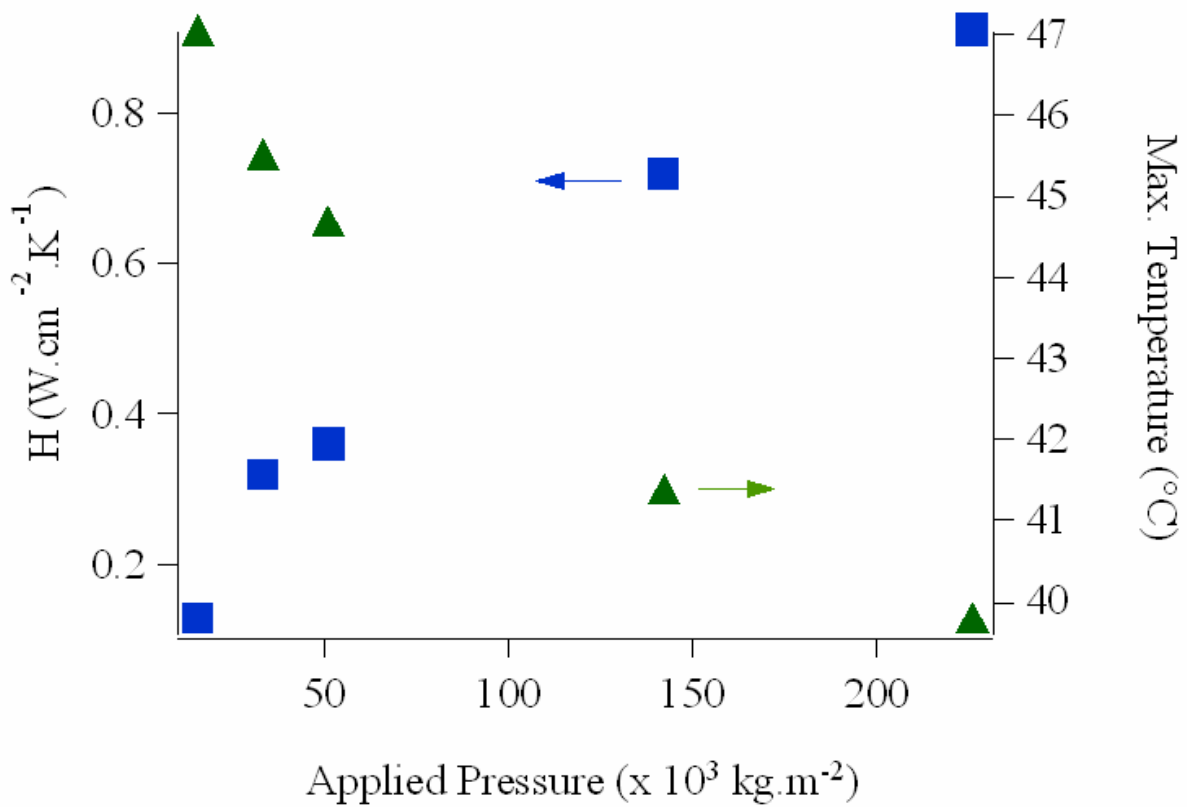

**Figure 13**: *Evolution of the heat transfer coefficient H (squares) and maximum temperature (triangles) versus applied pressure for indium-wrapped crystals.*

The most dramatic change in heat transfer coefficient is obtained with heat sink grease (see table 3). The temperature gap drops down to 1°C and H reaches 2 W.cm$^{-2}$.K$^{-1}$. The heat contact is here independent on the applied pressure. This better heat transfer coefficient is achieved while the thermal grease has a much lower thermal conductivity than indium (0.62 W/m.K for CircuitWorks ® CT40-5 thermal grease *vs*. 82 W/m.K for pure indium). This is an illustration, in conjunction with the data about the variation of H with the applied pressure, of the idea that achieving a good H is first and foremost a question of decreasing the thermal resistance at the interface (eliminating air gaps, maximizing the surface contact…)

**Table 3:** *Table of measured H coefficients for different contacts. $T_{max}$ is the temperature at the center of the pumped region; $T_e$ is the temperature at the edge of the crystal (averaged on the 4 sides if not symmetrical), and $T_m$ is the copper mount temperature near the crystal.*

| *Contact* | **$H$ ($W.cm^{-2}.K^{-1}$)** | $T_{max}$ (°C) | $T_e$ (°C) | $T_e - T_m$ (°C) |
|---|---|---|---|---|
| Bare | **0.25** | 49.8 | 33.5 | 10.7 |
| Graphite layer | **0.28** | 46.5 | 30.5 | 8.7 |
| Indium foil *(applied pressure : 22 kg/cm²)* | **0.9** | 40.0 | 25.1 | 4.9 |
| Heat sink grease | **2.0** | 37.0 | 21.6 | 1.5 |

# III. Thermal lensing effects: theory

## *III.1. Introduction*

The previous chapter was dedicated to the calculation and measurement of the temperature distribution, which is the first essential step for the study of thermal effects. The appearance of thermal gradients causes the crystal to be under stress. The presence of inhomogeneous temperature, stress, and strain distributions is responsible of many deleterious effects for laser action: the most radical effect is fracture, observed when the hoop (tangential) stress at the periphery of the crystal exceeds the so-called tensile stress. More subtle effects arise from the stress-induced modification of the optical indices of refraction: alteration of the stability domains of the cavity, depolarization, losses and degradation in beam quality, all of these four phenomena being largely intermixed. In this paper we designate by "thermal lensing" effects all the phenomena resulting in a phase change of a beam passing through a pumped crystal; in other words we do not restrict this expression to an ideal spherical thin thermal lens, we also include its aberrations and its polarization-dependant aberrations. This chapter presents a general and synthetic scope of these effects, and points out how they are related to each other. We base our discussion on analytical simple scaling relationships, and we point out the validity of these formulas.

In this review, we come back to well-established theories that have been exposed many times in the past [60, 69, 70], but we also bring some new insights, to our knowledge, on some points of practical interest.

In particular, we will point out several inaccuracies generally reported about the values of the photoelastic constants in YAG, which are the result of an incorrect use of the Hooke Law; we also present what is (still to our knowledge) the first derivation of the photoelastic constants that have to be used for *end-pumped* crystals, that is in other words when the calculation is made using the *plane stress* rather than the plane strain approximation.

We will eventually point out that the use of the *dn/dT* coefficient (temperature derivative of the refractive index) is very confusing. In one hand, the classical formula (which reveals the existence of three contributions: the "dn/dT" part, the bulging of end faces, and the photoelastic effect) which is used since decades is correct *provided that the dn/dT appearing in this expression is understood as a partial derivative taken at constant strain*. In the other hand, the experimentalist can measure a quantity which is closer to a partial derivative *at constant stress*, and the partial derivatives are obviously not equal. The dn/dT parameter is then not actually the correct parameter to be used in order to estimate the thermal lens focal length: this subtlety means in particular that one cannot, in general, make use of a value of *dn/dT* readily found in handbooks to estimate the magnitude of the thermal lens of an operating laser, because the experimental measurement conditions are in the two cases mutually inconsistent. We'll see however that when the dn/dT is large *and* positive, the difference can be ignored.

We will conclude this review by a synthetic diagram showing all the thermal effects and how they are connected together.

Given that thermo-optical properties pertain more to a crystal host than to a doping ion, this section is more general than the others and does not restrict to the case of Ytterbium-doped materials.

## III.2. Stress and strain calculations

Once the temperature field has been computed, the next step is to calculate the stress and strain distributions inside the crystal, obtained from the so-called "generalized" Hooke law, because it includes the thermal expansion term [49]:

$$\varepsilon_{ij} = S_{ijkl}\sigma_{kl} + \alpha_{T\,ij}\Delta T \tag{III.1}$$

where i, j, k, l = 1,2,3 and the Einstein summation convention is used. $\Delta T$ is the temperature shift with respect to equilibrium (no strain), ($S_{ijkl}$) is the compliances tensor, ($\sigma_{kl}$) is the stress tensor, ($\varepsilon_{ij}$) is the strain tensor, and ($\alpha_{T_{ij}}$) is the thermal expansion coefficients tensor.

The analytical formulations of thermal stress and strain distributions in end-pumped lasers require a large amount of approximations, thoroughly discussed in Cousins's reference paper published in 1992 [60].

In order to obtain an *analytical* solution to the stress problem, an additional approximation is required, which consists in considering the problem in two dimensions. This is either the *plane strain* approximation (valid for long and thin rods) or the *plane stress* approximation (valid for thin disks). Interestingly, Cousins [60] pointed out that the plane stress approximation remained valid (within approximately 10%) for aspect ratios up to $L/2r_0 = 1.5$, providing that the stresses were considered as *mean* values integrated along the whole thickness of the rod. In the previous section, when we derived the temperature distribution, we had considered the crystal as a stacking of thin slices, so that the temperature could be calculated in a single thin slice as if the surrounding material did not exist. It is not possible to use this approach for the stress distribution, because a given slice is under the mechanical influence of the slices located on both sides, and cannot be considered as independent. This is why an attempt to take into account absorption saturation effects and pump divergence (as far as thermal stresses are concerned) inevitably requires a finite element analysis.

As far as diode end-pumping is concerned, the plane stress approximation is then the most meaningful approximation that can be done. However, the exact calculation remains possible for a given crystal (using FEA codes) provided that all the compliances and thermal expansion coefficients are known. To the best of our knowledge, these coefficients have been measured for a very restricted number of laser crystals up to now: we may readily find these data for YAG, sapphire, YLF and $Y_2O_3$ [71]: other data are available in the *Handbook of Optics* [71] but for crystals which are not commonly used in laser applications. The data for other materials are almost inexistent.

The analytical solution of the generalized Hooke law (eq. III.1) can be found, for example in [60] under the plane stress approximation. In this paper, the discussion was restricted to isotropic materials (at a mechanical rather than an optical point of view, that is when the compliances can be reduced to only two parameters, the Young modulus and the Poisson ratio), absorption saturation was not considered, and the divergence of the pump beam was not taken into account either.

Once this has been calculated, it is possible to study thermal fracture issues. It is generally admitted that fracture occurs when the maximum hoop (tangential) stress $\sigma_{max}$ at the surface periphery of the crystal exceeds the tensile stress $\sigma_{TS}$. The latter depends on both the fracture toughness of the material and on its surface flatness. These aspects have been studied in detail by Marion [61-63]. Data about fracture toughness of materials can be readily found in the literature for YAG, fluoroapatites, sapphire, yttrium orthosilicate YSO and some phosphate glasses [39, 63, 71, 72].

For a qualitative discussion of fracture issues in Yb-doped materials, the reader is invited to refer to a previous publication [73].

### III.3. How can we take into account the photoelastic effect ?

We now consider how the temperature, stress and strain fields inside the crystal alter the phase of the cavity beam, all these effects being referred as "thermal lensing" phenomena.

The appearance of stresses in the crystal causes the linear optical indicatrix (related to the linear indices of refraction) to change its shape, its size and its orientation. This photoelastic effect is accounted by the 4$^{th}$ rank elasto-optical tensor ($p_{ijkl}$):

$$\Delta B_{ij} = p_{ijkl}\, \varepsilon_{kl} \tag{III.2}$$

Where ($B_{ij}$) is the dielectric impermeability 2$^{nd}$ rank tensor. This expression if obviously valid in the linear optical regime only, and when piezoelectric effect is neglected [49].

The complete computation of thermal effects in a given material requires that we know everything about the tensors ($S_{ijkl}$), ($p_{ijkl}$) and ($\alpha_{T_{ij}}$). The minimum number of independent terms of each tensor depends on the crystal symmetry, as discussed by Nye [49]. For instance, let us consider crystals like KGW or KYW [74], GdCOB [75], YCOB [76] or YSO [43] which are of particular interest for Ytterbium doping. These crystals belong to the monoclinic crystal system: this means that the compliances can be "reduced" to 13 independent parameters, and the elasto-optical tensor, once the redundant coefficients have been identified, appears to have 20 independent coefficients [49]. Adding the 3 thermal expansion coefficients, this means that *we need to know no less than 36 coefficients before to be able to draw the new index ellipsoid at a given point of the crystal*. Obviously these parameters are not known (for any monoclinic crystal, in fact, to the best of our knowledge), which means that a rigorous calculation even with a FEA code is just not possible. This simple remark highlights the importance of experimental measurements of thermal effects in such crystals, and shows the interest as well as the inherent limitations of a simple analytical model.

## *III.4. Simplified account of photoelastic effect in isotropic crystals*

Now we are aware of these difficulties, we focus our discussion on a simpler study case, actually the only case where *analytical expressions* are obtainable, that is:

- we consider isotropic crystals only, and more particularly the widespread YAG crystal, for which all the parameters previously evoked are well known. Generally, cubic crystals belonging to the space groups $\bar{4}3m, 432, m3m$ (like YAG) require 3 independent elastic coefficients ; however the remarkable isotropic mechanical properties of YAG enable to think of only two mechanical coefficients, that is the Young modulus and the Poisson ratio. In the end, 6 coefficients only are needed for YAG.

- The plane stress approximation is used;

- the pump profile is still axisymmetric;

- we consider what occurs inside the pump volume, that is for r < w$_p$.

- the pump divergence inside the crystal is neglected.

- temperature, stress, strain, index are considered integrated along the whole thickness of the rod. For a physical quantity A(r,z), we shall note :

$$\langle A(r) \rangle = \int_0^L A(r,z)\, dz$$

the integrated value of A(r,z) along the rod.

According to the Neumann-Curie theorem [49], under these assumptions the principal axes of all the involved tensors (stress, strain, index ellipsoid) are *radial* and *tangential*. The notations used in the following are depicted in figure 14.

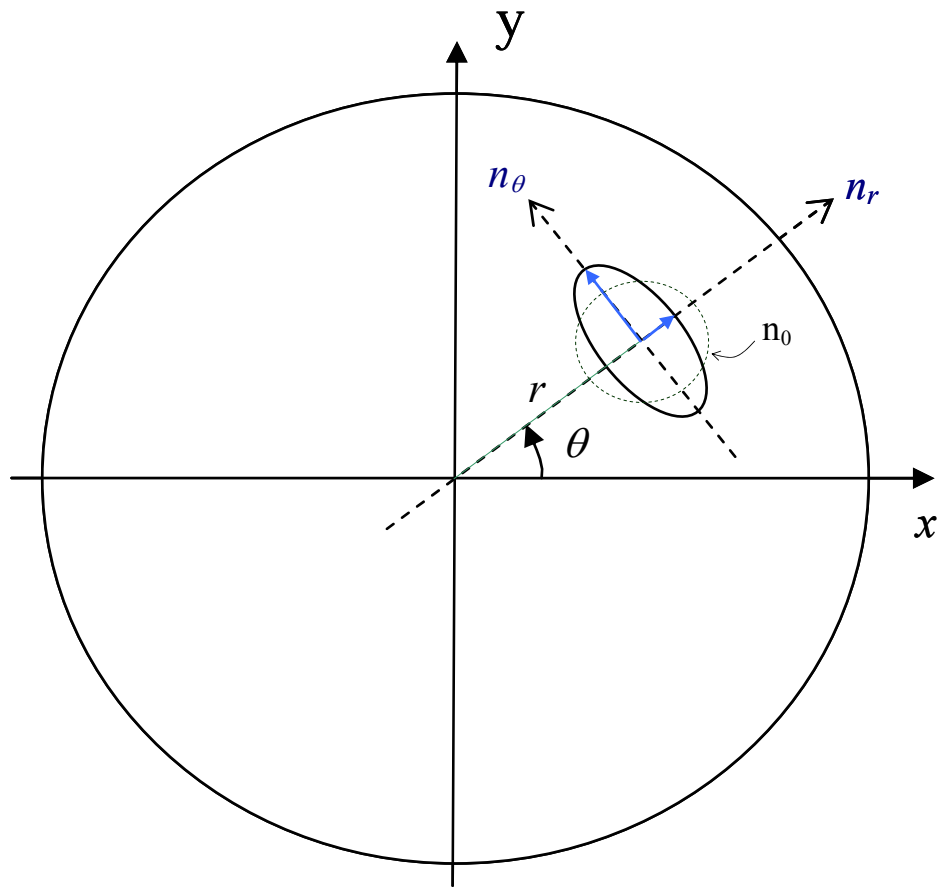

**Figure 14**: *Orientation of the indices' ellipsoid in an isotropic crystal under thermal stress.*

The shift of the principal indices $n_r$ and $n_\theta$ are related to the diagonal coefficients of the optical indicatrix by:

$$\Delta B_{r,\theta} = \Delta\left(\frac{1}{n_{r,\theta}^2}\right) = \frac{-2\Delta n_{r,\theta}}{n_0^3} \tag{III.3}$$

We can also write the indices variation as a function of the strains as follows:

$$\langle \Delta n_{r,\theta} \rangle = -\frac{1}{2} n_0^3 \langle \Delta B_{r,\theta} \rangle = \frac{\partial n_{r,\theta}}{\partial \varepsilon_r}\langle \varepsilon_r \rangle + \frac{\partial n_{r,\theta}}{\partial \varepsilon_\theta}\langle \varepsilon_\theta \rangle + \frac{\partial n_{r,\theta}}{\partial \varepsilon_z}\langle \varepsilon_z \rangle \tag{III.4}$$

The six coefficients $\frac{\partial n_i}{\partial \varepsilon_j}$ (i= r, θ; j=r, θ, z) can be calculated from the $p_{ijkl}$ coefficients by a correct change of coordinates.

The complete solutions for the strains $\varepsilon_r$, $\varepsilon_\theta$ and $\varepsilon_z$ can be found, for YAG, in many papers and textbooks [60, 69] and can also be found in the appendix.

Inside the pump volume, it can be readily shown that stresses and strains have a parabolic dependence, like the temperature distribution. Since the indices of refraction are linear combinations of strains, it turns out that radial and tangential index distributions must also be parabolic.

The (integrated) shift of refractive index may be written as:

$$\langle \Delta n_{r,\theta}(r) \rangle = \sum_{j=r,\theta,z} \frac{\partial n_{r,\theta}}{\partial \varepsilon_j} \int_0^L \varepsilon_j(r,z)\,dz \ = \langle \Delta n_{r,\theta}(r=0)\rangle - \frac{\eta_h\, P_{abs}}{2\pi K_c} n_0^3\, \alpha_T\, C_{r,\theta}^{(')} \frac{r^2}{w_p^2} \tag{III.5}$$

where $C_r$ and $C_\theta$ (or $C'_r$ and $C'_\theta$) are constants which will be called, following the pioneering work of Koechner, the "photoelastic constants". Their calculation and expression is given in the appendix. We would like to point out two important clarifications about these coefficients (and also justify the presence of this appendix in this review):

- ♦ W. Koechner published incorrect values of these coefficients in his reference book [69] because the temperature term in the Hooke law has been omitted; this omission has first been highlighted by Cousins [60], but the expression of the photoelastic constants remained uncorrected in the following editions of this book, and nowadays still remains used under this form in many papers.

- ♦ Secondly, the derivation of the photoelastic constants requires turning the 3D problem into a 2D problem, as discussed in the previous section. Only the *plane strain* case was considered by Koechner. However, we saw that the plane stress case is closer to reality in end-pumped rods. Here we denote as $C_r$ and $C_\theta$ the photoelastic constants valid for long and thin rods (the "Koechner case", that is when the plane strain approximation is valid), and $C'_r$ and $C'_\theta$ the photoelastic constants derived within the framework of the plane stress approximation. Since we are only interested in end-pumping, we only consider the $C'_{r,\theta}$ constants in the following.

The above-mentioned relations are derived by making the assumption that the pump beam radius is constant through the crystal thickness; however we don't have to assume a particular absorption regime, so that they stay valid in presence of absorption saturation, under lasing as well as under nonlasing conditions.

As we will see in the next section, the most interesting feature for the laser scientist is the index shift between the center and the edge of the pumped zone, since it yields the contribution to the global thermal lens. From (II.1.10), and using the bracket notation for z-integrated values, we can write (III.5) under the form:

$$\langle \Delta n_{r,\theta}(0) - \Delta n_{r,\theta}(r) \rangle = 2 n_0^3 \, \alpha_T \, C'_{r,\theta} \langle T(0) - T(r) \rangle \qquad \text{(III.6)}$$

### III.5. A consequence of strain-induced birefringence: depolarization losses.

The birefringence of a crystal submitted to thermal stress has two main consequences for a light beam passing through it: both its state of polarization and its phase will be altered. Before examining in detail the effect on phase (*i.e.* thermal lensing effects), let's first examine the influence on polarization.

We use the restrictions exposed in the previous subsection, given that the following can be readily extended to uniaxial crystals provided that the optical axis lies parallel to the propagation axis. In these cases, if no polarizing element is added into the cavity, the laser output is not polarized, and stress-induced birefringence has no net effect.

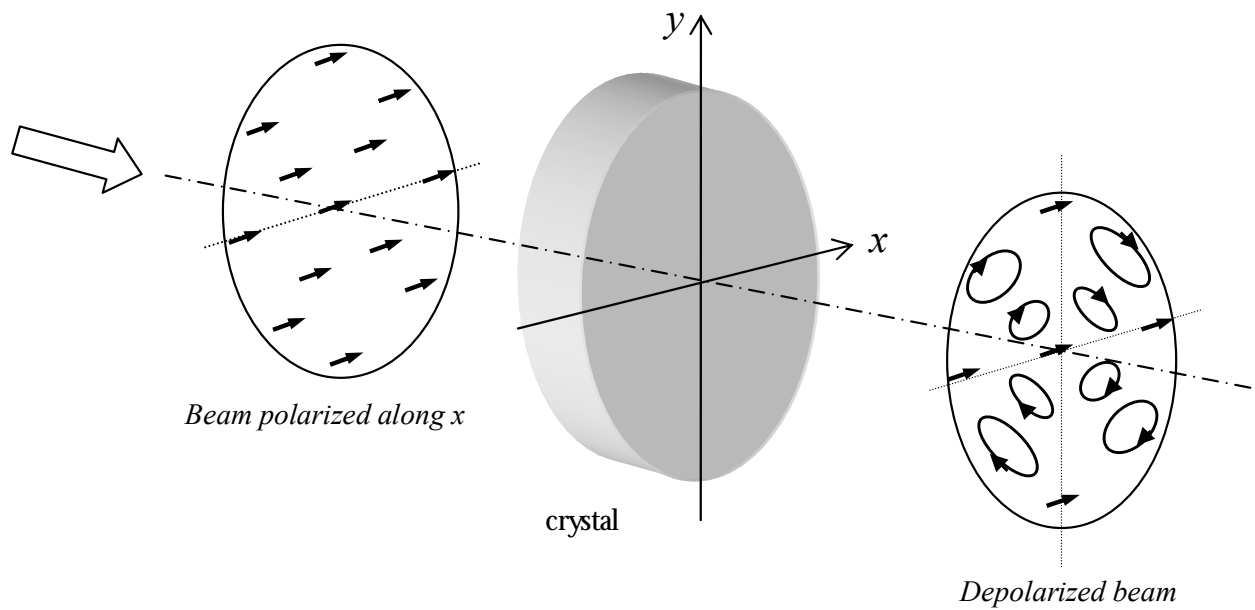

**Figure 15**: *Depolarization of a polarized beam passing through an isotropic crystal under thermal stress.*

For many applications however a polarized output is desirable: the situation is depicted schematically in figure 15. An incident beam (polarized along the x direction) will have its polarization modified differently for every single ray: For a ray crossing the (Ox) or (Oy) axis for instance, the polarization is not modified, for all the other rays the polarization becomes elliptical, with principal axis that are radial and tangential.

At every roundtrip in the laser cavity, the beam meets a polarizing element (such as a plate at Brewster angle) and this depolarizing element, yielding to the so-called "depolarization losses".

Another effect, consequence of the latter, is a modification of the beam spatial profile: since the beam is not altered along the (Ox) and (Oy) directions and suffers losses elsewhere, it tends to take the shape of a cross; this aspect is chiefly discussed in Koechner [69].

In biaxial crystals (or uniaxial crystals with optical axis normal to the direction of propagation), the output is naturally polarized along the crystallophysic (not to be confused with cristallographic) axis along which the emission cross section is the highest. In this case, stress-induced birefringence does not generally twist the index ellipsoid enough to significantly modify the polarization state.

Finally, clever solutions have been imagined to compensate for depolarization losses in isotropic crystals: for example the use of two rods with a Faraday rotator inbetween [78, 79], or even a simple quarter waveplate [70]. This last technique is however limited to a few configurations [80].

In the following, we present results obtained with Yb-doped crystals which are either isotropic (and naturally not polarized), or naturally birefringent, so that this problem was not encountered.

## *III.6. Thermally-induced optical phase shift*.

### *III.6.1. Expression of the optical path*

We present now this derivation with some detail, because it makes appear a trivial yet fundamental difference between these results and what is currently reported in the literature. The derivation is largely reproducing Cousins's work [60].

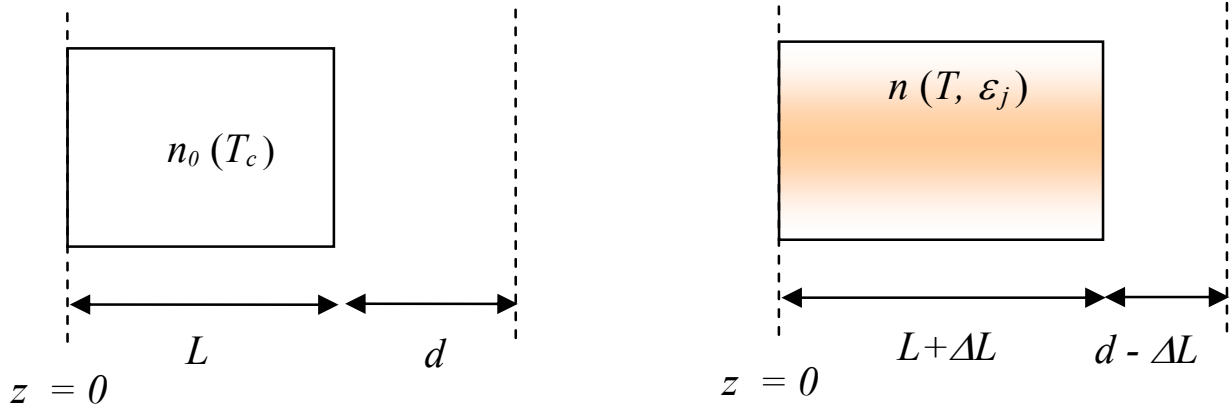

**Figure 16**: *Notations used for the calculation of the optical path*

Consider a crystal whose length is $L$ and index $n_0$ at temperature $T_c$ (temperature of the heat sink) and in absence of strain. We consider the optical path of a straight ray (running parallel to the crystal axis $z$) between a plane $z = 0$ (taken at the entrance face of the crystal) and a plane $z = L+d$ (see figure 16).

In absence of temperature and stress fields (pump off), the optical path is:

$$\delta^{\text{off}} = n_0(T_c) \times L + d \tag{III.7}$$

With the pump on, the optical path is dependant on the lateral shift $r$ of the ray with respect to the crystal axis, but also on the direction of polarization. As a consequence there will be two distinct thermal lens focal lengths, one for a (virtual) radially polarized beam and another for a tangentially polarized beam, what is commonly named "bifocussing".

We may write the optical path with the pump beam "on" as (see figure 16):

$$\delta^{\text{on}}_{r,\theta}(r) = \int_0^{L+\Delta L} n_{r,\theta}(T, \varepsilon_j)\, dz + d - \Delta L(r) \tag{III.8}$$

where $\Delta L(r)$ is the crystal length shift due to the inner compression, responsible for the bulging of the end faces.

Assuming small variations in the refractive index, one may expand $n_{r,\theta}(T, \varepsilon_j)$ in Taylor series, and discard the second and higher-order terms:

$$\delta_{r,\theta}^{on}(r) = \int_0^{L+\Delta L}\left[ n_0(T_c) + \left(\frac{\partial n_{r,\theta}}{\partial T}\right)_\varepsilon (T(r,z) - T_c) + \sum_{j=r,\theta,z}\left(\frac{\partial n_{r,\theta}}{\partial \varepsilon_j}\right)_T \varepsilon_j(r,z)\right] dz + d - \Delta L(r) \qquad (III.9)$$

Note that the temperature derivative of the refractive index appearing in this equation is a *partial* derivative calculated *at constant strain*. As we will discuss in the next subsection, this is not the usual *dn/dT* parameter.

The rod length change $\Delta L(r)$ can be written as a function of the axial strain $\varepsilon_z$ and equals

$$\Delta L(r) = \int_0^L \varepsilon_z(r,z)dz = \langle \varepsilon_z(r) \rangle \qquad (III.10)$$

From the strain-stress relationships featured in the appendix, it can be easily shown that the axial strain, under the plane stress approximation, is equal to:

$$\langle \varepsilon_z(r) \rangle = -\alpha_T (1+\nu)\langle T(r=0) - T(r) \rangle \qquad (III.11)$$

Given that the first-order terms in the integral appearing in (III.9) are much smaller than $n_0(T_c)$ and given that $\Delta L \ll L$, we can write that for the first order terms $\int_0^{L+\Delta L}(...) \approx \int_0^L(...)$

The relative optical path is then:

$$\begin{aligned}\delta_{r,\theta}^{rel}(r) &= \delta_{r,\theta}^{on}(r) - \delta^{off} \\ &= \left(\frac{\partial n_{r,\theta}}{\partial T}\right)_\varepsilon \langle T(r) - T_c \rangle + \sum_{j=r,\theta,z}\left(\frac{\partial n_{r,\theta}}{\partial \varepsilon_j}\right)_T \langle \varepsilon_j(r) \rangle + (n_0 - 1)(1+\nu)\alpha_T \langle T(r) - T(0) \rangle \end{aligned} \qquad (III.12)$$

### III.6.2. The thermal lens focal length

The thermal lens is related to the *optical path difference* (OPD or $\Delta$ in the following) between an on-axis central ray ($r = 0$) and an outer parallel ray passing inside the pumped region, defined by a radius $r < w_p$.

We note :

$$\Delta_{r,\theta}(r) = \delta^{rel}(0) - \delta^{rel}_{r,\theta}(r) \tag{III.13}$$

The expression of the optical path difference is, from (III.6) and (III.12):

$$\Delta_{r,\theta}(r) = \left[\left(\frac{\partial n_{r,\theta}}{\partial T}\right)_{\varepsilon} + 2n_0^3\, \alpha_T\, C'_{r,\theta} + (n_0 - 1)\alpha_T(1+\nu)\right]\langle T(0) - T(r)\rangle \tag{III.14}$$

We assume that the thermal derivative of the refractive index is equal for radial and tangential index, so we can write:

$$\begin{aligned}\Delta_{r,\theta}(r) &= \left[\left(\frac{\partial n}{\partial T}\right)_{\varepsilon} + (n_0 - 1)(1+\nu)\alpha_T + 2n_0^3 \alpha_T C'_{r,\theta}\right]\langle T(0) - T(r)\rangle \\ &= \chi_{r,\theta}\,\langle T(0) - T(r)\rangle \end{aligned} \tag{III.15}$$

where $\chi_{r,\theta}$ is usually called the "thermo-optic" coefficient. We remind the reader that this expression is valid only under some restrictive conditions that have been presented in detail in the section 4 of this chapter.

Given that the integrated temperature shift is given by (from eq. II.1.11):

$$\langle T(0) - T(r)\rangle = \int_0^L (T(0,z) - T(r,z))\, dz = \frac{\eta_h P_{abs}}{4\pi K_c}\frac{r^2}{w_p^2} \tag{III.16}$$

it appears that the optical path difference also follows a quadratic dependence in r. This means that in the paraxial approximation, the pumped crystal acts as a thin lens whose focal length is given by :

$$f_{th(r,\theta)} = \frac{r^2}{2\Delta_{r,\theta}(r)} \tag{III.17}$$

In the following the difference between $f_r$ and $f_\theta$ (responsible for bifocussing) will be omitted (realistic if the photoelastic effect is negligible, or if the laser beam is not polarized.)

The thermal lens dioptric power is thus defined by:

$$D_{th} = \frac{1}{f_{th}} = \frac{\eta_h P_{abs}}{2\pi w_p^2 K_c}\left[\left(\frac{\partial n}{\partial T}\right)_\varepsilon + (n_0-1)(1+\nu)\alpha_T + 2n_0^3 \alpha_T C_{r,\theta}^{()}\right] = \frac{\eta_h P_{abs} \chi}{2\pi w_p^2 K_c} \quad \text{(III.18)}$$

where χ is a polarization-averaged thermo-optic coefficient.

Let's notice that this formula still holds in Yb-doped materials, with strong absorption saturation, since no assumption was made concerning absorption.

If the pump profile is gaussian (e.g. end-pumping by another laser), one can show [70] that the thermal lens dioptric power is twice as large as (III.18), meaning that the thermal load is more spiky around the center of the beam than in the case of an uniform "top hat" pump beam energy deposition.

To conclude, let's address some orders of magnitude. To be correct, the previous derivation has to be performed within the paraxial approximation, which means $f_{th} \gg L$.

Taking some typical values ($\eta_h \sim 0.1$, $\chi \sim 10^{-5}$, $K_c \sim 5$ W/m/K, $w_p = 100$ μm, et $L = 3$ mm), it appears that the paraxial conditions are met provided that $P_{abs} \ll 100$ W. This corresponds to most of practical cases.

## III.7. Discussion about the use of the "dn/dT" coefficient.

Let's start this discussion by the expression of the thermo-optic coefficient:

$$\chi_{r,\theta} = (n_0-1)(1+\nu)\alpha_T + 2n_0^3 \alpha_T C_{r,\theta}^{'} + \left(\frac{\partial n}{\partial T}\right)_\varepsilon \quad \text{(III.19)}$$

In this subsection, we would like to point out how this expression can be misleading if one carelessly uses, to evaluate the magnitude of a thermal lens, the so-called *dn/dT* instead of $\left(\frac{\partial n}{\partial T}\right)_\varepsilon$.
The error is especially important when all terms except the *dn/dT* are discarded for the sake of simplification. This represents, to our knowledge, a discussion that has never been published so far. The three contributions appearing in (III.19) may be understood as follow:

- The term $(n_0-1)(1+\nu)a_T$ is clearly related to the bulging of end faces, and is the direct consequence of the inner compression of the crystal, which causes the optical path to increase (if $\alpha_T > 0$). It is strictly true for an infinitely thin crystal, since plane stress approximation is used to derive it. It is reported by Cousins [60] that for a rod whose ratio length/diameter is 1.5, this term overestimates the actual bulging by around 35%. In general this can be taken as an *upper limit* for end faces bulging in DPSSLs.

- The term $2n_0^3 \alpha_T C'_{r,\theta}$ accounts for the photoelastic effect only, as already discussed in subsections III.3 and 4. It explains bifocussing, depolarization and polarization-dependant astigmatism.

- As for the first term (III.19), it represents the partial derivative of refractive index *at constant strain*, which is the thermo-optic coefficient of a virtual perfectly rigid crystal.

It is noteworthy that it is actually *not* the usual *dn/dT* parameter that one can measure and find easily in the literature. The "usual techniques" for measuring dn/dT are based either on geometrical optics (e.g. measurement of the minimum-of-deviation angle of a prism cut in the material under study [72]) or on interferometric techniques (e.g. measuring fringe patterns displacements [81]). In all cases, the sample is put into an oven and exposed to different well-known temperatures. The crystal is free to expand, and the temperature rise into the material is uniform, which is incidentally an essential condition for valuable measurements. In this case, obviously, the crystal experiences thermal expansion, a phenomenon which causes the index to change (to decrease, in general) in virtue of a pure photoelastic effect. In all these practical circumstances, the strain tensor relates directly to the temperature shift by the thermal expansion tensor, in other words the stress terms in equation III.1 are zero. The coefficient measured experimentally can then be regarded as a partial derivative at constant stress:

$$\left(\frac{dn}{dT}\right)_{measured} = \left(\frac{\partial n}{\partial T}\right)_{\sigma} \tag{III.20}$$

On the other hand, the reality experienced by the laser crystal while optically pumped is radically different. We know that in all cases (transverse as well as end pumping, thin disks as well as long and thin rods) *the pumped area inside the crystal is under compression* (negative stresses and strains), which is true as soon as the thermal expansion coefficient is positive. It can be explained qualitatively by saying that the central region of the crystal, yet hotter than the edges, is prevented from expanding by the expanding (cooler) outer parts of the rod, which eventually causes the central region to be under compression.

As a result, we see that if we consider the measured *dn/dT* instead of the partial derivative at constant strain in eq. (III.19), the photoelastic contribution to the thermal lens, already fully taken into account with the term $2n_0^3 \alpha_T C'_{r,\theta}$, which accounts for thermal expansion if any, is partially cancelled by the photoelastic (thermal expansion) term hidden in the measured *dn/dT*.

In a first attempt to correct for this, we thus have to evaluate precisely the thermal expansion contribution to the measured *dn/dT*. It can be done in a simple way by considering the Clausius-Mossotti model for refractive index, which writes:

$$\frac{n^2-1}{n^2+2} = \rho(T)\alpha_e(T,\rho(T))\frac{N_a}{\varepsilon_0 M} \tag{III.21}$$

where ρ is the specific mass (density), M the molecular weight, $N_a$ the Avogadro number, and $\alpha_e$ the polarizability. This expression is valid strictly speaking for isotropic ionic crystals (for covalent crystals the local field correction is smaller and atomic polarizabilities loose their meaning due to the very nature of covalent bonding**)**

The modification of polarizability with temperature results from the change in thermal occupancies and spectra of the energy levels. Tsay *et al*. [82] have developed a two-oscillator model where they consider the contribution of both electronic and lattice vibration terms. The discussion about the different origins of *dn/dT* is beyond the scope of this review and will not be exposed in detail here.

We can differentiate (III.20), assuming that changes in density only originate from isotropic thermal expansion. This is consistent with the experimental procedure used to measure this coefficient.

$$\left(\frac{dn\{\rho(T),\alpha_e(T,\rho(T))\}}{dT}\right)_{measured} = \left(\frac{\partial n}{\partial T}\right)_\sigma = \left(\frac{\partial n}{\partial \rho} + \frac{\partial n}{\partial \alpha_e}\frac{\partial \alpha_e}{\partial \rho}\right)\frac{\partial \rho}{\partial T} + \frac{\partial n}{\partial \alpha_e}\frac{\partial \alpha_e}{\partial T} \tag{III.22}$$

From (III.20) and isotropic expansion assumption we obtain:

$$\left(\frac{\partial n}{\partial T}\right)_\sigma = \frac{(n^2+2)(n^2-1)}{6n}\left[\frac{1}{\alpha_e}\frac{\partial \alpha_e}{\partial T} - 3\alpha_T\left(1 + \frac{\rho}{\alpha_e}\frac{\partial \alpha_e}{\partial \rho}\right)\right] \tag{III.23}$$

This expression puts into the light three contributions to the measured *dn/dT*:

- a pure effect of thermal expansion : $-\dfrac{\alpha_T(n^2+2)(n^2-1)}{2n}$ (III.24)

- the influence of thermal expansion on polarizability,

$$\frac{-\alpha_T (n^2+2)(n^2-1)}{2n} \left[ \frac{\rho}{\alpha_e} \frac{\partial \alpha_e}{\partial \rho} \right] \quad \text{(III.25)}$$

- a thermal expansion-*independent* contribution, which can be assimilated to the partial derivative at constant strain: $\left( \frac{\partial n}{\partial T} \right)_\varepsilon = \frac{(n^2+2)(n^2-1)}{6n} \left[ \frac{1}{\alpha_e} \frac{\partial \alpha_e}{\partial T} \right]$ (III.26)

Since the second and third terms do not breakdown easily into a set of available material physical parameters, no general formula based on the measured *dn/dT* can be derived for the last partial derivative. We see also that thermal expansion appears in both (III.24) and (III.25), so that we cannot in a straightforward way dissociate "pure" thermal expansion from strain-related polarizability effects.

This formulation brings some questions (rather than answers) about the interpretation of some experimental results, in particular those obtained with materials whose measured *dn/dT* is negative.

It is often reported that in such materials (e.g. LiCAF [72, 83], FAP [39], YLF [84]), the thermal lens is weak or even divergent (case observed in YLF crystals) because the negative *dn/dT* counterbalances (or even surpasses) the other positive terms in the expression of the thermo-optic coefficient.

Although, because of the lack of data about these materials, the photoelastic term is just supposed to be positive, or evaluated from other materials whose properties are believed to be similar (Woods *et al.* [72] use data from $CaF_2$ to evaluate photoelastic constants of LiCAF, Payne *et al.* [39] used data from LG-750 phosphate glass to approximate that of FAP). In some cases (e.g. thermal lensing measurements in Nd:YLF crystals [84]), discrepancies are reported between theory and experiment, which do not occur with YAG which has a positive measured *dn/dT*.

A large and negative *dn/dT* coefficient means that the thermal expansion is dominating polarizability contribution for an unstressed crystal; it is observed that such behaviour is generally associated with a large thermal expansion coefficient. This also means that the photoelastic term, proportional to $\alpha_T$, can be expected to be greater. In contrast, we can say nothing about the sign of this term, whose knowledge requires that we know all of the $p_{ijkl}$ coefficients of the crystal.

*This means that there is no obvious relationship between the sign and magnitude of the measured dn/dT and the sign and magnitude of the thermal lens.*

To go further, the only term which is truly always positive is the end faces bulging term. The polarizability dependence on temperature (eq. III.26) is mostly positive too [82]. We can then assess that negative thermal lensing is more likely to be explained by negative photoelastic terms, and/or possibly, by a negative $\frac{\partial \alpha_e}{\partial \rho}$ term.

In conclusion, the crystals with negative measured *dn/dT* have to be considered very carefully as far as simulations are concerned: photoelastic terms must not be neglected. However, it remains that photoelastic contributions, whatever calculated or measured, tend to be small in many crystals. This means that the rude approximation made by replacing $\chi$ by the measured *dn/dT* (+ the end bulging term) will be all the more close to reality that the *dn/dT* is large and positive.

## *III.8. A novel definition for thermo-optic coefficient based on experimentally measurable parameters*.

In the precedent subsection, we saw that there are many problems related to the definition of the thermo-optic coefficient, which are above all related to the abusive use of the parameter (dn/dT) in a context where it is not relevant.

Actually, a better way to describe the phenomena is to start from measurable data that are relevant as far as solid-state lasers are concerned. The partial derivative at constant strain is a formal parameter which has not a real physical meaning since it is impossible to prevent the crystal from any strain, compression or thermal expansion. Furthermore, photoelastic and polarizability effects are so strongly intermingled, that one cannot imagine easily an experiment that could separate clearly the effect of one from the other. That's why we propose to base solid state laser thermal characterization for high power applications on measurable data and separate the thermo-optic coefficient $\chi$ in three truly independent contributions, as follows:

$$\chi = \chi_n + \chi_{bulging} + \chi_{birefringence} \tag{III.27}$$

where

$$\chi_n = \left(\frac{\partial n}{\partial T}\right)_\varepsilon + n_0^3 \alpha_T \left(C_r' + C_\theta'\right) \tag{III.28}$$

$$\chi_{bulging} = (n_0 - 1)(1+\nu)\alpha_T \tag{III.29}$$

$$\chi_{birefringence} = \pm n_0^3 \alpha_T \left(C_r' - C_\theta'\right) \tag{III.30}$$

$\chi_{bulging}$ accounts for curvature of end faces, and is measurable by performing, for example, interferometric or wavefront measurements on a probe beam reflected on each side of the crystal, as done by Baer *et al.* [85] or Kleine *et al.* [86]. The expression given in (III.29) applies to the thin disk ideal model, but its real value can be computed depending on every special geometry, quite easily with a finite element code, since this is a pure thermomechanical problem, where data are more readily accessible or measurable.

$\chi_{birefringence}$ accounts for strain-induced birefringence ("+" for radial polarization, "-" for tangential). It can be measured separately by performing measurements of polarization-dependant astigmatism, for instance, thanks to a wavefront measurement method sensitive to aberrations (see next section).

Eventually, the term $\chi_n$ which accounts for all the refractive index variations with temperature, is not rigorously calculable for all the reasons exposed above, but its exact value can be deduced for each material and for each pumping configuration, from the separate measurement of $\chi$ (global thermo-optic coefficient), $\chi_{birefringence}$ and $\chi_{bulging}$.

To conclude, let us give some orders of magnitude of different terms in widespread YAG. The measured value of the global thermo-optic coefficient (see next section of this article for details) for this material under diode pumping is $10 \, 10^{-6} \, K^{-1}$.

The different contributions calculated from tabulated data are:

$$\left(\frac{dn}{dT}\right)_{measured} = 9.10^{-6} \, K^{-1}$$

$$\left(\frac{dn}{dT}\right)_{measured} + \frac{\alpha_T (n^2 + 2)(n^2 - 1)}{2n} = 31.5 \, 10^{-6} K^{-1}$$

$$\chi_{bulging} = (n_0 - 1)(1 + \nu)\alpha_T = 7.2 \, 10^{-6} \, K^{-1}$$

$$2 n_0^3 \alpha_T \, C_r = +0.27 \, 10^{-6} \, K^{-1}$$
$$2 n_0^3 \alpha_T \, C_\theta = -0.93 \, 10^{-6} \, K^{-1}$$

(see appendix for the detail of the calculation of photoelastic constants within the plane stress approximation).

The thermo-optic coefficient are small, but the bulging term is far from being negligible. Weber *et al.* [87] have shown that in Nd:YAG, bulging represented 30% of the global thermal lens. We calculated also for information a "corrected" *dn/dT*, which is obtained after subtraction of the inappropriate thermal expansion term (III.24): this coefficient turns out to be very large here.

In diode-end-pumped Nd:YLF and Nd:YVO$_4$, Baer et al. [85] and Kleine et al. [86] have shown experimentally that the bulging term represented half of the total thermo-optic coefficient. A finite-element analysis, performed by Peng et al. [88], leads to a similar conclusion for a Nd:YVO$_4$ crystal.

### III.9. The aberrations of the thermal lens.

In conclusion to this chapter, we will introduce the thermal lens higher-order distortions *id est* the thermal lens aberrations.

For a perfectly parabolic distortion of the wave front or equivalently a pure thermal lens the thermal distortion can be easily compensated by addition in the laser cavity of the opposite divergent lens or by adjusting the distance of the different cavity elements. But, if aberrations are present the compensation is very difficult and requires complex systems [89]. While uncorrected, these aberrations lead to degradation in beam quality (brightness), and also to losses due to diffraction of the beam high spatial frequencies [90].

The aberrations are present when the wavefront distortions induced by the absorbed pump beam are not perfectly parabolic. This occurs when the longitudinal pump beam has not a true top-hat profile, for example in the case of a Gaussian pump beam profile [91]. Moreover, if the laser beam size is larger than the pumped area, the aberrations also become important as shown in figure 17. The rays, far from the pumped area are almost not deviated. This is the signature of spherical aberration (which is sometimes referred as a "thermal lens varying with radius r".)

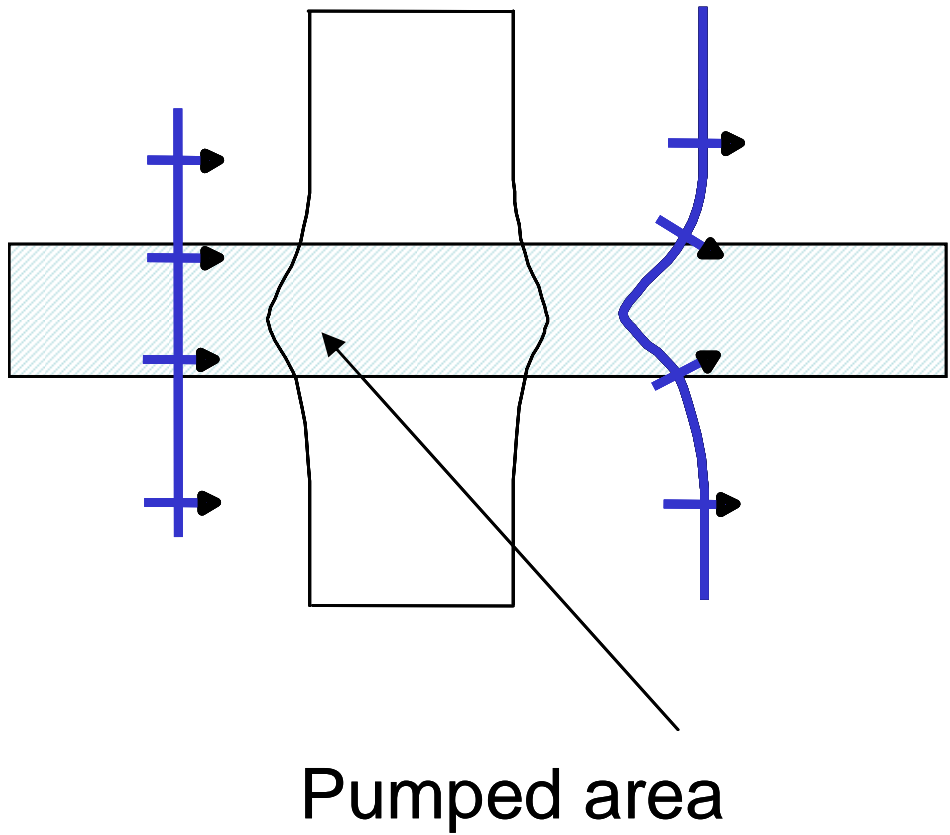

**Figure 17**: *When the laser beam is larger than the pumped area, spherical aberration is observed.*

In general the aberrations affect the laser modes of the cavity in a way that to degrade the beam quality. This degradation can be evaluated by the $M^2$ factor or the Strelh ratio for exemple [92]. We are just giving here an exemple of the influence of aberrations on the beam quality, some results given by Clarkson [70].

Let us consider that only 3$^{rd}$ order spherical aberration is present. In this case, the optical path difference $\Delta(r)$ can be written as:

$$\Delta(r) = \frac{r^2}{2 f_{th}} - C_4 r^4 \qquad (III.31)$$

if the laser initial beam $M^2$ factor is $M_i^2$, the $M^2$ with the added aberations is :

$$M_f^2 = \sqrt{\left(M_i^2\right)^2 + \left(M_q^2\right)^2} \qquad (III.32)$$

with

$$M_q^2 = \frac{8\pi C_4 w_L^4}{\lambda\sqrt{2}}$$ (III.33)

where λ is the wavelength and $w_L$ the laser beam waist.

One can show that obviously $C_4=0$ for a top-hat pump profile provided that $w_L < w_p$. But for a gaussian pump-beam profile (beam waist $w_p$) we obtain [70]:

$$M_q^2 = \frac{2\chi\eta_h P_{abs}}{K_c \lambda\sqrt{2}} \left(\frac{w_L}{w_p}\right)^4$$ (III.34)

In that case, the $w_L/w_p$ is present to the power of 4, which means a strong increase of $M^2$ even for small mode mismatch.

Another classical aberration to be considered is the thermal astigmatism. In this case, the analytical solution is not as simple as with the 3$^{rd}$ order spherical aberration and we will only focus on the qualitative approach, answering this simple question: in which conditions does the thermal lens exhibit astigmatism? In practice it will occur whenever:

- the thermal conductivity is anisotropical ;
- the thermal expansion tensor $a_{T_{ij}}$ does not reduce to a scalar quantity;
- the cooling is inhomogeneous;
- the laser beam is *polarized*, even for an isotropic material. This is the so-called polarization-dependant astigmatism, which can be used a sa probe to evaluate the strain-induced birefringence (see figure 18.)

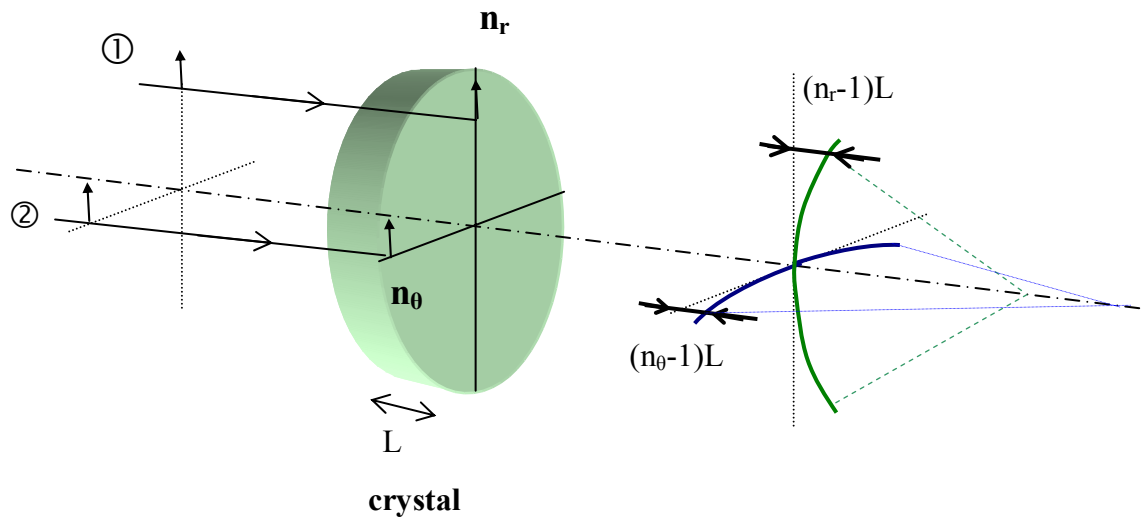

**Figure 18**: *Illustration of polarization-dependant astigmatism. Here a vertically polarized strikes the crystal from the left. The ray # ① sees the radial index of refraction $n_r$ while ray # ② sees the tangential index of refraction $n_\theta$. If for the sake of simplicity the indices are considered constant over the whole crystal length L, the astigmatism is $(n_r - n_\theta)L$.*

As a conclusion for this part, we present a schematic diagram (fig. 19) summing up the different thermal effects arising in a solid state laser medium, and how they are related to each other.

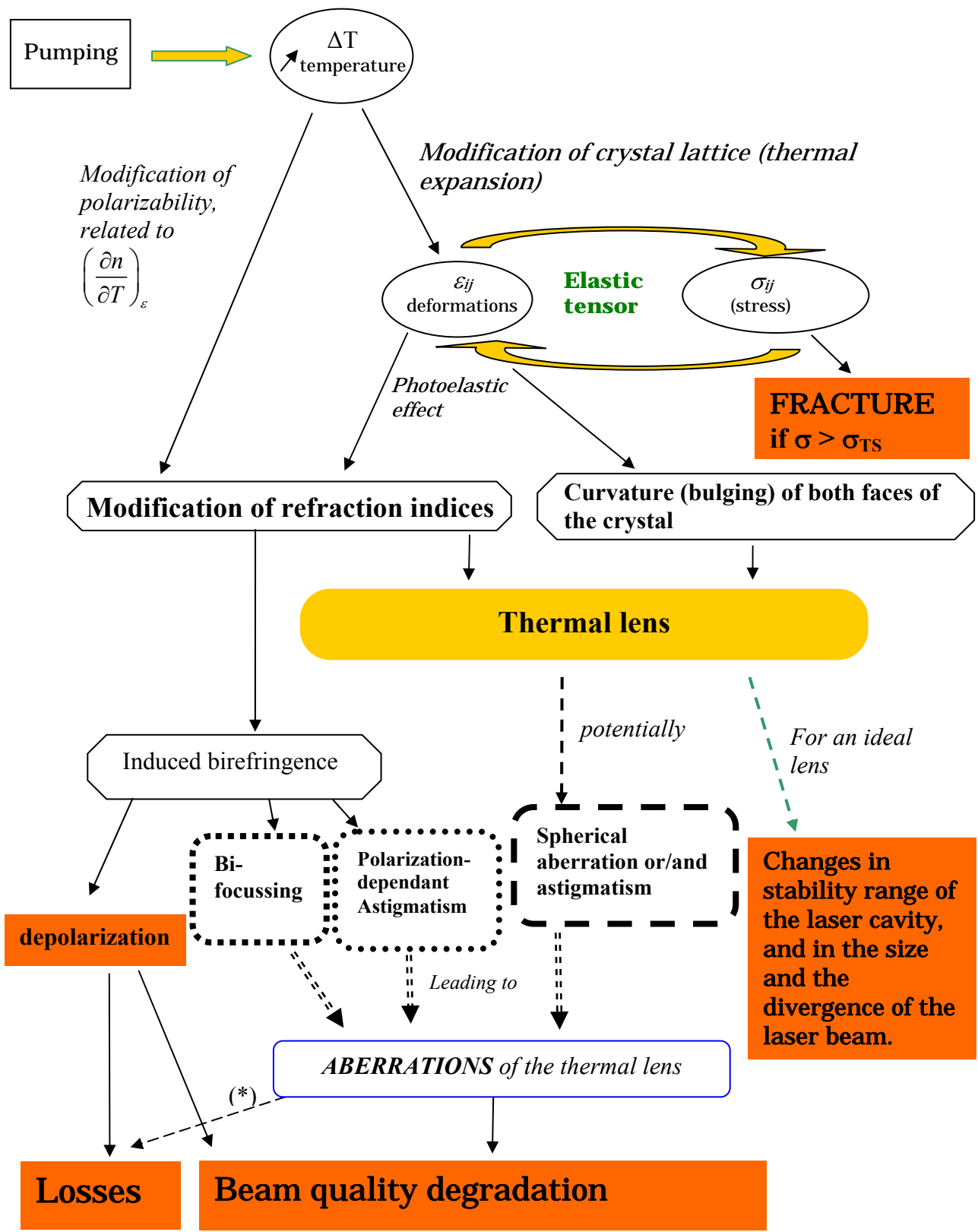

**Figure 19**: *Summary of the thermal effects in solid-state lasers. The observable consequences observable are presented in full rectangles. The aberrations can be split in two classes: the ones*

*that do not depend on the polarization (lined rectangles) and the ones that depend on polarization (dotted rectangles) which come from the strain-induced birefringence.*

*(\*): 2 types of losses are induced by aberrations: the diffraction losses (associated with degradation in beam quality) and the losses induced by the eventual presence of a diaphragm in the cavity to prevent from oscillation of higher order laser modes.*

# IV. Thermal lensing techniques

## IV.1. Introduction

The first evidence of thermal effects in lasers were demonstrated in 1965 by Gordon, Leite and Whinnery [93] working at Bell Labs on He-Ne lasers for Raman spectroscopy applications. The use of liquids to Q-switch the laser lead to the observation of unexpected effects such as relaxation oscillation of jump of modes. The exceptionally long time constant of this phenomenon (several seconds) lead to the conclusion that a thermal lens was at the origin of the observed effects. This lens was created by the small absorption occurring in the liquids. The first application proposed by the authors was then to use it to measure very small absorption coefficients down to $10^{-4}$ cm$^{-1}$. As a matter of fact, the so-called "thermal lens method" allows nowadays to measure absorption coefficient lower than $10^{-7}$ cm$^{-1}$ [94]. Since 1965, the photothermal-methods panel available to physicians has grown in diversity [95]. Actually, due to the complexity of simulating thermal effects in lasers (see previous subsections), the experimental determination is often the only accurate method.

Since the 70's, numerous attempts have been done to measure the thermal lens in solid-state laser media. They can be classified in three categories: the geometrical methods based on the deflexion of a beam, the methods based on the properties of cavity modes, and the methods based on the wavefront measurements.

## IV.2. Geometrical methods

These methods are probably the simplest methods to measure the thermal lenses. They can be separated in two sub-groups: one can exploit either the defocussing or the deflection of a beam passing through the pumped medium

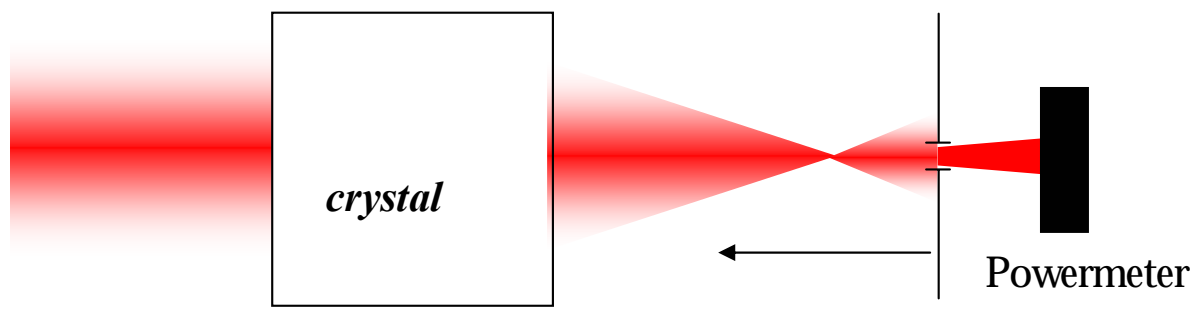

**Figure 20 :** *Example of thermal lens measurement based on the displacement of the focal point.*

The principle of the first category of methods is very simple. Considering a probe beam going through the crystal, the measurement of the axial shift of the focal point position allows to retrieve directly the thermal lens using simple geometrical optics. When the rod is relatively large, one can use a collimated beam of comparable size to directly measure the position of this focal point by Z-scan measurement for example as shown in figure 20 (a small aperture is longitudinally translated in order to find the maximum of probe-beam transmission [96].) This method is based on the assumption that the lens is perfect (without aberrations) and then is especially suited to transverse pumping with large-size materials . In fact, the method is not easily applicable to end-pumped lasers because of the very large depth-of-focus associated with a very small and low solid-angle–probe beam, and obviously does not yield the thermal lens aberrations.

Moreover, for small beam diameters (10-100 µm), this method needs to be generalized using Gaussian beam optics. Hu and Whinnery [97] described a simple method to evaluate the thermal lens with a probe beam whose size is comparable to the cavity beam in end-pumping schemes. This last method is described in figure 21: it can be based either on the measurement of the divergence of the probe beam [97], or in the measurement of the beam diameter in an appropriate plane.

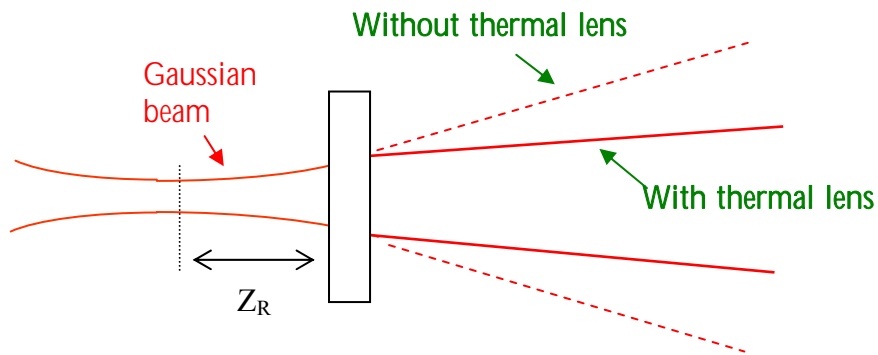

**figure 21 :** *Example of measurement based on focal point displacement for tightly focused probe beam according to Hu and Whinnery's method.*

To conclude on the methods based on the focal point displacement, we can say that their main drawback is their low accuracy. The relative precision is only 20-30% on the focal lens measurement.

We can use geometrical optics in another way by measuring the deflection of a probe beam. Instead of using a beam covering the whole pumped area, we can measure the deflection of a small beam slightly off-axis as shown in figure 22.

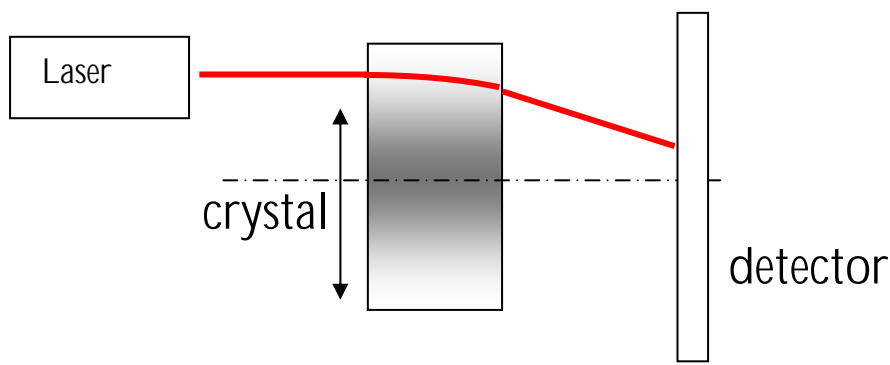

**figure 22** : *Example of measurement based on deflection measurement [98].*

The whole surface can be scanned in order to measure not only the focal lens but also the aberrations due to the thermal effects [98]. However this method is complex and it can only be used for large transversally-pumped crystals (even if the resolution can be lower than the probe beam size [99]). Moreover this method can be considered as a point-to-point measurement of the simpler Shack-Hartmann technique described later.

### *IV.3. Methods based on the properties of cavity eigenmodes*

After the development of end-pumped lasers (particularly thanks to the increasing performance of laser-diodes), alternative methods appeared. These less straightforward methods are based on the properties on the laser cavity eigenmodes, in particular on the fact that the thermal lens affects the stability zones of a laser cavity. All these methods are based on the theory of paraxial beam propagation theory in cavities presented by Kogelnik and Li [100] and can be formalized using the ABCD matrices. An example of this influence of the thermal lens on the stability of a cavity is given in figure 23. As a direct consequence, these methods consider ideal lens (aberration-free) and do not give any information on the thermally-induced aberrations. Nevertheless these methods remain easy to implement since there is no probe beam (the laser itself is used to measure the thermal lens). In counterpart, it is impossible to measure the thermal lens in absence of laser extraction (with pumping but no laser emission). Here are three different examples of these methods based respectively on the work of *Frauchiger and al., Neuenschwander and al.* and *Ozygus et al.*

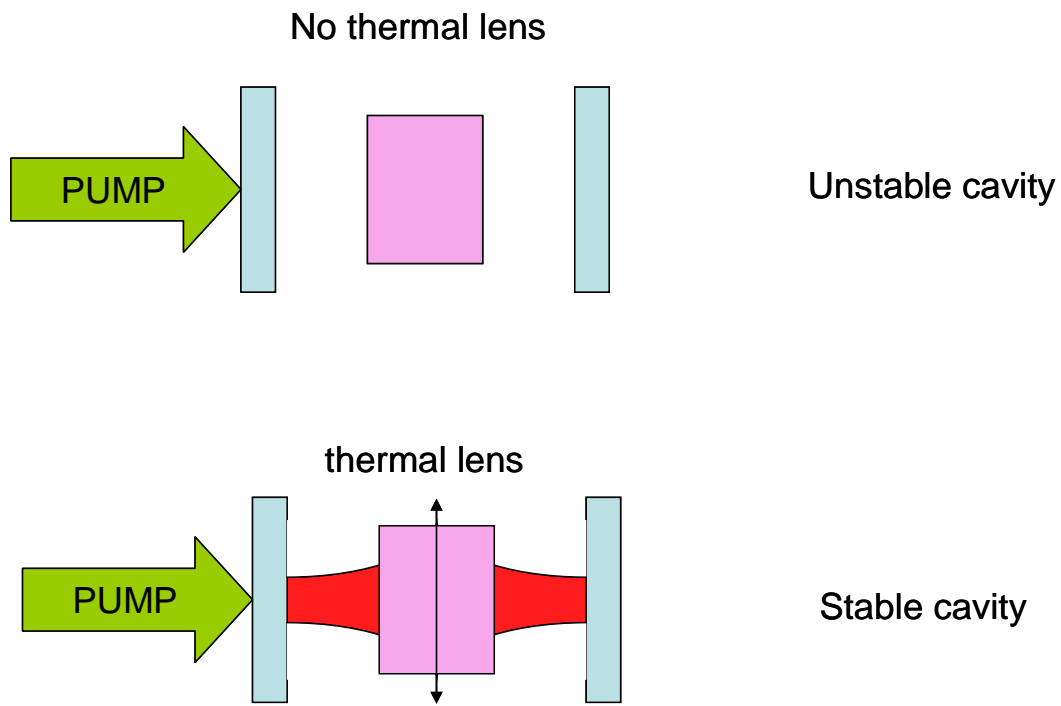

**Figure 23:** *example of this influence of the thermal lens on the stability using a plano-plano cavity [101].*

- Frauchiger *and al.* [102] measured the divergence of the output laser beam in a diode-pumped Nd:YAG and retrieved the thermal lens by a paraxial calculation. This example allows us to perceive the limitation of these techniques using the properties of cavity modes. Indeed, they are very sensitive to the beam quality. The measured divergence is actually directly proportional to the $M^2$ factor of the beam. If the laser mode is not perfectly $TEM_{00}$ the reliability of the method is strongly affected.

- Neuenschwander *and al.* [101] used a plano-plano laser cavity stabilised by the thermal lens (figure 23). Adding two extra-cavity lenses and measuring the beam diameter in different longitudinal positions they found the waist in the cavity and therefore the thermal lens. This method is interesting because it allows the simultaneous measurement of the $M^2$ factor which reduces the limitations due to the beam quality. In that case, the limitation is due to the precision on the distances between the different optical components which is not always perfectly known.

- Ozygus *et al.* [103] proposed an alternative method that consisted in using the frequencies of different transverse modes. In fact, the frequencies of the different transverse modes (in a plano-concave laser cavity) depend not only on the length of the cavity but also on the radius of curvature of the mirrors (considering the flat mirror and the crystal as another concave mirror). If one achieves to have two modes lasing simultaneously in the cavity, the measurement of the beating frequency with an optical spectrum analyser allows retrieving the thermal lens. Another upgraded technique based on the same effect was also presented by Ozygus *et al.* later in 1997 [104]. In the last one, one translates one of the mirrors to find the positions of the spectral degeneracy (when two eigenmodes have the same frequency). The advantage of this method is its capacity to be used for measuring long-focal-length thermal lenses. It allows the measurement of focal lengths as long as 5 m.

In conclusion, the methods based on the properties of the cavity modes allows to have relative precisions on the thermal focal lens of 15 % for a TEM00 beam but down to 60 % for an non-diffraction limited beam [101].

## *IV.4. Methods based on wavefront measurements*

In this part, we will distinguish three wavefront measurement techniques: "classical" interferometry (based on fringe measurements on Michelson, Fizeau, … interferometers), shearing interferometry, and Shack-Hartmann sensing.

### *IV.4.1. Classical interferometric techniques*

69 /115

As far as "classical" interferometry is concerned, one can consider equal-thickness fringes between the parallel end faces of the rod (this is the so-called Fizeau interferometer); one can also insert the rod under study in an arm of a Michelson-type [105] or Mach-Zehnder -type interferometer [106] for example. The first type of method is simpler since it does not require a second interferometrically adjusted arm.

These methods are particularly well suited for large amplifier rods but in counterpart, they are not convenient for end-pumping. Indeed, even if there is no fundamental contraindication to use this method for small spots, in practice in this case the number of fringes is too small to obtain an exploitable interferogram. As an example, if we consider a 200-µm diameter probe beam, a focal length of 5 cm only induces a phase shift between the centre and the edge of the beam of $\Delta = h^2/2f = \lambda/6$ (with $h$ the radius of the beam, $f$ the focal length, and $\lambda$ the wavelength of the probe beam, here $\lambda =670$ nm). In this case, of course, no fringes are visible.

To overcome this problem, one can choose to take a probe beam larger than the pumped area, given that in these conditions, as shown in figure 17, some spherical aberration will be present, which requires an additional numerical model to fit the data and retrieve the thermal lens, as done by Pfistner et al. [107]. The weakness of this method relies precisely in the retrieving algorithm since the whole interferogram consists on only a few fringes. The precision is then a hard point and in the best case this precision is evaluated to λ/4. This work [107] has also been done on YAG, GSGG and YLF crystals.

### IV.4.2. Shearing interferometric techniques

A classical solution to obtain information on the phase "between two fringes" is phase shift interferometry. This technique has been used by Khizhnyak *et al* [108] with longitudinally-pumped Nd :YAG lasers. This method is easy to implement since it's based on commercial products, but it remains quite unused due to its important cost. One generally prefers lateral-shearing-interferometer methods.

Methods based on lateral shearing interferometers are particularly well suited to end pumping. The principle is the following: the beam is duplicated in several replicas, typically 2 [109-110], 3 [111] or 4 [112-113] (as presented in figures 24 and 25 with a tri-wave lateral shearing interferometer setup).

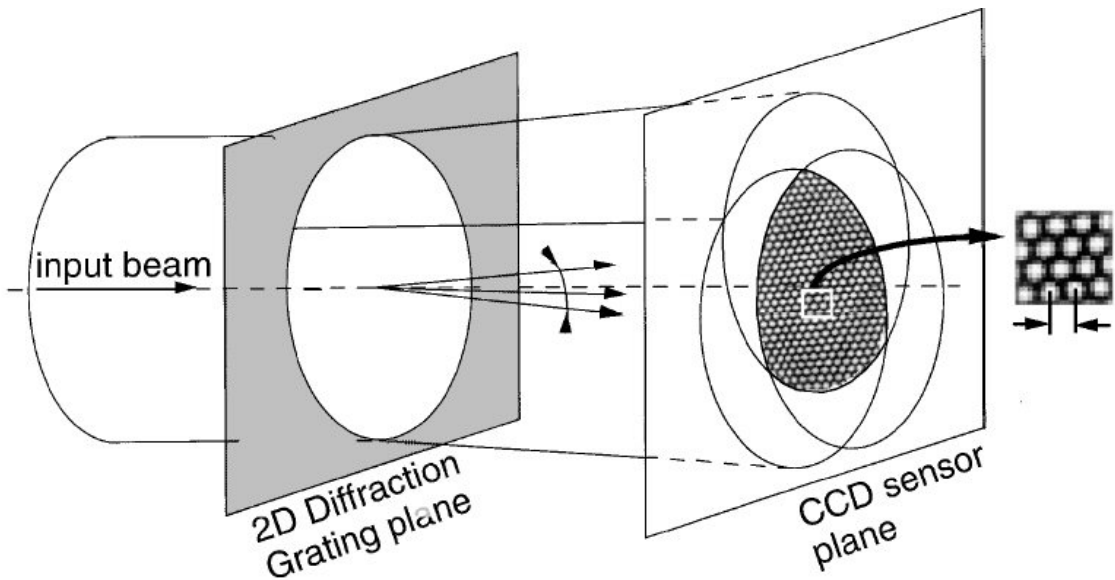

***Figure 24 :*** *Example of tri-wave lateral shearing interferometer (courtesy of J.C. Chanteloup).*

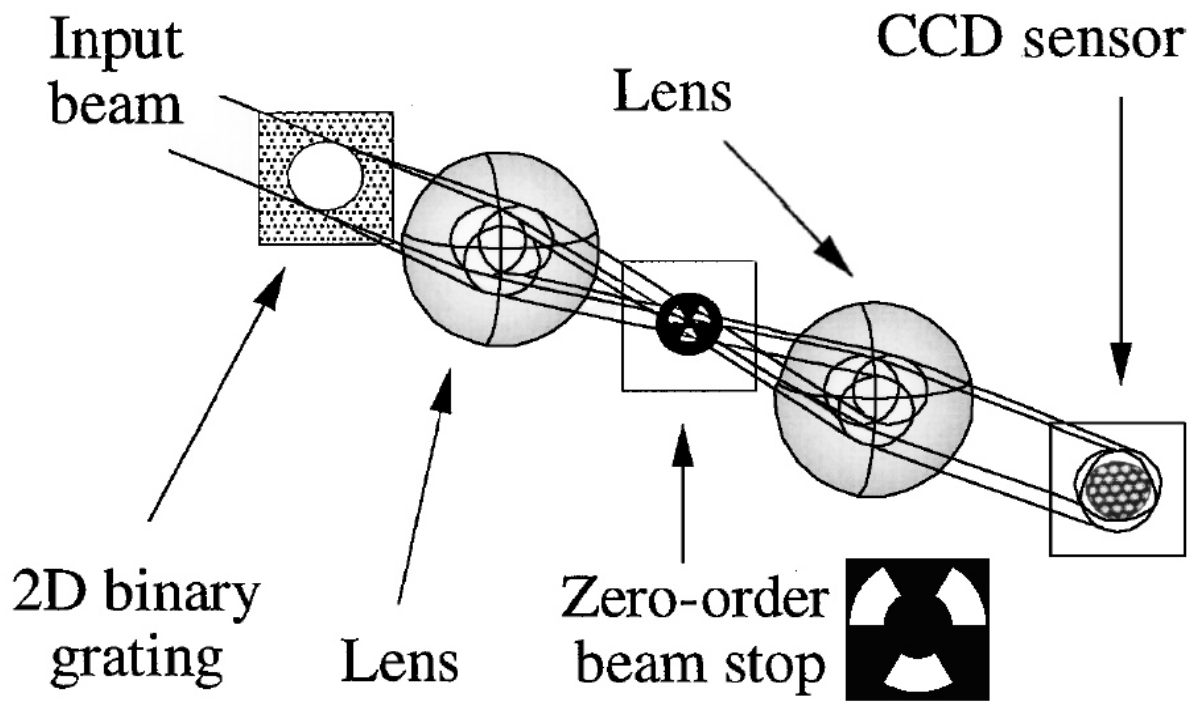

**Figure 25:** *Example of tri-wave lateral shearing interferometer setup (courtesy of J.C. Chanteloup).*

The replicas are slightly shifted from each other in the lateral direction which provides an interferogram whose fringes (for 2 waves) or dots (for 3, 4 waves) separation give information on the derivative of the wavefront. For example in the absence of wave-front distortion the lines are rectilinear for the 2-wave shearing interferometer, they form an homogenous honey-comb for the 3-wave shearing interferometer and perfect squares for the 4-wave shearing interferometer. This technique is more sensitive than classical interferometry since the sensitivity is tunable by adjusting the shearing distance. The larger the shift, the more precise the technique . The precision of this method is excellent since it is in the order of $\lambda/50$ [109] and can even reach $\lambda/200$ [111].

The method using 3 (or 4)-wave « trilateral shearing interferometry » [111] has the advantage over the 2-wave shearing interferometry to allow the cartography in the 2 dimensions in one acquisition. This method is simple to implement (figure 25) since the splitting of the beam can be realized readily with a 3D grating. Moreover the use of a grating makes this method totally achromatic which allows its use for broad-band lasers such as femtosecond lasers. In 1998,

Chanteloup *et al.* [112] reported on the wavefront distortions of a terawatt-class femtosecond laser system with an accuracy of λ/50.

This method was also used to characterize thermal lensing in Ytterbium-doped materials, namely in Yb:YAB by J.L. Blows *et al.* [110], who used the 2-wave shearing interferometer technique to measure thermal lenses, and then thermal conductivities and fractional thermal loadings. The experimental setup used is reproduced on figure 26.

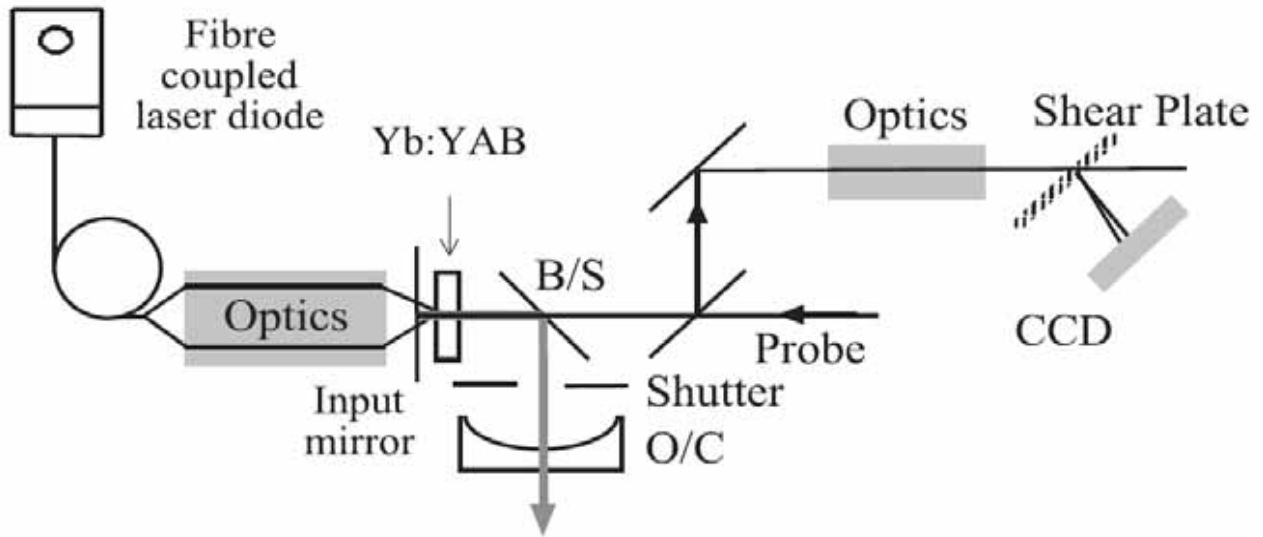

**Figure 26:** *setup for measuring thermal lensing in diode-pumped Yb:YAB crystal with lateral shearing interferometry from [110] (courtesy of J. Dawes, Centre for lasers and Applications, Sydney)*

The use of a probe beam at 530 nm allowed the thermal lens to be evaluated under lasing and nonlasing conditions (that is with a shutter inside the cavity), which is an essential requirement to perform radiative quantum efficiencies measurements.

### *IV.4.3. Methods based on Shack-Hartmann wavefront sensing*

Although it can be seen also as a multiple beam interferometric method, we separate this method from the previously mentioned ones since it can be understood easily with geometrical optics.

In 1900, Hartman proposed to use a drilled plate [115] to measure wavefronts. The principle is simple: since light rays run perpendicular to the wavefront, one can retrieve the local wavefront slope as soon as the direction of the ray is measured. The Hartmann plate is made of small apertures which scatter the beam into regularly-spaced diffraction patterns, and behind which is located a detector (typically a CCD camera nowadays). In 1971, Roland Shack and Ben Platt improved the Hartmann setup by replacing the array of holes by an array of microlenses (figures 27, 26).

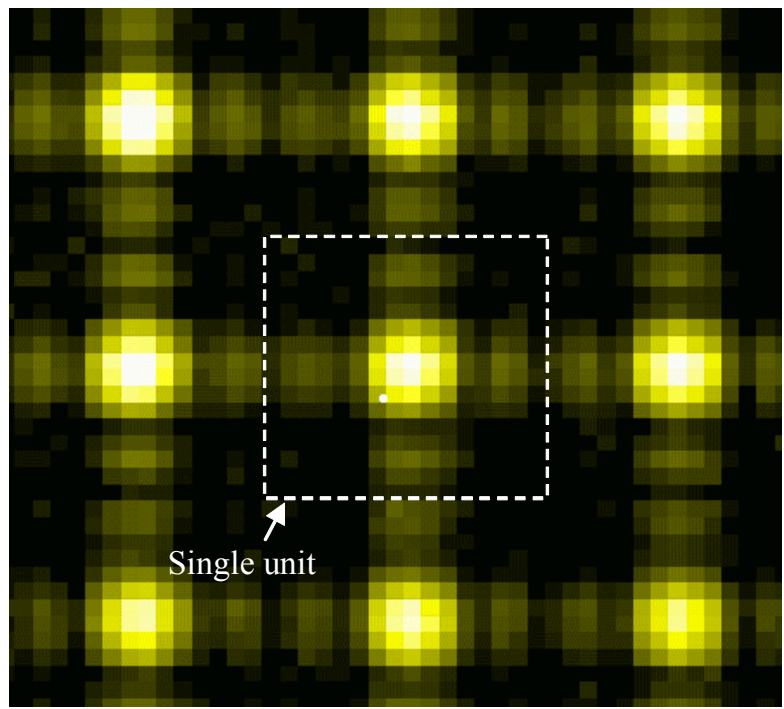

**Figure 27:** *example of intensity pattern observed in the focal plane of the Shack-Hartmann microlenses.*

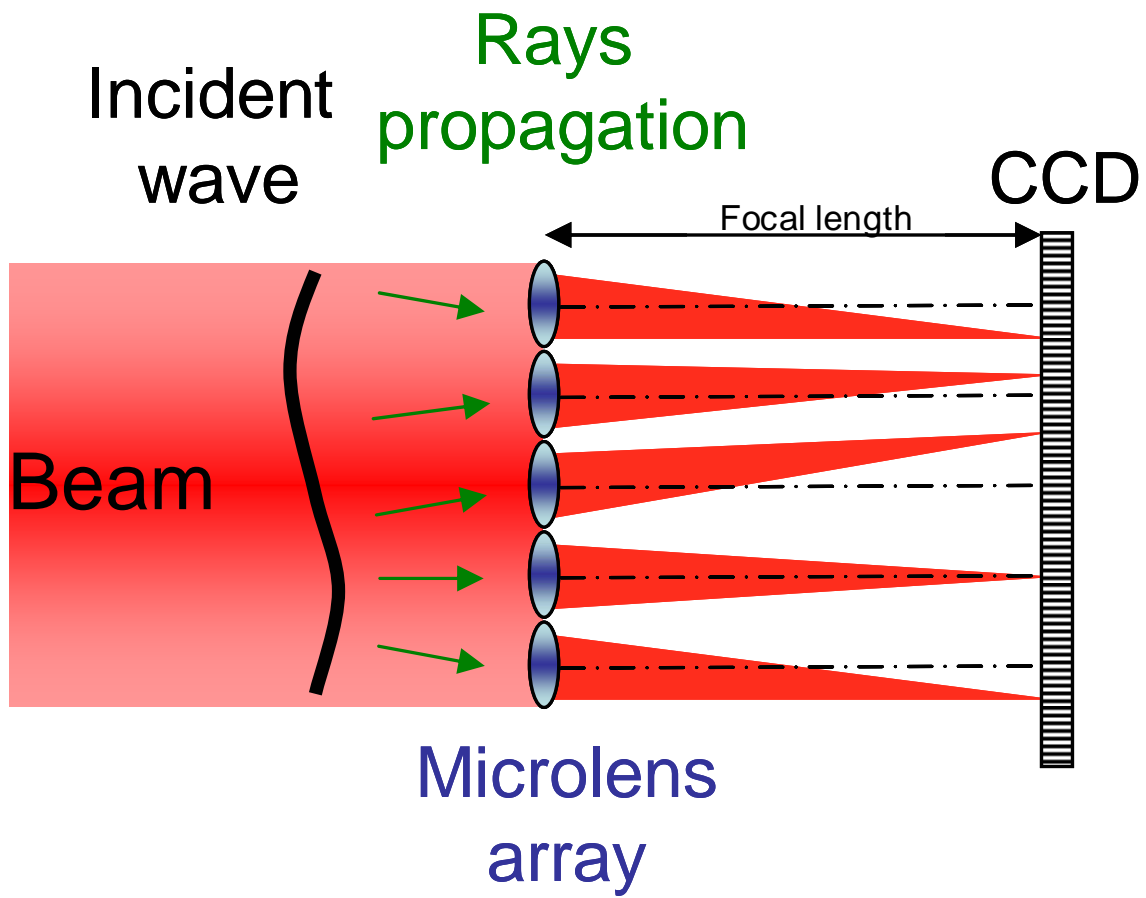

**Figure 28:** *Shack-Hartmann wavefront sensor setup.*

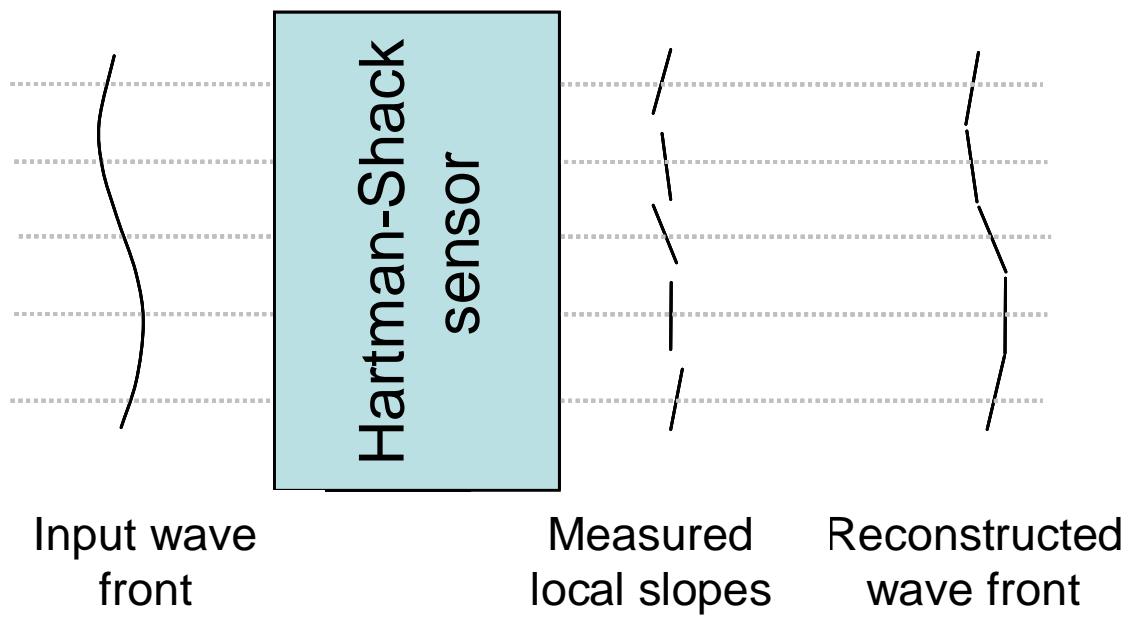

**Figure 29:** *Shack-Hartmann sampling.*

The small axial shift of the microlens diffraction pattern centroid (see figure 28) is directly proportional to the average local slope of the wavefront on the aperture of the micro-lenses. The displacement vectors allow having 2D information. The standard sensitivity of such systems is typically λ/100 RMS for commercial products, which settles this technique as a competitor of lateral shearing interferometry. One of the advantages of this method, compared to interferometric ones, is its insensitivity to mechanical vibrations or thermal fluctuations. In counterpart, the principal limitation of this kind of sensor is the discrete sampling that limits the transverse resolution (figure 29) and thus prevents from obtaining information about high-spatial-frequency phase distortions. Nevertheless, it's noteworthy that this point is not a problem for thermal lensing and aberration measurements because in virtue of the general heat equation (II.1.1), temperature variations in a crystal are smooth even if the thermal load exhibits sharp variations.

In 1998, Armstrong [116] reported the measurement of thermal lens in transversally-pumped Nd:YAG and Nd:YAP rods. More recently Ito *et al.* [117] and Pittman *et al.* [118] reported the use of a Hartmann-Shack sensor to measure the thermal lens in Ti:sapphire rods of terawatt-class femtosecond lasers.

In 2001, reports of temporal changes of thermal lens effects on high power pulsed Yb:glass lasers have been done by Nishimura *et al.* using a Shack-Hartmann wavefront sensor [120]**.** Here the sensor was used for its ability to yield real-time (100 Hz sampling rate) estimation of Zernike coefficients of aberrations, and allowed to measure characteristic thermal relaxation times.

Our group reported recently [73] a derivation of Armstrong's work to measure thermal lensing in various diode-pumped Yb:doped crystals, under lasing or nonlasing conditions. The setup appears in figure 30. The probe beam was a laser diode at 670 nm coupled in a single mode fiber, chosen for its high spatial coherence (an essential feature to correctly define the "reference wavefront") and its low temporal coherence (necessary to avoid coherent cross talk, that is interference between two neighbouring microlens diffraction patterns.) After collimation by a microscope objective, the probe beam was focused onto the crystal and superimposed with the pump beam. The crystal was then imaged upon the microlens array using a magnifying relay imaging system. An uncoated glass plate was inserted in the pump beam path to reflect the probe beam towards the sensor. A selective interference filter at 670 nm was added in front of the sensor to eliminate any unwanted signal at the pump or laser wavelengths. A « reference wavefront » is recorded when the pump diode is turned off, which includes all static aberrations of the optical elements and of the cold crystal itself. It is then subtracted to the measured wavefront when the pump is on. Thus, only phase distortions originating in thermal effects are recorded. The phase front was then reconstructed by projection over the set of the orthogonal Zernike polynomials [70].

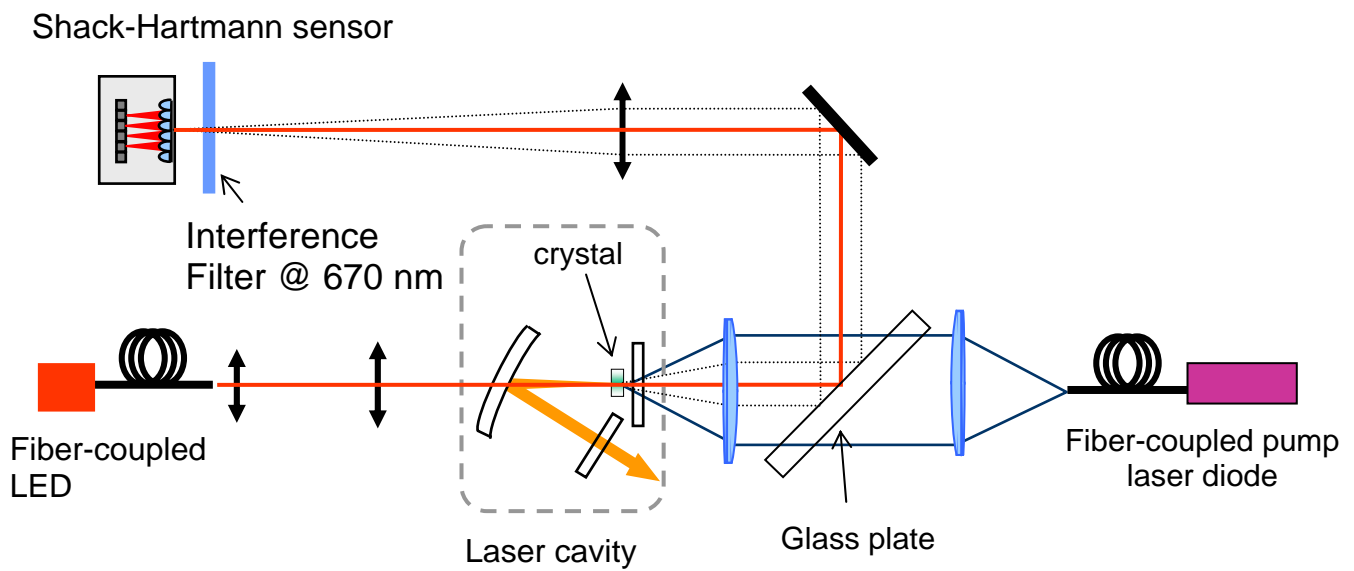

**Figure 30:** *Experimental setup used to measure thermal lensing in diode-pumped Yb-doped crystals with a Shack-Hartmann wavefront sensor, from [73].*

### *IV.4.4 Other techniques*

There exists some more marginal and less used techniques to measure thermal lensing. For example, it is possible to achieve phase reconstruction basing on Fourier optics. One can show that it is possible to retrieve the phase by knowing the intensity profile in different planes linked by Fourier transformation. For instance it is quite obvious that a uniform intensity over a circular aperture and an Airy pattern appearing in two Fourier-related planes implies that the wavefront propagating from one plane to the other is purely spherical. Nevertheless the general inverse problem is far from being obvious. Grossard *and al.* [114] proposed a technique for measuring thermal lensing aberrations in a diode-pumped Nd:YVO$_4$ crystal using intensity profiles in 3 planes and a complex phase retrieval algorithm derived from Gerchberg and Saxton's work.

## *IV.5. Conclusion*

In conclusion, we made a review of the main methods used to experimentally measure the thermal effects in lasers, putting in emphasis the advantages and the limitations of each method. A summary of this review is presented in table 4.

| Methods | Is it taking into account aberrations? | Is it adapted to end-pumping? | With AND without laser effects? | Under real laser conditions? | Narrow spectrum probe beam required? | Insensitive to vibration? | Real time acquisition? | Typical precision | implementation/ cost |
|---|---|---|---|---|---|---|---|---|---|
| **Based on focal point displacement** | no | Very difficult | yes | yes | yes | yes | yes | low $\frac{\Delta f}{f} \approx 30\%$ | Very easy / very cheap |
| **Based on cavity properties** | no | yes | Only with | Almost impossible | No probe | yes | yes | Low $\frac{\Delta f}{f} \approx 15 \text{ à } 60\%$ | Difficult in general / cheap |
| **« classical » Interferometry** | yes | Very difficult | yes in theory | no | no | no | yes | low $\sigma_\Delta \sim \lambda/4$ for the interferometry without phase shift. | Difficult / Cheap (except phase shift interf.) |
| **lateral shearing interferometry** | yes | yes | yes | yes | yes | yes | yes for the 3-4-wave shearing | high $\sigma_\Delta \sim \lambda/50$ to $\lambda/200$ | medium / expensive |
| **reconstruction de la phase from different position intensity profiles** | yes | yes | yes | yes | yes | yes | no | unknown | Difficult / no commercial |
| **Hartmann- Shack** | yes | yes | yes | yes | yes | yes | yes | high $\sigma_\Delta \sim \lambda/100$ | easy/ expensive |

***Table 4**: summary of the different techniques used to measure thermal lensing. $\sigma_\Delta$ denotes the reported RMS deviation on phase shift.*

## V. Thermal lensing measurements in ytterbium-doped materials: the evidence of a non radiative path

We will conclude this review by giving some examples of experimental thermal lensing measurements in diode-pumped Yb-doped crystals. Most of the examples are taken from our previous publications [73,121]: the reader interested in the technical details of the experiments, as well as by the theoretical considerations underneath, is invited to refer to these works.

## V.1. the thermal load in Yb-doped materials

What are the heat sources in a laser medium ? Following T. Y. Fan and using the same notations [**8**], the fractional thermal load $\eta_h$ (that is the fraction of the absorbed pump power converted into heat) can be written:

$$\eta_h = 1 - \eta_p \left[ (1-\eta_l)\eta_r \frac{\lambda_p}{\lambda_f} + \eta_l \frac{\lambda_p}{\lambda_l} \right] \tag{V.1}$$

Where:

- $\lambda_p$, $\lambda_l$, $\lambda_f$ are the pump wavelength, the observed laser wavelength, and the mean fluorescence spectrum wavelength, respectively;

- $\eta_p$ is the pump quantum efficiency, which is the fraction of absorbed pump photons contributing to inversion. Non-unity pump quantum efficiency accounts for residual absorption of the undoped crystal, or can be related to the presence of nonradiative sites.

- $\eta_r$ is the radiative quantum efficiency for the upper manifold: it represents the fraction of excited atoms that decay by a radiative path (in absence of stimulated emission). Non-unity radiative quantum efficiency can be related to multiphonon relaxation (although it is very unlikely since a large number of phonons are necessary to bridge the 10 000 cm$^{-1}$ gap separating the excited and ground manifolds) or more probably to concentration quenching. The latter phenomenon corresponds to the trapping of the energy by a color center, an impurity, or a lattice defect ($Yb^{2+}$, rare-earth impurities or hydroxyl groups have been evoked as possible "dark sites" [122-124]) after several transfers of excitation between neighbouring ions.

- $\eta_l$ is the laser extraction efficiency, defined as the fraction of excited ions that are extracted by stimulated emission. An expression of the laser extraction efficiency can be derived by writing the stimulated, spontaneous and nonradiative relaxation rates [73]. In most cases an approximate relation can be used by neglecting reabsorption at laser wavelength. In the latter case we obtain the simplified expression:

$$\eta_l \approx \frac{\sigma_{em}(\lambda_l) I}{\sigma_{em}(\lambda_l) I + \frac{1}{\eta_r \tau_{rad}}} \quad (V.2)$$

where I is the intracavity laser intensity, and $\sigma_{em}(\lambda_l)$ the emission cross section at laser wavelength. As shown by eq. (V.2) the laser extraction efficiency tends towards 1 for intracavity laser intensities that surpass the laser saturation intensity. Generally, cw oscillators based on Yb-doped materials work with high reflectivity output couplers: as a consequence the intracavity laser intensity is very high, at least one order of magnitude higher than the laser saturation intensity, so that $\eta_l$ is typically close to unity in an operating Yb laser. In this case, the thermal load becomes nearly independent on the radiative quantum efficiency, and is only given by the quantum defect. Nevertheless, the quantum efficiency directly affects the excited state population and has subsequently crucial importance for Q-switched lasers or low repetition rate amplifiers. Incidentally, the relation (V.2) also illustrates that the performance of an Yb-based *cw oscillator* becomes nearly independent of the emission cross section at laser wavelength, provided that the pump intensity is far higher than the pump saturation intensity.

## *V.2. Evidence of nonradiative effects in Yb-doped materials: the example of Yb:YAG*

During the past decade, it has been assumed and claimed many times that nonradiative effects did not exist in Yb-doped materials, in virtue of the very simple electronic structure that prevented deleterious effects that are well known with other ions, such as cross relaxation, excited state absorption or upconversion. Nevertheless, all the measurements performed in the past few years have all brought a contradiction to this statement. Blows *et al*. have demonstrated clear evidence of a nonradiative path in Yb:YAB [110], Ramirez et al. [**125**] in Yb:MgO:LiNbO$_3$ , and as for YAG, non-unity quantum efficiencies have been reported by Barnes *et al.* [**126**], Patel *et al.* **[127],** Ramirez *et al*. [**125**] and Chenais *et al.* **[121]**. This recent work has also shown the existence of nonradiative effects in Yb:GGG, Yb:YSO, Yb:KGW, Yb:YCOB, Yb:GdCOB.

An example of quantum efficiency measurement thanks to the thermal lens method is shown in figure 31 with an Yb:YAG crystal. A simple qualitative explanation is given in figure 32. The clear difference between the thermal lens dioptric power under lasing and nonlasing conditions can be modelled using eq. (V.1) and eq. (V.2), provided that laser power is measured simultaneously. Given some additional approximations, detailed in [121], one can retrieve both radiative quantum efficiency and thermo-optic coefficient $\chi$. The results recently reported for different crystals have been summarized in table 5.

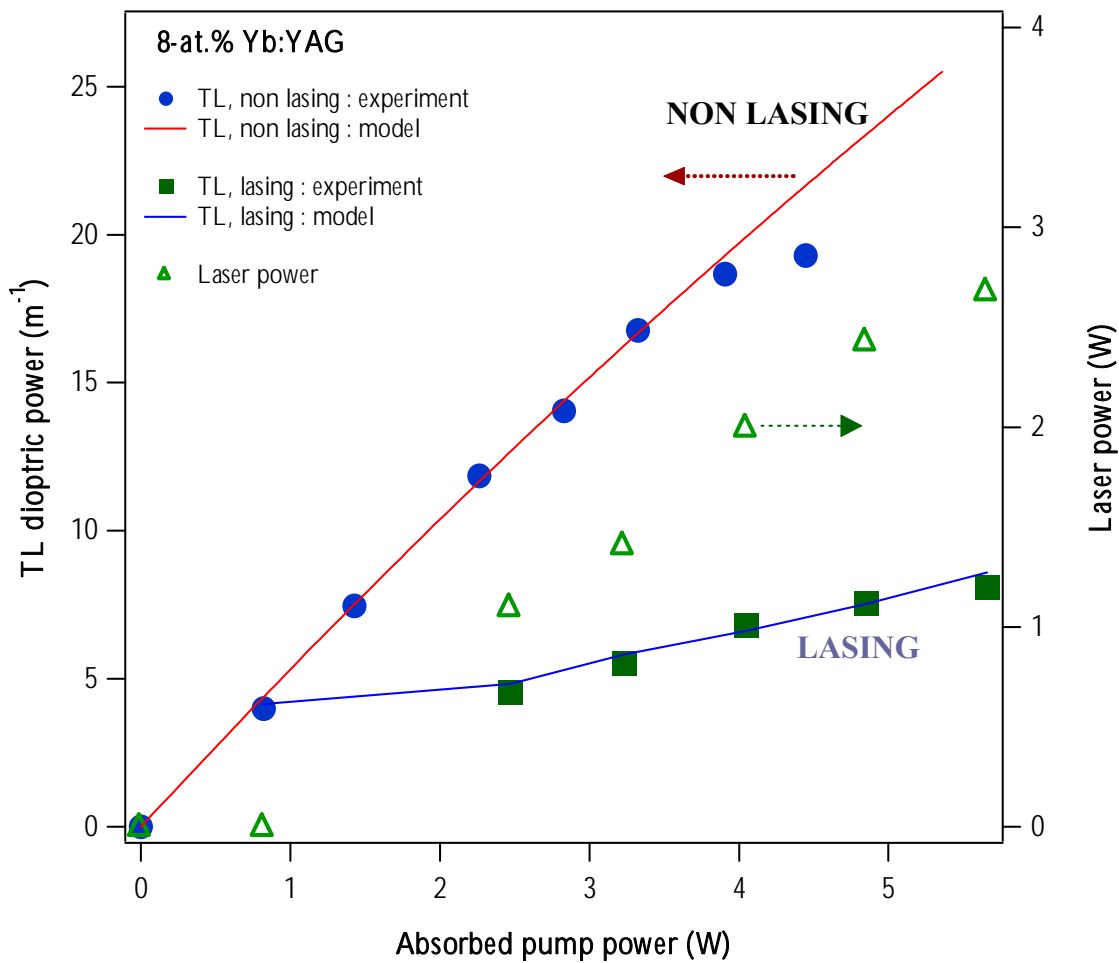

**figure 31**: *Thermal lens dioptric power (here, aberrations of the thermal lens were negligible) under lasing and nonlasing conditions. On the right (same graph) the measured laser power, useful to compute the laser extraction efficiency. (from [121])*

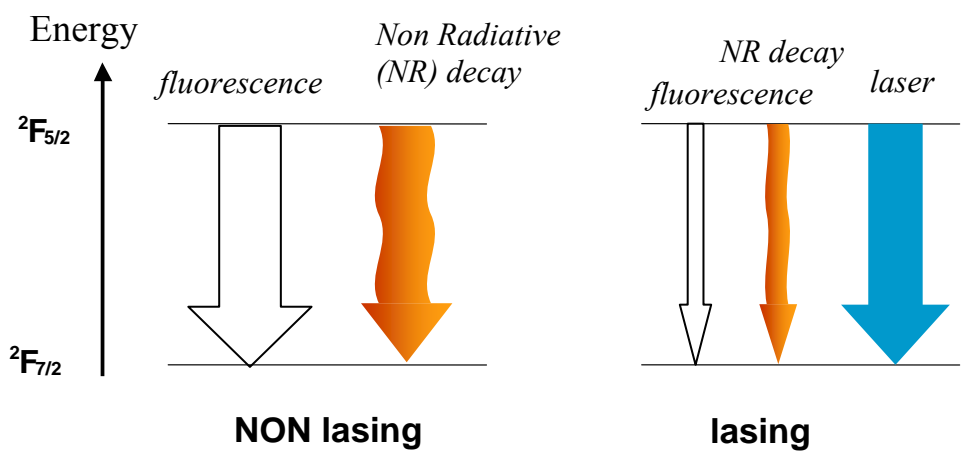

**figure 32** : *Simple qualitative explanation of the observed difference between thermal lens dioptric power under lasing and nonlasing conditions. When the laser is on, stimulated emission short-circuits the nonradiative path, causing the thermal load to be lower.*

It is seen that for a crystal like YAG, many different techniques have been used (for the details of each method, refer to the cited publications), and that a large dispersion of reported quantum efficiencies appears. This dispersion tends to assess the conjecture of concentration quenching as the nonradiative source in Yb-doped materials. This conjecture has been very alleviated in highly-doped Yb:YAG samples and was attributed to cooperative processes between two $Yb^{3+}$ ions towards $Yb^{2+}$ impurities [**122**]. Owing to the intrinsic nature of concentration quenching and the major role played by impurities, it is clear that the radiative quantum efficiency is a parameter that pertains to a single given sample, characterized by its doping concentration, the growth technique and its associated environment (in particular the nature of the crucible), and of course the degree of purity of the compounds.

**Table 5**: *Reported values of measured thermo-optic coefficients and radiative quantum efficiencies in the literature for different Yb-doped materials.*

| crystal | Mean fluorescence wavelength $\lambda_f$ (nm) | Thermo-optic coefficient ($\times 10^{-6}$ K$^{-1}$) *(measured, under diode-pumping conditions)* | Radiative quantum efficiency $\eta_r$ *(measured)* | Method used *(for quantum efficiency measurement)* | Reference |
|---|---|---|---|---|---|
| Yb:YAG | *1007* | *10.0 (from [121])* | *0.70* | *Thermal lensing (Shack-Hartmann)* | *[121]* |
| | | | *0.874* | *photometric* | *[126]* |
| | | | *0.835* | *calorimetric* | *[126]* |
| | | | *0.97* | *lifetime* | *[127]* |
| | | | *0.85* | *Direct temperature measurement* | *[125]* |
| Yb:GGG | *1013* | *31* | *0.90* | *Thermal lensing (Shack-Hartmann)* | *[121]* |
| Yb:GdCOB | *1011* | *6.5* | *0.71* | *idem* | *[121]* |
| Yb:YCOB | *1035* | *17* | *0.90* | *idem* | *[121]* |
| Yb:KGW | *993* | *7.5* | *0.96* | *idem* | *[121]* |
| Yb:YSO | *1001* | *15* | *0.89* | *idem* | *[121]* |
| Yb:YAB | *996* | *Non reported* | *0.88* | *Thermal lensing (lateral shearing)* | *[110]* |

### *V.3. Laser wavelength dependence on the thermal load in Yb-doped broadband materials: the example of Yb:Y$_2$SiO$_5$*

We saw that the fractional thermal load (eq. V.1) is dependant on the operating wavelength of the laser. In most practical circumstances however, this dependence is hidden by the fact that the laser extraction efficiency also greatly depends on the laser wavelength, since it is linked to the emission cross section. The Yb:$Y_2SiO_5$ (Yb:YSO [128]) crystal exhibits two maxima of comparable amplitude in its emission spectrum, at 1042 and 1058 nm respectively. In addition the output is naturally linearly polarized along the crystallophysic axis Y for both wavelengths. It is thus a good candidate to put clearly into evidence the influence of laser wavelength on thermal lensing. To perform the experiment, we added a SF6 dispersive prism cut at Brewster angle in the collimated arm of the three-mirror cavity appearing in figure 30, so that identical laser efficiencies were achieved at both wavelengths (2.1 Watts were obtained for 8.5 Watts of absorbed pump power). The results are shown in figure 33. It appears clearly that the thermal lens is weaker when the laser oscillates at 1042 nm, as expected since quantum defect is lower at this wavelength. The theoretical curve derived for the 1058-nm laser oscillation was obtained from the 1042-nm curve by just modifying the wavelength and the emission cross section in eq. (V.1) and (V.2), without any adjustable parameter (see [121]). This simple experiment shows the interest of multiwavelength thermal lensing measurements in broadband materials. Indeed, we have considered here a simple formulation of the fractional thermal loading (given by eq. V.1) which gives here satisfactory results; but it is possible, from the work of Patel *et al.* [**127**] for instance, to derive a more accurate expression of the fractional thermal load, which takes into account the probability of excitation transfer to a neighbouring ion **[129].** If measurements are performed versus absorbed power at different (more than 2) laser wavelengths, for which the laser extraction efficiency is also known, this means that we have the possibility to infer other spectroscopic parameters (such as the transfer probability to a neighbouring ion) involved in the expression of the thermal load.

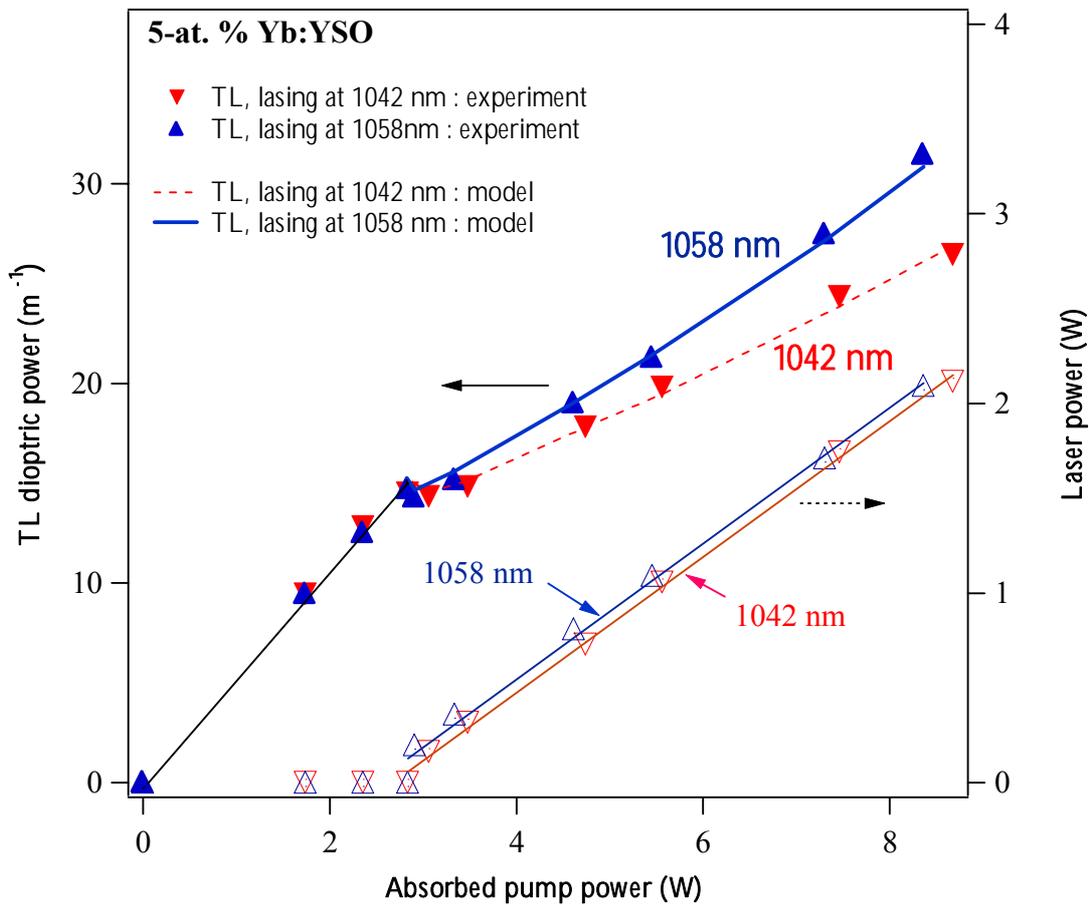

**Figure 33**: *Thermal lens dioptric power at 1042 and 1058 nm (left) and laser power (right in Yb:YSO. The effect of wavelength on quantum defect is clearly visible for this material.* (from [121])

## IV.4. The influence of the mean fluorescence wavelength on the thermal load: an illustration with Yb:KGW

The fact that the mean fluorescence wavelength affects the fractional thermal load is a very interesting feature of broadband Yb-doped materials. There are even some materials whose mean fluorescence wavelength is below the tail of the absorption spectrum. More precisely, according to eq (V.1) it is clear that if we may find a pump wavelength verifying:

$$\lambda_f < \eta_p \eta_r \lambda_p \qquad (V.3)$$

then the thermal load will be negative (in absence of laser extraction) and "radiative cooling" will be achieved. Such radiative cooling has been reported in Yb-doped ZBLANP glass in 1995 [130] and in a KGW crystal by Bowman *et al.* in 2002 [131]. The latter author has also proposed to use this phenomenon to realize a thermal load-free (radiation-balanced) laser [132]. The key idea is to correctly adjust the laser intensity so that the spontaneous emission rate, source of cooling providing the pump wavelength is long enough, exactly balances the stimulated emission rate, source of heating.

We show here how the mean fluorescence wavelength plays a key role in the interpretation of the results obtained when measuring the thermal lens in a Yb:KGW crystal. KGW is now a well-known crystal, suited for ultrafast laser applications [32-33, 17, 46, 74]. When used with wavevector k//c, the polarization-averaged mean fluorescence wavelength is 993 nm while the observed laser wavelength is 1030 nm and the pump wavelength tuned to the zero-line absorption peak, i.e. 979 nm. Results are shown in figure 34, where it can be seen that unlike previously reported results, thermal lensing is actually stronger under lasing conditions. A simple explanation is given with the schematic picture appearing in figure 35. The simple model suggested above fits well with experimental data and yields a high quantum efficiency for this sample (0,96), consistent with the fact that tungstate crystals are grown with the flux method, which is known to carry less impurities during growth than the Czochralski technique.

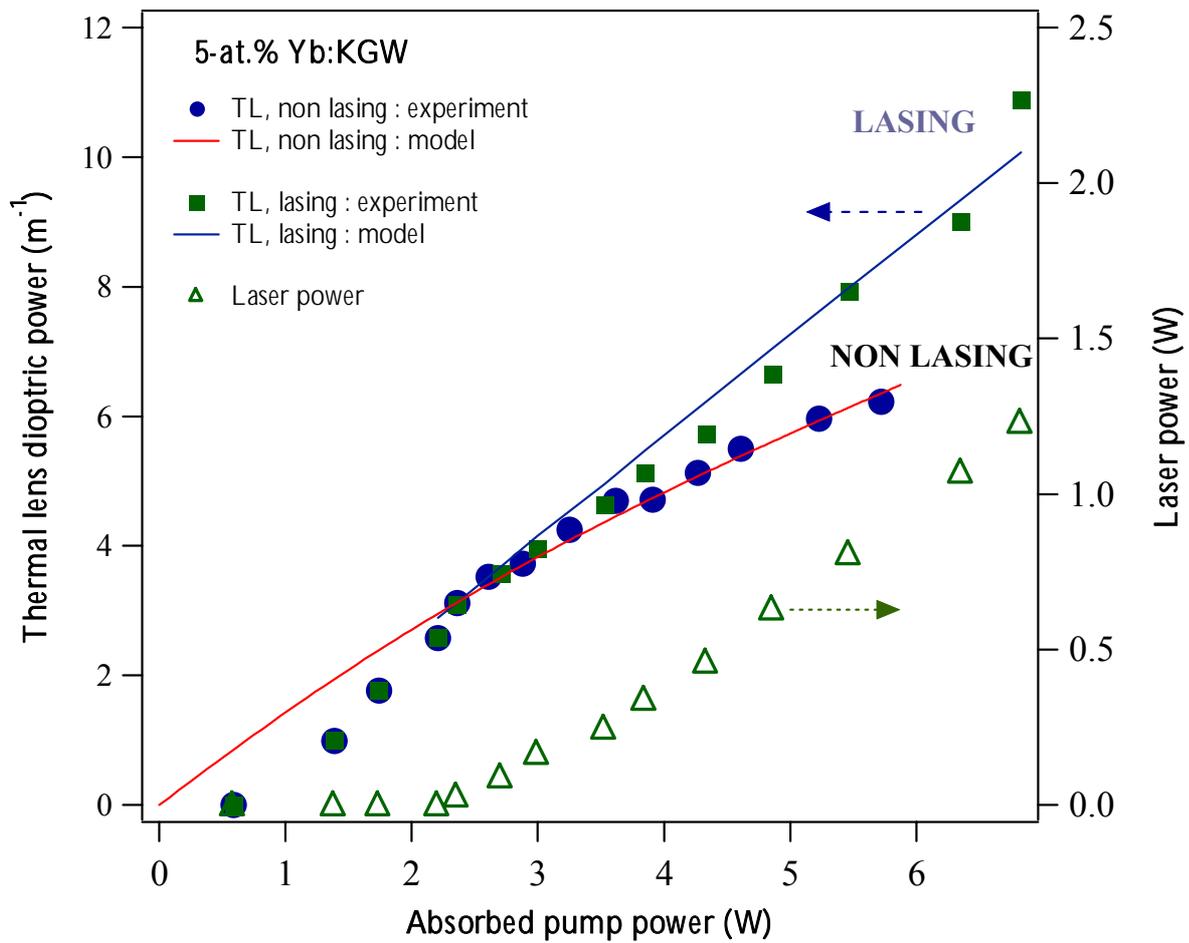

**figure 34**: *Thermal lensing measurements in Yb:KGW (from [121])*

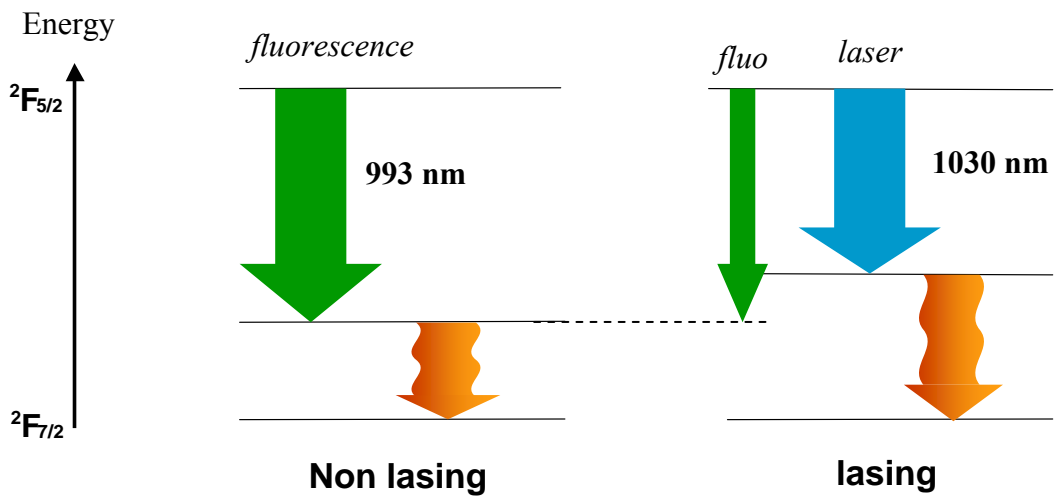

**Figure 35**: *Qualitative explanation of the paradoxical behaviour of the Yb:KGW crystal. Under nonlasing conditions (left), the thermal defect is low due to a low mean fluorescence wavelength (993 nm). When laser extraction at 1030 nm becomes the dominant way down for excitation, then the quantum defect is higher. This simple picture assumes that $\eta_r = 1$.*

## IV.5. Conclusion

As a conclusion for this final part of this paper, we have shown some examples of thermal lensing measurements, which allowed us to highlight several points of interest pertaining to Yb-doped laser media :

- all measurements show a difference between thermal lensing dioptric power with and without laser action: this provides the proof that in all the crystals under investigation (here YAG, GGG, YSO, GdCOB, YCOB, KGW, YAB) there is a nonradiative return path for excited state population. This is all the more detrimental for laser performance in pulsed regime, since in cw oscillators the laser extraction efficiency can be large enough to dissimulate this effect.

- Since Yb-doped materials have broader spectra than their Nd-doped counterparts, thermal lensing has two specific properties, which have been illustrated by two experiments:

1) the thermal load depends on the laser oscillation wavelength: the lower the wavelength and the lower the quantum defect;

2) the mean fluorescence wavelength $\lambda_f$ plays a key role. Materials exhibiting a low $\lambda_f$ will be less sensitive to heating under nonlasing conditions, and in KGW we saw that the thermal lens was even higher under lasing action.

The peculiarities of Yb-doped materials are not limited to these two latter facts. Chenais *et al.* [121] have reported the appearance of a roll-off in the thermal lens dioptric power at high pumping densities, under nonlasing conditions, with several different materials [121]. Furthermore, it has been observed by several groups ([110], [124]) a green luminescence that can be related to cooperative luminescence.

The detailed interpretation of these phenomena and their connection to the thermal load are still works in progress.

# VI. Conclusion

Ytterbium-doped materials have brought new prospects and deep changes in the area of high power solid state lasers. Associated to new and clever pumping concepts (fiber lasers, thin disk lasers, spinning disk lasers...) they are now well-established competitors of Nd-doped materials for high power applications. This paper has been an effort to make a review of the recently published works dealing with thermal effects in solid state lasers, with a particular scope on the special case of Yb-doped crystals.

In the first section we have presented the general properties of Ytterbium-doped media, and pointed out the crucial role of the matrix host on the properties of the laser material. The part II was devoted to a detailed presentation of the temperature distribution in a diode-pumped Yb-doped crystal: how to calculate it and how to measure it. We pointed out the importance of boundary conditions, and gave some practical information about the role of the thermal contacts in the temperature profile. We have shown that it was possible to easily include pump absorption saturation effects and pump beam divergence inside the crystal, exploiting the fact that the heat transfer coefficient towards end faces was far smaller than towards edge faces.

In the third part of this review we focussed on the thermo-optical properties and made a quite detailed study of the so-called thermal lensing phenomenon. A comprehensive understanding of this aspect requires a good knowledge about both thermo-mechanical and thermo-optical properties of the materials under consideration. This lead us to point out several inaccuracies reported in previous works, concerning the calculation of photoelastic constants for isotropic materials, and more generally the abusive employment of the dn/dT parameter when it is used to estimate the magnitude of the thermal lens: we have shown that the partial derivative of index with temperature taken at constant strain is the most appropriate figure, this is because the dn/dT is classically measured under experimental conditions that are not consistent with the usual situation of a diode-pumped crystal under mechanical stress. We proposed an alternative way to split the thermo-optic coefficient into three truly independent terms, and addressed in conclusion a schematic diagram of thermal effects showing how all the different apparent consequences (lensing, depolarization, strain-induced birefringence, astigmatism, fracture...) are connected to each other.

Given the high complexity of these thermo-optical phenomena, and the unfeasibility of precise calculations as far as all the properties of a crystal are not known (that is the case for the majority of laser crystals), we then focussed our attention on thermal lensing measurement techniques, which was the topic of part IV.

We presented in that section a review of what are, to our knowledge, the main different techniques that can be employed to measure thermal lensing in end-pumped laser media, and discussed their relative advantages and drawbacks.

Finally, we presented some examples of thermal lensing measurements that have been reported recently in Yb-doped crystals. All these measurements agree to find non-unity radiative quantum efficiencies for the Yb-doped materials under investigation. This non-ideal behaviour, presumably related to concentration quenching, provides contradiction to the general consideration that Yb-doped materials are totally free of deleterious nonradiative effects. We concluded this review by giving some examples of the influence of the laser operating wavelength on the thermal load, as well as the influence of the mean fluorescence wavelength.

**Acknowledgements:** We wish to acknowledge all the people involved in the realization of this review: at first Romain Gaumé (Stanford University) for fruitful discussions about issues related to dn/dT and characterization of materials; Bruno Viana, Gerard Aka and Daniel Vivien from the *Laboratoire de Chimie Appliquée de l'Etat Solide* (Ecole Nationale Supérieure de Chimie Paris, France), Alain Brenier and Georges Boulon from the *Laboratoire de Physico-Chimie des Materiaux Luminescents* (Lyon, France) and at last Bernard Ferrand from *CEA-LETI* (Grenoble, France), for growing the crystals used in most of the experimental results reported here. We gratefully acknowledge the *Ecole Supérieure d'Optique* (Orsay, France) for loaning us the Shack-Hartmann wavefront sensor as well as the infrared camera. The experiments related to our previous works have been done under the auspices of the CNRS and the *Délégation Générale de l'Armement* (DGA). We also thank Jean-Christophe Chanteloup and Judith Dawes for allowing us to reproduce some figures taken from their works.

# VII. <u>**Appendix:**</u> Calculation of the photoelastic constants $C_{r,\theta}$ and $C'_{r,\theta}$ using plane strain *and plane stress* approximations.

The $C_{r,\theta}$ constants are useful parameters to account for photoelastic effects in the most simple cases (optically isotropic crystals or glasses, and parabolic dependence of strain inside the crystal). When Koechner published for the first time the derivation of these constants in 1970, the term accounting for thermal dilatation in the generalized Hooke law was omitted [51]. In 1992, Cousins pointed out the mistake but did not correct, at that time, the expressions of the constants $C_{r,\theta}$, whose derivation however requires the use of the generalized Hooke law. In the last version of Koechner's reference book (*Solid State Laser engineering,* Springer Verlag ed.), the faulty equations were actually removed or corrected, but the formulation of the photoelastic constants $C_{r,\theta}$ remained unchanged since the first edition.

Because the generalized Hooke law is used to infer an expression of the constants, it means that their expression depends on which assumption has been made about stresses and strains. In Koechner's calculation, the plane strain approximation was made. In the case of end-pumping, it is well known however that the plane stress approximation is closer to reality. In this appendix we derive the expression of the photoelastic constants, within the framework of the two approximations.

## VII.1. Basic equations

*We restrict our discussion to cubic crystals. In this case, only the elasto-optical coefficients $p_{11}$, $p_{12}$, $p_{44}$ are nonzero. These coefficients are given in the [100] system linked to the crystal. In optically isotropic materials, the principal axes of strain are given by the cylindrical coordinate system axes (radial, tangential, axial). After a change of coordinates one obtains [77] :*

$$\frac{\partial n_r}{\partial \varepsilon_r} = -\frac{n_0^3}{12}\left[3p_{11} + 3p_{12} + 6p_{44}\right]$$

$$\frac{\partial n_r}{\partial \varepsilon_\theta} = -\frac{n_0^3}{12}\left[p_{11} + 5p_{12} - 2p_{44}\right] \quad\quad (A.1.)$$

$$\frac{\partial n_r}{\partial \varepsilon_z} = -\frac{n_0^3}{12}\left[2p_{11} + 4p_{12} - 4p_{44}\right]$$

for the radial index, and :

$$\frac{\partial n_\theta}{\partial \varepsilon_r} = -\frac{n_0^3}{12}[p_{11} + 5p_{12} - 2p_{44}]$$

$$\frac{\partial n_\theta}{\partial \varepsilon_\theta} = -\frac{n_0^3}{12}[3p_{11} + 3p_{12} + 6p_{44}] \quad (A.2.)$$

$$\frac{\partial n_\theta}{\partial \varepsilon_z} = -\frac{n_0^3}{12}[2p_{11} + 4p_{12} - 4p_{44}]$$

for the tangential index.

As stated by Shoji *et al.*[119], these expressions are valid only for propagation along the [111] axis.

For YAG we have :

$p_{11}$ = - 0.029

$p_{22}$ = + 0.0091

$p_{44}$ = - 0.0615.

The strains are related to stresses by the generalized Hooke law, which in our case (whatever the approximation we make) writes:

$$\varepsilon_r = \frac{1}{E}[\sigma_r - \nu(\sigma_\theta + \sigma_z)] + a_T(T - T_c)$$

$$\varepsilon_\theta = \frac{1}{E}[\sigma_\theta - \nu(\sigma_r + \sigma_z)] + a_T(T - T_c) \quad (A.3.)$$

$$\varepsilon_z = \frac{1}{E}[\sigma_z - \nu(\sigma_\theta + \sigma_r)] + a_T(T - T_c)$$

## *VII.2. Derivation of $C_{r,\theta}$ using the plane strain approximation (Koechner's case)*

We consider a homogeneously pumped rod. Only the radial dependence of stresses and temperature is considered, since the additional constants represent a constant phase shift which does not affect the phase profile.

We define:

$$S = \frac{a_T E}{16 K_c (1-\nu)}$$

and

$$Q = \frac{\eta_h P_{abs}}{\pi r_0^2 L}.$$

The stresses and temperature fields, within the plane strain approximation, are given by Koechner [69]:

$$\sigma_r(r) = QSr^2 + cte.$$
$$\sigma_\theta(r) = 3QSr^2 + cte \quad (A.4.)$$
$$\sigma_z(r) = 4QSr^2 + cte$$

$$T(r) = -\frac{Q}{4K_c} r^2 + cte. \quad (A.5.)$$

Omitting the constant terms, the generalized Hooke Law yields:

$$\varepsilon_z = \frac{1}{E}\left[\sigma_z - \nu(\sigma_\theta + \sigma_r)\right] + a_T T = \left[\frac{QS}{E} 4(1-\nu) - \frac{a_T Q}{4K_c}\right] r^2 = Q\left[\frac{a_T}{4K_c} - \frac{a_T}{4K_c}\right] r^2 = 0 \quad (A.6.)$$

which checks the plane strain condition. We have also:

$$\varepsilon_r = \frac{1}{E}\left[\sigma_r - \nu(\sigma_\theta + \sigma_z)\right] + a_T T = \left[\frac{QS}{E}(1-7\nu) - \frac{a_T Q}{4K_c}\right] r^2 = -3\frac{1+\nu}{E} QSr^2 \quad (A.7.)$$

$$\varepsilon_\theta = \frac{1}{E}\left[\sigma_\theta - \nu(\sigma_r + \sigma_z)\right] + a_T T = \left[\frac{QS}{E}(3-5\nu) - \frac{a_T Q}{4K_c}\right]r^2 = -\frac{1+\nu}{E}QSr^2 \qquad (A.8.)$$

The radial index variation is related to the strains by:

$$\Delta n_r = \frac{\partial n_r}{\partial \varepsilon_r}\varepsilon_r + \frac{\partial n_r}{\partial \varepsilon_\theta}\varepsilon_\theta + \frac{\partial n_r}{\partial \varepsilon_z}\varepsilon_z \qquad (A.9.)$$

and the $\frac{\partial n_i}{\partial \varepsilon_j}$ are given by (A.1) and (A.2).

Following Koechner, we can write the index variations under the form :

$$\Delta n_{r,\theta} = -\frac{1}{2}n_0^3 \frac{a_T Q}{K_c} C_{r,\theta}\, r^2 \qquad (A.10.)$$

With :

$$C_r = \frac{(1+\nu)(5p_{11} + 7p_{12} + 8p_{44})}{48(\nu - 1)} \qquad (A.11.)$$

$$C_\theta = \frac{(1+\nu)(p_{11} + 3p_{12})}{16(\nu - 1)} \qquad (A.12.)$$

For YAG, we find $C_r$ = + 0.020 et $C_\theta$ = + 1.77 $10^{-4}$. The valued computed by Koechner are: $C_r$ = 0.017 et $C_\theta$ = -0.0025 (with the same values taken for $p_{mn}$ and the same Poisson ratio: $\nu = 0.25$). The error made by Koechner is about 20%.

These coefficients have been extensively used for the purpose of evaluating depolarization losses in Nd:YAG. In this latter case, the relevant parameter is [60]:

$$C_B = \frac{1}{2}(C_\theta - C_r) \qquad (A.13.)$$

With (A.11) and (A.12), we find:

$$C_B = \frac{1+\nu}{48(1-\nu)}(p_{11} - p_{12} + 4p_{44}) \qquad (A.14.)$$

which is coincidentally the same formula obtained by Koechner from incorrect expressions of $C_r$ et $C_\theta$.

## VII.3. Derivation of $C'_{r,\theta}$ using the plane stress approximation (end pumping case)

We consider a thin disk, pumped by a top-hat pump beam profile of radius $w_p$ equal to the rod radius. This is also true in the conditions defined in Section III.4, provided that we are only concerned in the area $r < w_p$ and that integrated values of parameters along the crystal length are considered.

Let's define :

$$Q' = \frac{\eta_h P_{abs}}{\pi w_p^2}$$

$$S' = \frac{a_T E}{16 K_c}$$

The temperature distribution is given by:

$$\langle T(r) \rangle = -\frac{Q'}{4K_c} r^2 + cte. \tag{A.15.}$$

where we use Cousins' notation (see Part III.4), meaning that the bracketed quantity is integrated along the rod length L, and is then homogeneous to the said quantity times a length.

We have:

$$\begin{aligned} \langle \sigma_r(r) \rangle &= Q'S'r^2 + Cte \\ \langle \sigma_\theta(r) \rangle &= 3Q'S'r^2 + Cte \\ \langle \sigma_z(r) \rangle &= 0 \quad (plane\ stress) \end{aligned} \tag{A.16.}$$

From (A.3) we obtain:

$$\begin{aligned} \langle \varepsilon_r \rangle &= -3\frac{1+\nu}{E} Q'S'r^2 \\ \langle \varepsilon_\theta \rangle &= -\frac{1+\nu}{E} Q'S'r^2 \\ \langle \varepsilon_z \rangle &= -4\frac{1+\nu}{E} Q'S'r^2 \end{aligned} \tag{A.17.}$$

By analogy with (A.10.), the index shift can be written under the form:

$$\langle \Delta n_{r,\theta} \rangle = -\frac{1}{2} n_0^3 \frac{\alpha_T Q'}{K_c} C'_{r,\theta}\, r^2$$

which yields, using (A.1), (A.2), and (A.9):

$$C'_r = \frac{-(1+\nu)(9p_{11} + 15p_{12})}{48} \tag{A.18.}$$

$$C'_\theta = \frac{-(1+\nu)(7p_{11}+17p_{12}-8p_{44})}{48} \qquad (A.19.)$$

For YAG : **$C'_r$ = + 0.0032** et **$C'_\theta$ = - 0.011.**

The optical indicatrix does not deform the same way compared to the case of the long thin rod: here the tangential index becomes greater than $n_0$. An interesting feature is the stress-induced birefringence term, defined previously (eq. A.13):

$$C'_B = \frac{1+\nu}{48}(p_{11}-p_{12}+4p_{44}) \qquad (A.20.)$$

This expression differs from (A.14) only by a factor of 1-ν (=0.75 for YAG, but also for most materials).

It is of special interest to calculate the average value of the photoelastic constants:

$$\frac{C'_r + C'_\theta}{2} = -0.0039$$

which gives an order of magnitude of the role of photoelastic effect in the thermal lens.

Here, in the end pumping case, the thermal stresses yield a *divergent contribution to the thermal lens*, whereas it was convergent in the case of side-pumping.

See subsections 6 to 8 of Section III for a detailed discussion about the use of these parameters.

# References:

1. T.Y. Fan : "Diode-pumped Solid-State Lasers", in *Laser Sources and Applications*, Proc. of the 47[th] Scottish Universities Summer School in Physics, St Andrews, edited by A. Miller and D.Finlayson, SUSSP Publications & Institute of Physics Publishing, 1996.

2. D. Hanna and W. Clarkson : "A review of diode-pumped lasers", in *Advances in Lasers and applications,* Proc. of the 52[nd] Scottish Universities Summer School in Physics, St Andrews, edited by D. Finlayson and B. Sinclair, SUSSP Publications & Institute of Physics Publishing, 1999.

3. R.J. Keyes, T.M. Quist « Injection luminescent pumping of $CaF_2:U^{3+}$ with GaAs diode lasers », *Appl. Phys. Lett.* **4**, 50-52 (1964).

4. W. Streifer, D. Scifres, G. Harnagel, D. Welch, J. Berger, M. Sakamoto : "Advances in laser diode pumps", *IEEE J. Quant. Elec.* Vol. 24, pp. 883-894 (1988).

5. D. Brown : « Ultrahigh-average power diode-pumped Nd :YAG and Yb:YAG lasers », IEEE J. Quant. Elec., Vol. 33, No. 5, pp. 861-873 (1997).

6. W. Krupke : «Ytterbium Solid-State Lasers – The first decade», IEEE Journal On Sel. Topics in Quant. Elec. 6, 1287-1296 (2000).

7. P. Hardman, W. Clarkson, G. Friel, M. Pollnau, D. Hanna : "energy-transfer upconversion and thermal lensing in high-power end-pumped Nd:YLF laser crystals", *IEEE J. Quant. Elec*. , Vol. 35, No. 4, pp. 647-655 (1999).

8. T.Y. Fan : "Heat generation in Nd:YAG and Yb:YAG", *IEEE J. Quant. Elec*., Vol. 29, No. 6, pp. 1457-1459 (1993).

9. D. Sumida, A. Betin, H. Bruesselbach, R. Byren, S. Matthews, R. Reeder, M. Mangir : "Diode-pumped Yb:YAG catches up with Nd:YAG", *Laser Focus World*, Juin 1999, pp. 63-70.

10. L. Johnson, J. Geusic, L. Van Uitert : "Coherent oscillations from $Tm^{3+}$, $Ho^{3+}$, $Yb^{3+}$ and $Er^{3+}$ ions in yttrium aluminium garnet", *Appl. Phys. Lett.*, Vol. 7, No. 5, pp. 127-129 (1965).

11. R. Wynne, J. Daneu, T. Y. Fan : "thermal coefficients of the expansion and refractive index in YAG", *Appl. Opt.*, vol. 38, no. 15, pp. 3282-3284 (1999).

12. P. Lacovara, H. Choi, C. Wang, R. Aggrawal, T. Fan : « room-temperature diode-pumped Yb:YAG laser », *Opt. Lett.*, vol. 16 No. 14, pp.1089-1091 (1991).

13. G.A. Slack and D.W. Oliver : "Thermal conductivity of garnets and phonon scattering by rare-earth ions", *Phys. Rev. B*, **4**, 592-609 (1971).

14. F. Krausz, M.E. Fermann, T. Brabec, P. F. Curley, M. Hofer, M. H. Ober, C. Spielmann, E. Wintner, A. J. Schmidt, « Femtosecond Solid-State Lasers »,IEEE J. Quant. Elec. 28, 2097-2121 (1992).

15. U. Keller, « Ultrafast all-solid-state laser technology », Appl. Phys. B, 58, 347-364 (1994).

16. J. Aus der Au, D. Kopf, F. Morier-Genoud, M. Moser, U. Keller « 60-fs pulses from a diode-pumped Ndglass laser » Opt. Lett. 22 307 (1997).

17. F. Brunner, T. Sdmeyer, E. Innerhofer, F. Morier-Genoud, R. Paschotta, V. E. Kisel, V. G. Shcherbitsky, N. V. Kuleshov, J. Gao, K. Contag, A. Giesen, U. Keller «240-fs pulses with 22-W average power from a mode-locked thin-disk YbKY(WO$_4$)$_2$ laser», *Opt. Lett.* **27** 1162 (2002).

18. A. Courjaud, R. Maleck-Rassoul, N. Deguil, C. Hönninger, and F. Salin, "Diode pumped multikilohertz femtosecond amplifier", in OSA Trends in Optics and Photonics, Advanced Solid-State Lasers, 68, 121-123 (2002).

19. F. Druon, G.J. Valentine, S. Chénais, P. Raybaut, F. Balembois, P. Georges, A. Brun, A.J. Kemp, W. Sibbett, S. Mohr, D. Kopf, D.J.L. Birkin, D. Burns, A. Courjaud, C. Hönninger, F. Salin, R. Gaumé, A. Aron, G. Aka, B. Viana, C. Clerc, H. Bernas, «Diode-pumped femtosecond oscillator using ultra-broad Yb-doped crystals and modelocked using low-temperature grown or ion implanted saturable-absorber mirrors», *Euro. Phys. J.* **20**, 177-182 (2003).

20. F. Druon, F. Balembois, P. Georges, A. Brun, A. Courjaud, C. Honninger, F. Salin, A. Aron, F. Mougel, G. Aka, D. Vivien «Generation of 90-fs pulses from a mode-locked diode-pumped Yb$^{3+}$: Ca$_4$GdO(BO$_3$)$_3$ laser» *Opt. Lett.* **25**, 423-425 (2000).

21. F. Druon, S. Chénais, P. Raybaut, F. Balembois, P. Georges, R. Gaum, G. Aka, B. Viana, S. Mohr, D. Kopf «Diode-pumped Yb:Sr$_3$Y(BO$_3$)$_3$ femtosecond laser» *Opt. Lett.* **27**, 197-199 (2002).

22. F. Druon, S. Chénais, P. Raybaut, F. Balembois, P. Georges, R. Gaumé, P. H. Haumesser, B. Viana, D. Vivien, S. Dhellemmes, V. Ortiz, C. Larat «Apatite-structure crystal, Yb$^{3+}$:SrY$_4$(SiO$_4$)$_3$O, for the development of diode-pumped femtosecond lasers», *Opt. Lett.* **27**, 1914-1916 (2002).

23. F. Druon, F. Balembois, P. Georges « Laser crystals for the production of ultra-short laser pulses » *Ann. Chim. Mat.* **28**, 47-72 (2003).

24. C. Honninger, F. Morier-Genoud, M. Moser, U. Keller, L. R. Brovelli, and C. Harder, «Efficient and tunable diode-pumped femtosecond Yb:glass lasers», *Opt. Lett.* **23**, 126-128 (1998).

25. C. Honninger, R. Paschotta, M. Graf, F. Morier-Genoud, G. Zhang, M. Moser, S. Biswal, J. Nees, A. Braun, G. Mourou, I. Johannsen, A. Giesen, W. Seeber, and U. Keller, «Ultrafast ytterbium-doped bulk laser amplifiers» *Appl. Phys. B*, **69**, 3 (1999).

26. E. Innerhofer, T. Sdmeyer, F. Brunner, R. Hring, A. Aschwanden, R. Paschotta, C. Honninger, M. Kumkar, U. Keller «60-W average power in 810-fs pulses from a thin-disk YbYAG laser» *Opt. Lett.* **28**, 367-369 (2003).

27. Junji Kawanaka, Koichi Yamakawa, Hajime Nishioka, Ken-ichi Ueda «30-mJ, diode-pumped, chirped-pulse YbYLF regenerative amplifier » *Opt. Lett.* **28**, 2121-2123 (2003).

28. Peter Klopp, Valentin Petrov, Uwe Griebner, Klaus Petermann, Volker Peters, Götz Erbert « Highly efficient mode-locked Yb:Sc$_2$O$_3$ laser » *Opt. Lett.* 29, 391-393 (2004).

29. P. Klopp, V. Petrov, U. Griebner, and G. Erbert, "Passively mode-locked Yb:KYW laser pumped by a tapered diode laser," Opt. Express 10, 108-113 (2002). http://www.opticsexpress.org/abstract.cfm?URI=OPEX-10-2-108

30. Peter Klopp, Valentin Petrov, Uwe Griebner, Klaus Petermann, Volker Peters, Götz Erbert « Highly efficient mode-locked Yb:Sc$_2$O$_3$ laser» *Opt. Lett.* 29, 391-393 (2004).

31. H. Liu, S. Biswal, J. Paye, J. Nees, G. Mourou, C. Hnninger, U. Keller « Directly diode-pumped millijoule subpicosecond Ybglass regenerative amplifier » , *Opt. Lett.* 24, 917-919 (1999).

32. H. Liu, J. Nees, G. Mourou « Diode-pumped Kerr-lens mode-locked Yb:KY(WO$_4$)$_2$ laser» *Opt. Lett.* 26, 1723-1725 (2001).

33. Hsiao-Hua Liu, John Nees, Grard Mourou, « Directly diode-pumped Yb:KY(WO4).2 regenerative amplifiers » *Opt. Lett.* 27, 722-724 (2002).

34. P. Raybaut, F. Druon, F. Balembois, P. Georges, R. Gaumé, G. Aka, B. Viana «Directly diode-pumped oscillators and regenerative amplifiers for ultra-short pulse generation » Invited paper CThFF5 CLEO 2004.

35. Pierre Raybaut, Frederic Druon, Francois Balembois, Patrick Georges, Romain Gaumé, Bruno Viana, Daniel Vivien « Directly diode-pumped Yb$^{3+}$:SrY$_4$(SiO$_4$)$_3$O regenerative amplifier » *Opt. Lett.* 28, 2195-2196 (2003).

36. A. Shirakawa, K. Takaichi, H. Yagi, J. -. Bisson, J. Lu, M. Musha, K. Ueda, T. Yanagitani, T. S. Petrov, and A. A. Kaminskii, "Diode-pumped mode-locked Yb$^{3+}$:Y$_2$O$_3$ ceramic laser," *Opt. Express* 11, 2911-2916 (2003).

37. L. DeLoach, S. Payne, L. Chase, L. Smith, W. Kway, W. Krupke : "Evaluation of absorption and emission properties of $Yb^{3+}$-doped crystals for laser applications", *IEEE J. Quant. Elec.*, vol. 29, no. 4, pp. 1179-1091 (1984).

38. S.A. Payne, L.D. DeLoach, L.K. Smith W.L. Kway, J.B. Tassano, W.F. Krupke, B.H.T. Chai, G. Louts : "Ytterbium-doped apatite-structure crystals : a new class of laser materials ", *J. Appl. Phys* **76,** 497-503 (1994)

39. S. Payne, L. Smith, L. DeLoach, W. Kway, J. Tassano, W. Krupke : "Laser, optical, and thermomechanical properties of Yb-doped fluorapatite", *IEEE. J. Quant. Elec.*, vol. 30, No.1, pp. 170-179 (1994).

40. F. Mougel, G. Aka, A. Kahn-Harari, H. Hubert, J.M. Benitez, D. Vivien : "Infrared laser performance and self-frequency doubling of $Nd^{3+}$:$Ca_4GdO(BO_3)_3$ (Nd:GdCOB)", *Opt. Mat.* **8**, pp.161-173 (1997).

41. F. Druon, F. Augé, F. Balembois, P. Georges, A. Brun, A. Aron, F. Mougel, G. Aka, D. Vivien : « Efficient, tunable, zero-line diode-pumped, continuous-wave $Yb^{3+}$:$Ca_4LnO(BO_3)_3$ (*Ln* = Gd, Y) lasers at room temperature and application to miniature lasers », *J. Opt. Soc. Am. B,* Vol. 17, No. 1 (2000).

42. P.-H. Haumesser, R. Gaumé, B. Viana, D. Vivien : "Determination of laser parameters of ytterbium-doped oxide crystalline materials", *J. Opt. Soc. Am. B,* Vol. 19, No. 10, pp. 2365-2375 (2002).

43. R. Gaumé, P.-H. Haumesser, B. Viana, G. Aka, D. Vivien, E. Scheer, P. Bourdon, B. Ferrand, M. Jacquet, N. Lenain : « Spectroscopic properties and laser performance of $Yb^{3+}$:$Y_2SiO_5$, a new infrared laser material », OSA TOPS vol. 34, *Advanced Solid State Lasers*, p. 469, 2000.

44. S. Chénais, F. Druon, F. Balembois, P. Georges, R. Gaumé, B. Viana, D. Vivien, A. Brenier, G. Boulon : « Diode-pumped Yb:GGG laser : comparison with Yb:YAG », *Optical Materials* **22** pp.99-106 (2003).

45. N. V. Kuleshov, A. A. Lagatsky, V. G. Shcherbitsky, V. P. Mikhailov, E. Heumann, T. Jensen, A. Diening, G. Huber : "CW laser performance of Yb and Er, Yb doped tungstates", *Appl. Phys. B* **64**, pp.409-413 (1997).

46. A.A. Lagatsky, N.V. Kuleshov, V.P. Mikhailov: « diode-pumped CW lasing of Yb:KYW and Yb:KGW », *Opt. Comm.* **165**, pp.71-75 (1999)

47. A. Lucca, M. Jacquemet, F. Druon, F. Balembois, P. Georges, P. Camy,, J.L. Doualan, R. Moncorgé « high power diode-pumped CW laser operation of Yb :CaF$_2$ » post-deadline CPD-A1 CLEO (2004).

48. S. Jiang, M. Myers, D. Rhonehouse, U. Griebner, R. Koch, H. Schonnagel : "Ytterbium doped phosphate laser glasses", *Proceedings of SPIE* "Solid State Lasers VI", Vol. 2986 (1997).

49. J.F. Nye : *physical properties of crystals,* Clarendon Press, Oxford, 1985.

50. D. Brown : « Ultrahigh-average power diode-pumped Nd :YAG and Yb:YAG lasers », *IEEE J. Quant. Elec.*, Vol. 33, No. 5, pp. 861-873 (1997).

51. W. Koechner : « Absorbed Pump Power, Thermal Profile and Stresses in a cw Pumped Nd :YAG Crystal », *Appl. Opt.*, Vol. 9, No. 6, June 1970, 1429-1434

52. M. Schmid, T. Graf, and H. P. Weber, "Analytical model of the temperature distribution and the thermally induced birefringence in laser rods with cylindrically symmetric heating," *J. Opt. Soc. Amer. B*, vol. V17, no. 8, pp. 1398–1404, 2000.

53. U. Farrukh, A. Buoncristiani, C. Byvik : « an analysis of the temperature distribution in finite solid-state laser rods », *IEEE J. Quant. Elec.*, vol. 34, no. 11, pp. 2253-2263 (1988)

54. M. Innocenzi, H. Yura, C. Fincher, R. Fields : « Thermal modeling of continuous-wave end-pumped solid-state lasers », *Appl. Phys. Lett.* **56** (19), 7 may 1990, pp. 1831-1833

55. Y. Chen, T. Huang, C. Kao, C. Wang, S. Wang : « optimization in scaling fiber-coupled laser-diode end-pumped lasers to higher power : influence on thermal effect », *IEEE J. Quant. Elec.*, Vol. 33, No. 8, pp. 1424-1429 (1997)

56. F. Sanchez, M. Brunel, K Aït-Ameur : « Pump-saturation effects in end-pumped solid-state lasers », *J. Opt. Soc. Am. B*, Vol. 15, No. 9, pp. 2390-2394 (1998).

57. F. Augé, F. Druon, F. Balembois, P. Georges, A. Brun, F. Mougel, G. P. Aka, and D. Vivien, "Theoretical and experimental investigations of a diode-pumped quasi-three-level laser: The Yb –doped $Ca_4GdO(BO_3)_3$ (Yb:GdCOB) laser," *IEEE J. Quantum Electron.*, vol. 36, pp. 598–606, May 2000.

58. H.S. Carslaw, J.C. Jaeger : *conduction of heat in solids*, 2nd edition, Clarendon Press, Oxford, 1986.

59. Jacobs and Starr, *Rev. Sci. Instr.* **10**, 140 (1941).

60. A. Cousins: « temperature and thermal stress scaling in finite-length end-pumped laser rods» , *IEEE J. Quant. Elec.* , vol. 28, no. 4, pp.1057-1069 (1992).

61. J. Marion: "strengthened solid-state laser materials", *Appl. Phys. Lett.* **47** (7), pp. 94-96 (1985).

62. J. Marion: "Fracture of solid-state laser slabs", *J. Appl. Phys.* **60** (1), pp. 69-77 (1986).

63. J. Marion: "appropriate use of the strength parameter in solid-state slab laser design", *J. Appl. Phys.* **62** (5), pp. 1595-1604 (1987).

64. S. Ferré: « Caractérisation expérimentale et simulation des effets thermiques d'une chaîne laser ultra-intense à base de saphir dopé au titane », *thèse de doctorat de l'école polytechnique*, 2002.

65. H. Glur, R. Lavi, T. Graf: "Reduction of thermally induced lenses in Nd:YAG rods with low temperatures, IEEE J. of Quant. Elec., Vol. 40, No. 5, 499-504 (2004)

66. D. Kopf, K. Weingarten, G. Zhang, M. Moser, M. Emanuel, R. Beach, J. Skidmore, U. Keller :" High average power diode-pumped femtosecond Cr:LiSAF lasers", *Appl. Phys. B* **65**, pp. 235-243 (1997).

67. M. Tsunekane, N. Taguchi, H. Inaba : " reduction of thermal effects in a diode-end-pumped, composite Nd:YAG rod with a sapphire end", *Applied Optics*, vol. 37, no. 15, pp. 3290-3294 (1998).

68. A. Giesen, H. Hügel, A. Voss, K. Wittig, U. Brauch, H. Opower :"scalable concept for diode-pumped hgh-power solid-state lasers", *Appl. Phys. B* **58**, 365-372 (1994).

69. W. Koechner : *Solid State laser engineering*, 5$^{th}$ version, Springer, 1999.

70. W.A. Clarkson : « thermal effects and their mitigation in end-pumped solid-state lasers », *J. Phys. D : Appl. Phys.* **34** pp. 2381-2395 (2001)

71. *Handbook of Optics*, 2$^{nd}$ ed., vol. II (devices, measurements and properties), edited by M. Bass, E. Stryland, D. Williams, W. Wolfe, McGRAW-HILL, Inc., 2001.

72. B. Woods, S. Payne, J. Marion, R. Hugues, L. Davis : "thermomechanical and thermo-optical properties of the LiCaAlF$_6$:Cr$^{3+}$ laser material", *J. Opt. Soc. Am. B* Vol. 8, No. 5, pp. 970-977 (1991).

73. S. Chenais, F. Balembois, F. Druon, G. Lucas-leclin, P. Georges, "thermal lensing in Diode-pumped Ytterbium Lasers – Part I : Theoretical analysis and wavefront measurements", IEEE *J. Quant. Elec.* Vol 40, No. 9, 1217-1234, (2004)

74. A.A. Lagatsky, N.V. Kuleshov, V.P. Mikhailov: « diode-pumped CW lasing of Yb:KYW and Yb:KGW », *Opt. Comm.* **165**, pp.71-75 (1999)

75. F. Augé, F. Druon, F. Balembois, P. Georges, A. Brun, F. Mougel, G. P. Aka, D. Vivien : « Theoretical and experimental investigations of a diode-pumped quasi-three-level laser : the Yb$^{3+}$-doped Ca$_4$GdO(BO$_3$)$_3$ (Yb:GdCOB) laser » , *IEEE J. Quant. Elec.*, Vol. 36, No.5, pp. 598-606 (2000).

76. A. Aron, G. Aka, B. Viana, A. Kahn-Harari, D. Vivien, F. Druon, F. Balembois, P. Georges, A. Brun, N. Lenain, M. Jacquet : "Spectroscopic properties and laser performances of Yb:YCOB and potential of the Yb:LaCOB material", *Opt. Mat.* **16**, pp.181-188 (2001).

77. W. Koechner, D.K. Rice : "effect of birefringence on the performance of linearly polarized YAG:Nd lasers," *IEEE J. Quant. Elec.*, vol. 6, pp.557-566 (1990).

78. Scott W.C., De Wit M., *Appl Phys Lett.* vol. 18, pp.3-4 1971

79. Lu Q., Kugler N., Weber H., Dong S., Muller N. and Wittrock U., *Opt Quant Electron*, vol 28, pp. 57-69 (1996)

80. R. Fluck, M. Hermann, L. Hackel : "energetic and thermal performance of high-gain diode-side pumped Nd:YAG rods", *Appl. Phys. B* **70**, pp. 491-498 (2000).

81. S De Nicola, A Finizio, G Pierattini and G Carbonara "Interferometric method for concurrent measurement of the thermo-optic coefficients of quartz retarders" Pure Appl. Opt. 3 209-213 (1994)

82. Y-F. Tsay, B. Bendow, S. Mitra : "theory of the temperature derivative of the refractive index in transparent crystals", *Phys. Rev. B*., Vol. 8, No. 6, pp. 2688-2696 (1973).

83. J. Eichenholz, M. Richardson, : "measurement of thermal lensing in Cr$^{3+}$-doped colquirites", *IEEE J. Quant. Elec.*, vol. 34, no. 5, pp. 910-919 (1988)

84. C. Pfistner, R. Weber, H. Weber, S. Merazzi, R. Gruber : " Thermal beam distortions in end-pumped Nd:YAG, Nd:GSGG, and Nd:YLF Rods ", *IEEE. J. Quant. Elec.* vol. 30, no. 7, pp. 1605-1615 (1994).

85. T. Baer, W. Nighan, M. Keierstead : "modeling of end-pumped, solid-state lasers", in *Conference on Lasers and Electro-Optics*, Vol. 11 of 1993 OSA Technical Digest Series (Optical society of America, Washington D.C., 1993), p. 638.

86. K. Kleine, L. Gonzalez, R. Bhatia, L. Marshall,, D. Matthews : "High brightness Nd:YVO$_4$ laser for nonlinear optics", *Advanced Solid State Lasers*, M. Fejer, H. Injeyan and U. Keller eds., Vol. 26 of OSA Trends in Optics and Photonics Series (Optical Society of America, Washington D.C., 1999), pp. 157-158.

87. R. Weber, B. Neuenschwender, M. Macdonald, M.B. Roos, H.P. Weber : "Cooling schemes for longitudinally diode-laser pumped Nd:YAG rods", *IEEE J. Quant. Elec.* **34**, pp. 1046-1053 (1998).

88. X. Peng, A. Asundi, Y. Chen, Z. Xiong : "study of the mechanical properties of Nd:YVO$_4$ crystal by use of laser interferometry and finite-element analysis", *Appl. Opt.*, Vol. 40, No. 9, pp. 1396-1403 (2001).

89. A. Brignon, J.-P. Huignard, M. H. Garrett, and I. Mnushkina, "Spatial beam cleanup of a Nd:YAG laser operating at 1.06 μm with two-wave mixing in Rh:BaTiO3," Appl. Opt. 36, 7788-7793 (1997).

90. W. Xie, S. Tam, Y. Lam, Y. Kwon : "Diffraction losses of high power solid state lasers", *Opt. Comm*. **189**, pp. 337-343 (2001).

91. J. Frauchiger, P. Albers, H. Weber : "modeling of thermal lensing and higher order ring mode oscillation in end-pumped CW Nd:YAG lasers", *IEEE J. Quant. Elec*. , vol. 28, no. 4, pp. 1046-1056 (1992).

92. F. Druon, G. Chériaux, J. Faure, G. Vdovin, J.C. Chanteloup, J. Nees, M. Nantel, A. Maksimchuk, G. Mourou, *"Wave-front correction of femtosecond terawatt lasers using deformable mirrors", Optics Letters*, Vol. 23 No 1, pp. 1043-45, 1998

93. J. Gordon, R. Leite, R. Moore, S. Porto, J. Whinnery : "long-transient effects in lasers with inserted liquid samples", *J. Appl. Phys.* Vol. 36, no. 1, pp. 3-8 (1965).

94. D. Fournier, A.C. Boccara, N. Amer, R. Gerlach : « sensitive *in situ* trace-gas detection by photothermal deflection », *Appl. Phys. Lett.* **37** (6), pp. 519-521 (1980).

95. S. Bialkowski : "Photothermal Spectroscopy Methods for Chemical Analysis" *Volume 134 Chemical Analysis: A Series of Monographs on Analytical Chemistry and Its Applications,* , J. D. Winefordner, Series Editor, John Wiley & Sons, Inc. 1996

96. D. Burnham : "simple measurement of thermal lensing effects in laser rods", *Appl. Opt.* vol. 9, no. 7, pp.1727-1728 (1970).

97. C. Hu, J.R. Whinnery : "New thermooptical measurement method and a comparison with other methods", *Appl. Opt.* Vol. 12, No. 1, pp. 72-79 (1973).

98. R. Paugstadt, M. Bas : "method for temporally and spatially resolved thermal-lensing measurements", *Appl. Opt.* , vol. 33, no. 6, pp. 954-959 (1994).

99. R. Misra, P. Banerjee : "theoretical and experimental studies of pump-induced probe deflection in a thermal medium", *Appl. Opt.* , vol. 34, no. 18, pp.3358-3365 (1995).

100. H. Kogelnik, T. Li : "laser beams and resonators", *Appl. Opt.* vol. 5, pp. 1550 (octobre 1966).

101. B. Neuenschwander, R. Weber, H. Weber : "determination of the thermal lens in solid-state lasers with stable cavities", *IEEE. J. Quant. Elec.* , vol. 31, no. 6, pp. 1082-1087 (1995).

102. J. Frauchiger, P. Albers, H. Weber : "modeling of thermal lensing and higher order ring mode oscillation in end-pumped CW Nd:YAG lasers", *IEEE J. Quant. Elec.* , vol. 28, no. 4, pp. 1046-1056 (1992).

103. B. Ozygus, J. Erhard : "thermal lens determination of end-pumped solid-state lasers with transverse beat frequencies", *Appl. Phys. Lett.* **67** (10), pp. 1361-1362 (1995).

104. B. Ozygus, Q. Zhang : "thermal lens determination of end-pumped solid-state lasers using primary degeneration modes", *Appl. Phys. Lett.* **71** (18), pp. 2590-2592 (1997).

105. A. Cabezas, L. Komai, R. Treat : "dynamic measurements of phae shifts in laser amplifiers", *Appl. Opt.*, Vol. 5, No. 4, pp. 647-651 (1966).

106. H. Welling, C. Bickart : "spatial and temporal variation of the optical path length in flash-pumped laser rods", *J. Opt. Soc. Am*. Vol. 56, No. 5, pp. 611-618 (1966).

107. C. Pfistner, R. Weber, H. Weber, S. Merazzi, R. Gruber : " Thermal beam distortions in end-pumped Nd:YAG, Nd:GSGG, and Nd:YLF Rods ", *IEEE. J. Quant. Elec.* vol. 30, no. 7, pp. 1605-1615 (1994).

108. A. Khizhnyak, G. Galich, M. Lopiitchouk : "characteristics of thermal lens induced in active rod of cw Nd:YAG laser", *Semiconductor Physics ; Quantum Electronics & Optoelectronics (SQO),* Vol. 2, No. 1, pp. 147-152 (1999).

109. J. Blows, J. Dawes, T. Omatsu : "thermal lensing measurements in line-focus end-pumped neodymium yttrium aluminium garnet using holographic lateral shearing interferometry", *J. Appl. Phys.* , Vol. 83, No. 6, pp. 2901-2906 (1998).

110. J.L. Blows, P. Dekker, P. Wang , J.M. Dawes , T. Omatsu, "Thermal lensing measurements and thermal conductivity of Yb:YAB" Appl. Phys. B 76, 289-292 (2003)

111. J. Primot : "three-wave lateral shearing interferometer", *Appl. Opt.* , Vol 32, No. 31, pp.6242-6249 (1993). J. Primot : "three-wave lateral shearing interferometer", *Appl. Opt.* , Vol 32, No. 31, pp.6242-6249 (1993).

112. J-C. Chanteloup: « Multiple-wave lateral shearing interferometry for wave-front sensing », *Opt. Lett*., Vol. 23, No. 8, pp. 621-623 (1998)

113. J-C. Chanteloup, F. Druon, M. Nantel, A. Maksimchuk, G. Mourou : « single-shot wave-front measurements of high-intensity ultrashort laser pulses with a three-wave interferometer », *Applied Optics*, Vol. 44, No. 9, pp. 1559-1571 (2005)

114. L. Grossard, A. Desfarges-Berthelemot, B. Colombeau, C. Froehly : "iterative reconstruction of thermally induced phase distortion in a $Nd^{3+}$:$YVO_4$ laser", *J. Opt. A : Pure Appl. Opt.* **4** pp. 1-7 (2002).

115. J. Hartmann : "Bemerkungen uber den Bau und die Justirung von Spectrographen", *Zeitung Instrumentenkd.*, **20** p.47 (1900).

116. D. Armstrong, J. Mansell, D. Neal : "how to avoid beam distortion in solid-state laser design", *Laser Focus World*, décembre 1998, pp. 129-132.

117. S. Ito, H. Nagaoka, T. Miura, K. Kobayashi, A. Endo, K. Torizuka : " measurement of thermal lensing in a power amplifier of a terawatt Ti:sapphire laser", *Appl. Phys. B* **74**, pp. 343-347 (2002).

118. M. Pittman, S. Ferré, J.P. Rouseau, L. Notebaert, J.P. Chambaret, G. Chériaux : « design and characterization of a near-difraction-limited femtosecond 100-TW 10-Hz high-intensity laser system », *Appl. Phys. B* **74**, pp. 529-535 (2002).

119. I. Shoji, T. Taira : "Drastic reduction of depolarization resulting from thermally induced birefringence by use of a (110)-cut YAG crystal", OSA TOPS *Advanced Solid State Lasers* Vol. 68, pp.521-525, 2002.

120. A. Nashimura, K. Akaoka, A. Ohzu, T. Usami, "Temporal changes of thermal lens effects on highly-pumped Ytterbium glass by wavefront measurements", Journal of Nuclear Science and Technology, Vol. 38, No. 12, p. 1043-1047 (2001)

121. S. Chénais, F. Balembois, F . Druon, G. Lucas-Leclin, P. Georges, "Thermal Lensing in Diode-Pumped Ytterbium Lasers  - Part II: evaluation of quantum efficiencies and thermo-optic coefficients." IEEE J. Quantum Electronics Vol. 40 No 9 September, 1235-1243, 2004

122. D. F. de Sousa, N. Martynyuk, V. Peters, K. Lunstedt, K. Rademaker, K. Petermann, and S. Basun, "Quenching behavior of highly doped Yb:YAG and YbAG," in Conf. Lasers Electro-Optics Europe, Tech. Dig., Conf. Ed., 2003, CG1-3.

123. H. Yin, P. Deng, and F. Gan, "Defects in Yb:YAG," J. Appl. Phys., vol. 83, no. 8, pp. 3825–3828, 1998.

124. P. Yang, P. Deng, and Z. Yin, "Concentration quenching in Yb:YAG," *J. Lumin.*, pp. 51–54, 97.

125. M. O. Ramirez, D. Jaque, L. E. Bausa, J. A. S. Garcia, and J. G. Solé, "Thermal loading in highly efficient diode pumped ytterbium doped lithium niobate lasers," presented at the Conf. Lasers and Electro-Optics Europe 2003 (CLEO Europe), Münich, Germany, 2003.

126. N. Barnes, B. Walsh : "Quantum efficiency measurements of Nd:YAG, Yb:YAG, and Tm:YAG", OSA TOPS *Advanced Solid State Lasers* Vol. 68, pp.284-287, 2002.

127. F. Patel, E. Honea, J. Speth, S. Payne, R. Hutcheson, R. Equall : "Laser demonstration of $Yb_3Al_5O_{12}$ (YbAG) and materials properties of highly doped Yb:YAG", *IEEE. J. Quant. Elec.*, vol. 37, no. 1, pp. 135-144 (2001).

128. Jacquemet, M.; Jacquemet, C.; Janel, N.; Druon, F.; Balembois, F.; Georges, P.; Petit, J.; Viana, B.; Vivien, D.; Ferrand, B. "Efficient laser action of Yb:LSO and Yb:YSO oxyorthosilicates crystals under high-power diode-pumping", Applied Physics B, Volume 80, Issue 2, pp.171-176 (2005)

129. R. Gaumé, "Relations structure-propriétés dans les lasers solides de puissance à l'ytterbium. élaboration et caractérization de nouveaux matériaux et de cristaux composites soudés par diffusion," Ph.D. dissertation, Pierre et Marie Curie Univ., Paris, VI, France, 2002

130. R. Epstein *et. al* : « Observation of laser-induced fluorescent cooling of a solid », *Nature*, vol. 377, pp.500-503, 12 oct. 1995.

131. S. Bowman, N. Jenkins, B. Feldman, S. O'Connor: "Demonstration of a radiatively cooled laser", in *Proceedings of Conference on Lasers and Electro-Optics (CLEO 2002)*, Long Beach, CA, June 2002, p. 180.

132. S. Bowman : " lasers without internal heat generation", *IEEE J. Quant. Elec.*, vol. 35, no. 1, 115-122 (1999)